\documentclass[twocolumn]{aastex701}

\DeclareMathSymbol{\ast}{\mathbin}{symbols}{"03}

\newcommand{\OIII}{\hbox{{\rm [O}\kern 0.1em{\sc iii}{\rm ]}}}

\usepackage{xspace}
\let\oldAA\AA
\renewcommand{\AA}{\text{\oldAA}\xspace}

\begin{document}
%\title{The hodgepodge of physical properties of little red dots determined by JWST/NIRSpec spectroscopy}
%\title{The same but different: diverse spectroscopic flavors (and natures) of photometrically selected Little Red Dots}
%\title{The same but different: diverse flavors (and natures) of  Little Red Dots based on NIRSpec prism spectroscopy and MIRI imaging}
%\title{Little Red Dots: one observational tag, diverse spectroscopic flavors with distinct stellar and nuclear activity features}
\title{Little Red Dots: One Photometric Tag Concealing Diverse Spectroscopic Flavors of Massive Star Formation and Black Hole Activity}
\shorttitle{The same but different: NIRSpec+MIRI diversity of LRDs}

\author[0000-0003-4528-5639]{Pablo G. P\'erez-Gonz\'alez}
\affiliation{Centro de Astrobiolog\'{\i}a (CAB), CSIC-INTA, Ctra. de Ajalvir km 4, Torrej\'on de Ardoz, E-28850, Madrid, Spain}
\email{pgperez@cab.inta.csic.es}

\author[0000-0001-6813-875X]{Guillermo Barro}
\affiliation{University of the Pacific, Stockton, CA 90340 USA}
\email{gbarro@pacific.edu}

\author[0000-0002-6719-380X]{Stefano Carniani}
\affiliation{Scuola Normale Superiore, Piazza dei Cavalieri 7, I-56126 Pisa, Italy}
\email{stefano.carniani@sns.it}

\author[0000-0003-2388-8172]{Francesco D'Eugenio}
\affiliation{Kavli Institute for Cosmology, University of Cambridge, Madingley Road, Cambridge, CB3 0HA, UK}
\affiliation{Cavendish Laboratory, University of Cambridge, 19 JJ Thomson Avenue, Cambridge, CB3 0HE, UK}
\email{fd391@cam.ac.uk}

\author[0000-0003-2303-6519]{George H. Rieke}
\affiliation{Steward Observatory, University of Arizona, 933 North Cherry Avenue, Tucson, AZ 85721, USA}
\email{grieke@as.arizona.edu}

\author[0000-0002-9909-3491]{Roberta Tripodi}
\affiliation{INAF – Osservatorio Astronomico di Roma, via Frascati 33, 00078, Monteporzio Catone, Italy}
\email{roberta.tripodi@inaf.it}

\author[0000-0002-8651-9879]{Andrew J.\ Bunker}
\affiliation{Department of Physics, University of Oxford, Denys Wilkinson Building, Keble Road, Oxford OX1 3RH, UK}
\email{Andy.Bunker@physics.ox.ac.uk}

\author[0000-0002-1660-9502]{Xihan Ji}
\affiliation{Kavli Institute for Cosmology, University of Cambridge, Madingley Road, Cambridge, CB3 0HA, UK}
\affiliation{Cavendish Laboratory, University of Cambridge, 19 JJ Thomson Avenue, Cambridge, CB3 0HE, UK}
\email{xj274@cam.ac.uk}

\author[0000-0001-8442-1846]{Rui Marques-Chaves}
\affiliation{Department of Astronomy, University of Geneva, 51 Chemin Pegasi, 1290, Versoix, Switzerland}
\email{Rui.MarquesCoelhoChaves@unige.ch}

\author[0000-0001-7144-7182]{Daniel Schaerer}
\affiliation{Department of Astronomy, University of Geneva, 51 Chemin Pegasi, 1290, Versoix, Switzerland}
\email{daniel.schaerer@unige.ch}

\author[0000-0001-8349-3055]{Giacomo Venturi}
\affiliation{Scuola Normale Superiore, Piazza dei Cavalieri 7, I-56126 Pisa, Italy}
\email{giacomo.venturi1@sns.it}

\author[0009-0005-5968-8553]{Flor Ar\'evalo-Gonz\'alez}
\affiliation{INAF – Osservatorio Astronomico di Roma, via Frascati 33, 00078, Monteporzio Catone, Italy}
\email{flor.arevalogonzalez@inaf.it}

\author[0000-0001-7997-1640]{Santiago Arribas}
\affiliation{Centro de Astrobiolog\'{\i}a (CAB), CSIC-INTA, Ctra. de Ajalvir km 4, Torrej\'on de Ardoz, E-28850, Madrid, Spain}
\email{arribas@cab.inta-csic.es}

\author[0000-0002-5104-8245]{Pierluigi Rinaldi}
\affiliation{Space Telescope Science Institute, 3700 San Martin Drive, Baltimore, Maryland 21218, USA}
\email{prinaldi@stsci.edu}

\author[0000-0001-5171-3930]{Bruno Rodríguez Del Pino}
\affiliation{Centro de Astrobiolog\'{\i}a (CAB), CSIC-INTA, Ctra. de Ajalvir km 4, Torrej\'on de Ardoz, E-28850, Madrid, Spain}
\email{brodriguez@cab.inta-csic.es}

\author[0000-0002-7595-121X]{Joris Witstok}
\affiliation{Cosmic Dawn Center (DAWN), Copenhagen, Denmark
}
\affiliation{Niels Bohr Institute, University of Copenhagen, Jagtvej 128, DK-2200, Copenhagen, Denmark}
\email{joris.witstok@nbi.ku.dk}

\author[0000-0003-0883-2226]{Rachana Bhatawdekar}
\affiliation{European Space Agency (ESA), European Space Astronomy Centre (ESAC), Camino Bajo del Castillo s/n, 28692 Villanueva de la Cañada, Madrid, Spain}
\email{Rachana.Bhatawdekar@esa.int}

\author[0000-0002-3952-8588]{Leindert A. Boogaard} \affiliation{Leiden Observatory, Leiden University, PO Box 9513, NL-2300 RA Leiden, The Netherlands}
\email{boogaard@strw.leidenuniv.nl}

\author[0000-0003-3458-2275]{Stephane Charlot}
\affiliation{Sorbonne Universit\'e, CNRS, UMR 7095, Institut d'Astrophysique de Paris, 98 bis bd Arago, 75014 Paris, France}
\email{charlot@iap.fr}

\author[0000-0002-7636-0534]{Jacopo Chevallard}
\affiliation{Department of Physics, University of Oxford, Denys Wilkinson Building, Keble Road, Oxford OX1 3RH, UK}
\email{chevallard@iap.fr}

\author[0000-0001-6820-0015]{Luca Costantin}
\affiliation{Centro de Astrobiolog\'{\i}a (CAB), CSIC-INTA, Ctra. de Ajalvir km 4, Torrej\'on de Ardoz, E-28850, Madrid, Spain}
\email{lcostantin@cab.inta-csic.es}

\author[0000-0002-2678-2560]{Mirko Curti}
\affiliation{European Southern Observatory, Karl-Schwarzschild-Strasse 2, 85748 Garching, Germany}
\email{mirko.curti@eso.org}

\author[0000-0002-9551-0534]{Emma Curtis-Lake}
\affiliation{Centre for Astrophysics Research, Department of Physics, Astronomy and Mathematics, University of Hertfordshire, Hatfield AL10 9AB, UK}
\email{e.curtis-lake@herts.ac.uk}

\author[0000-0002-3331-9590]{Emanuele Daddi}
\affiliation{CEA, IRFU, DAp, AIM, Universit\'e Paris-Saclay, Universit\'e Paris Cit\'e, Sorbonne Paris Cit\'e, CNRS, 91191 Gif-sur-Yvette, France}
\email{edaddi@cea.fr}

\author[0000-0001-8047-8351]{Kelcey Davis}
\affiliation{Department of Physics, 196A Auditorium Road, Unit 3046, University of Connecticut, Storrs, CT 06269, USA}
\affiliation{Los Alamos National Laboratory, Los Alamos, NM 87545, USA}
\email{kelcey.davis@uconn.edu}

\author[0000-0003-0699-6083]{Tanio D\'{\i}az-Santos}
\affiliation{Institute of Astrophysics, Foundation for Research and Technology--Hellas (FORTH), Heraklion, 70013, Greece}
\affiliation{School of Sciences, European University Cyprus, Diogenes Street, Engomi, 1516, Nicosia, Cyprus}
\email{tanio@ia.forth.gr}

\author[0000-0001-5414-5131]{Mark Dickinson}
\affiliation{NSF National Optical-Infrared Astronomy Research Laboratory, 950 North Cherry Ave., Tucson, AZ 85719, USA}
\email{mark.dickinson@noirlab.edu}

\author[0000-0002-7622-0208]{Callum T. Donnan}
\affiliation{NSF's National Optical-Infrared Astronomy Research Laboratory, 950 North Cherry Ave., Tucson, AZ 85719, USA}
\email{callum.donnan@noirlab.edu}

\author[orcid=0000-0002-6460-3682]{Fergus R. Donnan}
\affiliation{Department of Astrophysics, University of California San Diego, 9500 Gilman Drive, San Diego, CA 92093, USA}
\email[]{fdonnan@ucsd.edu}

\author[0000-0002-1404-5950]{James S. Dunlop}
\affiliation{Institute for Astronomy, University of Edinburgh, Royal Observatory, Edinburgh, EH9 3HJ, UK}
\email{James.Dunlop@ed.ac.uk}

\author[0000-0002-2929-3121]{Daniel J.\ Eisenstein}
\affiliation{Center for Astrophysics $|$ Harvard \& Smithsonian, 60 Garden St., Cambridge MA 02138 USA}
\email{deisenstein@cfa.harvard.edu}

\author[0000-0001-7113-2738]{Henry C. Ferguson}
\affiliation{Space Telescope Science Institute, Baltimore, MD, USA}
\email{ferguson@stsci.edu}

\author[0000-0002-7714-688X]{Rom\'an Fern\'andez Aranda}
\affiliation{Centro de Astrobiolog\'{\i}a (CAB), CSIC-INTA, Ctra. de Ajalvir km 4, Torrej\'on de Ardoz, E-28850, Madrid, Spain}
\email{rfernandez@cab.inta-csic.es}

\author[0000-0001-8519-1130]{Steven L. Finkelstein}
\affiliation{Department of Astronomy, The University of Texas at Austin, Austin, TX, USA}
\affiliation{Cosmic Frontier Center, The University of Texas at Austin, Austin, TX, USA}
\email{stevenf@astro.as.utexas.edu}

\author[0000-0001-7201-5066]{Seiji Fujimoto}
\affiliation{David A. Dunlap Department of Astronomy and Astrophysics, \ University of Toronto, 50 St. George Street, Toronto, Ontario, M5S 3H4, Canada}
\affiliation{Dunlap Institute for Astronomy and Astrophysics, 50 St. George Street, Toronto, Ontario, M5S 3H4, Canada}
\email{seiji.fujimoto@utoronto.ca}

\author[0000-0003-3248-5666]{Giovanni Gandolfi}
\affiliation{INAF – Osservatorio Astronomico di Roma, via Frascati 33, 00078, Monteporzio Catone, Italy}
\email{giovanni.gandolfi@inaf.it}

\author[0000-0002-7831-8751]{Mauro Giavalisco}
\affiliation{University of Massachusetts Amherst, 710 North Pleasant Street, Amherst, MA 01003-9305, USA}
\email{mauro@umass.edu}

\author[0000-0001-9440-8872]{Norman A. Grogin}
\affiliation{Space Telescope Science Institute, Baltimore, MD, USA}
\email{nagrogin@stsci.edu}

\author[0000-0001-9626-9642]{Mahmoud Hamed}
\affiliation{Centro de Astrobiolog\'{\i}a (CAB), CSIC-INTA, Ctra. de Ajalvir km 4, Torrej\'on de Ardoz, E-28850, Madrid, Spain}
\email{mhamed@cab.inta-csic.es}

\author[0000-0002-3301-3321]{Michaela Hirschmann}
\affiliation{Institute of Physics, Laboratory of Galaxy Evolution, Ecole Polytechnique Federale de Lausanne (EPFL), Observatoire de Sauverny, 1290 Versoix, Switzerland}
\email{michaela.hirschmann@epfl.ch}

\author[0000-0001-9187-3605]{Jeyhan S. Kartaltepe}
\affiliation{Laboratory for Multiwavelength Astrophysics, School of Physics and Astronomy, Rochester Institute of Technology, 84 Lomb Memorial Drive, Rochester, NY 14623, USA}
\email{jeyhan@astro.rit.edu}

\author[0000-0002-8360-3880]{Dale D. Kocevski}
\affiliation{Department of Physics and Astronomy, Colby College, Waterville, ME 04901, USA}
\email{dkocevsk@colby.edu}

\author[0000-0002-9393-6507]{Gene C. K. Leung}
\affiliation{MIT Kavli Institute for Astrophysics and Space Research, 77 Massachusetts Ave., Cambridge, MA 02139, USA}
\email{gckleung@mit.edu}

\author[0000-0003-0532-6213]{Cristina M. Lofaro}
\affiliation{Institute of Astrophysics, Foundation for Research and Technology–Hellas (FORTH), Heraklion, GR-70013, Greece}
\affiliation{Department of Physics, University of Crete, 70013, Heraklion, Greece}
\email{clofaro@ia.forth.gr}

\author[0000-0003-1581-7825]{Ray A. Lucas}
\affiliation{Space Telescope Science Institute, 3700 San Martin Drive, Baltimore, MD 21218, USA}
\email{lucas@stsci.edu}

\author[0000-0002-6221-1829]{Jianwei Lyu}
\affiliation{Steward Observatory, University of Arizona,
933 North Cherry Avenue, Tucson, AZ 85721, USA}
\email{jianwei@arizona.edu}

\author[0000-0003-4368-3326]{Derek J. McLeod}
\affiliation{Institute for Astronomy, University of Edinburgh, Royal Observatory, Edinburgh, EH9 3HJ, UK}
\email{derek.mcleod@ed.ac.uk}

\author[0000-0003-0470-8754]{Jens Melinder}
\affiliation{Department of Astronomy, Stockholm University, Oscar Klein Centre, AlbaNova University Centre, 106 91 Stockholm, Sweden}
\email{jens@astro.su.se}

\author[0000-0002-3005-1349]{G\"oran {\"O}stlin}
\affiliation{Department of Astronomy, Stockholm University, Oscar Klein Centre, AlbaNova University Centre, 106 91 Stockholm, Sweden}
\email{ostlin@astro.su.se}

\author[0000-0001-7503-8482]{Casey Papovich}
\affiliation{Department of Physics and Astronomy, Texas A\&M University, College Station, TX, 77843-4242 USA}
\affiliation{George P.\ and Cynthia Woods Mitchell Institute for Fundamental Physics and Astronomy, Texas A\&M University, College Station, TX, 77843-4242 USA}
\email{lucas@stsci.edu}

\author[0000-0001-8940-6768]{Laura Pentericci}
\affiliation{INAF, Osservatorio Astronomico di Roma, Via Frascati 33, 00078 Monteporzio Catone, Roma, Italy}
\email{laura.pentericci@inaf.it}

\author[0000-0002-0939-9156]{Borja P\'erez-D\'iaz}
\affiliation{INAF – Osservatorio Astronomico di Roma, via Frascati 33, 00078, Monteporzio Catone, Italy}
\email{borja.perezdiaz@inaf.it}

\author[0000-0002-7893-6170]{Marcia Rieke}
\affiliation{Steward Observatory, University of Arizona, 933 North Cherry Avenue, Tucson, AZ 85721, USA}
\email{mrieke@as.arizona.edu}

\author[0000-0001-6010-6809]{Jan Scholtz}
\affiliation{Kavli Institute for Cosmology, University of Cambridge, Madingley Road, Cambridge, CB3 0HA, UK}
\affiliation{Cavendish Laboratory, University of Cambridge, 19 JJ Thomson Avenue, Cambridge, CB3 0HE, UK}
\email{js2685@cam.ac.uk}

\author[0000-0003-4702-7561]{Irene Shivaei}
\affiliation{Centro de Astrobiolog\'{\i}a (CAB), CSIC-INTA, Ctra. de Ajalvir km 4, Torrej\'on de Ardoz, E-28850, Madrid, Spain}
\email{ishivaei@cab.inta-csic.es}

\author[0000-0002-6748-6821]{Rachel S. Somerville}
\affiliation{Center for Computational Astrophysics, Flatiron Institute, 162 5th Avenue, New York, NY 10010, USA}
\email{rsomerville@flatironinstitute.org}

 \author[0000-0002-0827-9769]{Thomas M. Stanton}
\affiliation{Institute for Astronomy, University of Edinburgh, Royal Observatory, Edinburgh, EH9 3HJ, UK}
\email{t.stanton@ed.ac.uk}

\author[0000-0001-5642-752X]{Struan D. Stevenson}
\affiliation{Institute for Astronomy, University of Edinburgh, Royal Observatory, Edinburgh, EH9 3HJ, UK}
\email{struan.stevenson@ed.ac.uk}

\author[0000-0002-8224-4505]{Sandro Tacchella}
\affiliation{Kavli Institute for Cosmology, University of Cambridge, Madingley Road, Cambridge, CB3 0HA, UK}
\affiliation{Cavendish Laboratory, University of Cambridge, 19 JJ Thomson Avenue, Cambridge, CB3 0HE, UK}
\email{st578@cam.ac.uk}

\author[0000-0002-1410-0470]{Jonathan R. Trump}
\affil{Department of Physics, 196A Auditorium Road, Unit 3046, University of Connecticut, Storrs, CT 06269, USA}
\email{jonathan.trump@uconn.edu}

\author[0000-0003-4891-0794]{Hannah \"Ubler}
\affiliation{Max-Planck-Institut f\"ur extraterrestrische Physik (MPE), Gie{\ss}enbachstra{\ss}e 1, 85748 Garching, Germany}
\email{hannah@mpe.mpg.de}

\author[0000-0002-9373-3865]{Xin Wang}
\affiliation{School of Astronomy and Space Science, University of Chinese Academy of Sciences (UCAS), Beijing 100049, China}
\affiliation{National Astronomical Observatories, Chinese Academy of Sciences, Beijing 100101, China}
\affiliation{Institute for Frontiers in Astronomy and Astrophysics, Beijing Normal University, Beijing 102206, China}
\email{xwang@ucas.ac.cn}

\author[0000-0003-2919-7495]{Christina C.\ Williams}
\affiliation{NSF National Optical-Infrared Astronomy Research Laboratory, 950 North Cherry Avenue, Tucson, AZ 85719, USA}
\email{christina.williams@noirlab.edu}

\author[0000-0001-9262-9997]{Christopher N.\ A.\ Willmer}
\affiliation{Steward Observatory, University of Arizona, 933 North Cherry Avenue, Tucson, AZ 85721, USA}
\email{cnaw@as.arizona.edu}

\author[0000-0003-3466-035X]{{L. Y. Aaron} {Yung}}
\affiliation{Space Telescope Science Institute, 3700 San Martin Drive, Baltimore, MD 21218, USA}
\email{yung@stsci.edu}

\author[0000-0003-3307-7525]{Yongda Zhu}
\affiliation{Steward Observatory, University of Arizona, 933 North Cherry Avenue, Tucson, AZ 85721, USA}
\email{yongdaz@arizona.edu}

\begin{abstract}
We compile JWST/NIRSpec prism and MIRI data for 249 Little Red Dots (LRDs) at $2.3<z<9.3$, forming a representative spectroscopic subset of NIRCam-selected
LRDs. We derive a median stacked spectrum covering rest-frame 0.09–1.2~$\mu$m, with MIRI photometry extending the spectral energy distribution to 4~$\mu$m\footnote{Downloadable from \href{https://doi.org/10.5281/zenodo.18679061}{this link}.}. Four additional stacks for subsamples defined by optical-to-UV luminosity ratios show that LRDs form a heterogeneous population spanning diverse continuum slopes and line properties. Assuming LRDs host super-massive black holes (BHs) surrounded by dense gas clouds, and stars accompany this core, we infer masses of $\mathrm{M}_\mathrm{BH}\sim10^{6.0-6.5}$~M$_\odot$ and $\mathrm{M}_\bigstar\sim10^{8.3}$~M$_\odot$, corresponding to BH-to-stellar mass ratios of 1–2\%. The stacks show ubiquitous UV and optical Fe\,II emission, indicating a direct view of the broad-line region and high (but sub-Eddington)
accretion ($\lambda_\mathrm{Edd}=0.6\pm0.2$).
We find a significant stellar contribution in the far-UV, reaching $\sim80$\% in the bluest systems. Possible Wolf–Rayet features (He\,II$\lambda$4687, nitrogen lines) are identified, tracing a young (3–7~Myr) compact starburst event. We also detect strong Balmer breaks and atypical Balmer, Paschen, [O\,III], and optical and near-infrared He\,I line ratios, and an absorption at $\sim4550$~\AA\ (probably linked to Fe\,II), all consistent with radiative-transfer effects in high-density gas with warm temperatures (4000-7000~K). We find a diversity of LRD flavors modulated by the luminosity ratio between a short ($\lesssim20$~Myr) and intense phase of BH activity, the most extreme stage lasting $\sim3-7$~Myr, characterized by near-Eddington-limit radiation, and a nuclear and compact starburst dominated by massive stars (even super-massive, $\mathrm{M}_\mathrm{SMS}\sim10^{5}$~M$_\odot$), all embedded in dense gas with modest dust content producing a variety of optical depths.
\end{abstract}

%Our analysis shows that the continuum diversity across subtypes is primarily driven by variations in the host-to-BH luminosity ratio, together with modest attenuation of the core, and differences in optical depth.

%(selection criteria: F277W-F444W$>$1~mag, F150W-F200W$<$0.5~mag, F444W$\leq$28~mag)

%and smaller linewidths for the H$\beta$ broad component compared to H$\alpha$'s (except for the reddest, brightest LRDs)

\keywords{Galaxy formation (595) --- Galaxy evolution (594) --- High-redshift galaxies (734) --- Stellar populations (1622) --- Broad band photometry (184) --- Galaxy ages (576) --- JWST (2291) --- Active galactic nuclei(16)}

\section{Introduction}
\label{sec:intro}

The James Webb Space Telescope (JWST, \citealt{2006SSRv..123..485G,2023PASP..135f8001G}) has probed galaxy evolution in the first billion years ($z\gtrsim5$) with unprecedented detail \citep[see][and references therein]{2024arXiv240521054A,2026enap....4..453S}. Based on the JWST observations gathered during the first four years of scientific operations, three closely related results of significant cosmological impact have emerged. First, many independent works identified a relatively large number, compared to expectations (see, e.g., \citealt{2023ApJ...946L..13F,2023ApJ...951L...1P,2025ApJ...991..179P, 2023ApJ...954L..46L,2024ApJ...964...71H}), of bright ultraviolet (UV) galaxies already in place in the first 500~Myr of cosmic time ($z\gtrsim10$; \citealt{2023ApJ...948L..14C,2023MNRAS.518.6011D,2022ApJ...940L..55F,2023ApJS..265....5H,2022ApJ...940L..14N}). Many of these candidates have been confirmed spectroscopically \citep{2023A&A...677A..88B,2023Natur.622..707A,2024ApJ...972..143C,2024Natur.633..318C,2023ApJ...951L..22A,2023NatAs...7..622C,2024ApJ...960...56H,2023ApJ...957L..34W}. These results  prove that luminous galaxies were already in place 300 million years after the Big Bang pushing theories for galaxy formation back to earlier epochs than previously expected. 

The resulting high UV luminosity density could arise from several physical phenomena. Some of these galaxies appear to be dominated by star formation \citep[e.g.,][]{2023A&A...677A..88B,2025arXiv251219695H}. Another possibility is that many of the UV photons are linked to nuclear activity arising from the material surrounding supermassive black holes (SMBHs, or intermediate-mass black holes, IMBH, defined here as smaller than $10^6$~M$_\odot$), i.e., due to active galactic nuclei (AGN). This relates to the second highly relevant cosmological result presented in the JWST mission early years: a wealth of observational evidence of the presence of accreting SMBHs in many high-redshift galaxies. Indeed, spectroscopic data have revealed broad components in the emission lines detected in the spectra for many galaxies \citep{2023ApJ...954L...4K,2023ApJ...959...39H,2024A&A...691A.145M,2024arXiv240906772T}. Complementary to this, optical line ratios also point to strong radiation fields in many high-redshift galaxies, consistent with photoionization from AGN \citep{2025A&A...697A.175S,2025A&A...700A..12M,2025arXiv251208490C}.

The high abundance of AGN in the early Universe \citep[e.g.,][]{2024A&A...691A.145M} may be related to another key result: the deviation of some of these early and very active galaxies from local relations between the masses of the SMBH and the host, with the former claimed to be overmassive (by, in some cases, more than a dex) compared to their stellar mass \citep{2024A&A...691A.145M,2023ApJ...957L...3P,2024ApJ...964..154P}. 

The high abundance of active (and possibly evolved) SMBHs in the early ($<1$~Gyr old) Universe includes a previously unknown type of source known as Little Red Dots (LRDs). The name refers to their compact, point-like morphology in the reddest NIRCam filters and their relative faintness in the bluest NIRCam bands. These two features have been used  for efficient selection of these objects, showing them to be ubiquitous in all JWST imaging surveys \citep[see, e.g.,][]{2023Natur.616..266L,2023ApJ...946L..16P,2024ApJ...963..128B,2025ApJ...991...37A,2025ApJ...986..126K}, with a space density making them a significant fraction of all quasars and AGNs \citep[e.g.,][]{2024ApJ...968...34W,ma2025,2025MNRAS.538.1921M}. Some LRDs were found to show broad emission lines  with NIRSpec/PRISM  spectra \citep{2024ApJ...964...39G} and NIRCam/WFSS \citep{2024ApJ...963..129M}, leading to the conclusion that they contained, and were perhaps powered predominantly by, AGNs. Further follow-up spectroscopic observations of  LRDs confirmed the presence of broad emission line components in a majority of LRDs \citep{2023ApJ...954L...4K,2025arXiv251000101D,2024A&A...691A..52K,2025arXiv251000103T,2025A&A...702A..57H,2025ApJ...989L...7T}.

Dust was initially thought to be responsible for the defining red colors of LRDs. However, the presence of dust in all LRDs is not unequivocally confirmed \citep{2024ApJ...968....4P, 2024ApJ...968...34W,2025ApJ...991L..10S,2025A&A...700A.231X}, although some individual examples have been detected in the far-infrared \citep{2024arXiv241201887B} and small amounts of dust have been claimed based on stacks \citep{2024ApJ...975L...4C}. And this is, in fact, the third highly relevant cosmological result provided by JWST: the small but non-negligible amounts of dust formed in the very early Universe, not only revealed in LRDs but also in the attenuation seen in the highest-redshift galaxies \citep[see, e.g.,][]{2023MNRAS.523.3119W,2025NatAs...9..729H,2025ApJ...993..224D}, as well as the floor in metallicity around 5--10\% solar observed in most galaxies at high redshift \citep[e.g.,][but some lower metallicity galaxies have also been discovered, \citealt{2025arXiv250503873H,2025arXiv250907073V,2025arXiv250710521M}]{2023ApJS..269...33N,2023MNRAS.518..425C,2025NatCo..16.9830T,2025arXiv251219695H,2025MNRAS.540.2176C}.

These three early results of JWST in the cosmological research area are not completely understood. We can identify four main aspects of the interpretations that need refining or even validation, and that particularly affect the comprehension of the nature of LRDs and might be biasing our view of the early Universe.

%For example,  the redshift confirmation of high-z galaxy candidates is limited to the brightest objects and has been hampered by mis-identifications of dusty galaxies at intermediate redshift with high-z sources \citep{2023ApJ...943L...9Z,2023Natur.622..707A,2025arXiv250202637G}.

First, the physical conditions of the gas in high redshift galaxies have been demonstrated not to be exactly the same as in local star-forming galaxies (SFGs) at redshifts as low as $z\sim2$. Consequently, the extreme line ratios observed for some high-z AGN candidates can still be linked to intense and concentrated bursty star formation \citep{2023MNRAS.523.3516C,2024MNRAS.535.1796B,2025ApJ...980..242S}. Indeed, for most JWST-detected AGN the classic rest-optical emission line ratio diagnostic diagrams do not extrapolate easily to high redshifts, so star-forming galaxies and AGN are mixed in the same region of the diagnostic diagrams due to the low gas metallicities and high ionization parameters \citep{2023ApJ...945...35T,2024ApJ...962..195B,2023ApJ...955...54S,2025ApJ...994..146C,2026arXiv260111092C}.

Directly related to the previous point, the AGN-nature of the broad components of LRDs have been challenged by alternative interpretations such as the presence of dense low-metallicity outflows linked to star formation \citep{2025ApJ...995...24K}, extreme line broadening due to gas kinematics in very compact starbursts \citep{2024ApJ...977L..13B}, or significant effects linked to collisional excitation in dense gas clouds in the cores of high redshift galaxies \citep{2025ApJ...980L..27I}. The proposal for dense gas clouds surrounding a compact ionizing source \citep[e.g.,][]{coughlin+begelman2024} to explain LRDs has now gained momentum in the so-called quasi-star, blackhole envelope (BHE) or blackhole star (BH$^{\star}$) models \citep[e.g.,][]{2025MNRAS.544.3407K,2025arXiv250316596N,2025arXiv251203130I,2025arXiv250709085B,2025A&A...701A.168D,2025ApJ...994..113L,sneppen+2026,2026arXiv260120929S}.

A second source of uncertainty in interpreting JWST results, particularly for LRDs and other AGN candidates, concerns the derivation of stellar and SMBH masses. Early analyses inferred very large stellar masses for some systems \citep{2023Natur.616..266L}, largely because the red rest-frame optical emission was modeled as purely stellar. Reproducing the observed colors under this assumption required substantial dust attenuation and high stellar masses, in some cases approaching or exceeding expectations from baryon abundances in
early halos within the $\Lambda$CDM paradigm
\citep{2022ApJ...939L..31H,2023NatAs...7..731B,2023MNRAS.523.3201D}. Such estimates depend critically on modeling choices, including the adopted initial mass function (IMF); for example, a top-heavy IMF could
further bias stellar masses high
\citep{2022MNRAS.514.4639C,2023ApJ...951L..40S,2025MNRAS.537..752B}, as the warmer cosmic background radiation heats the interstellar gas and inhibits  the formation of low-mass stars and brown dwarfs.  

Conversely, stellar masses may be underestimated if the rest-frame optical and near-infrared emission is attributed predominantly to dust-attenuated accretion disks, bright tori \citep{2023arXiv230607320L}, or a BH$^{\star}$-like component with little or no stellar contribution, potentially leading to the interpretation of LRDs as hosting over-massive SMBHs. Ultimately, the relative contributions of stars and AGN to the global spectral energy distribution (SED) of LRDs remain uncertain, and assumptions about star-formation histories and
dust attenuation can significantly affect the derived stellar and black hole masses
\citep[e.g.,][]{2024ApJ...963..128B,2024ApJ...968....4P}.

Similarly, SMBH mass estimates for LRDs introduce additional uncertainty. These are typically based on single-epoch calibrations derived from low-redshift AGN \citep{2005ApJ...630..122G,2015ApJ...813...82R}, whose applicability at high redshift is uncertain. Moreover, current spectroscopic samples are likely biased toward the most luminous and extreme systems
\citep{2025ApJ...981...19L,2025arXiv251119609B}, potentially skewing inferred black hole mass distributions.

A third open question for  LRDs is how much dust they have, and whether their red nature can be linked to it, or only to the effects of dense gas clouds (i.e., obscuration only due to gas). As mentioned before, there is not a consensus about the amount of dust in LRDs. In fact, the latest modeling efforts in terms of BH$^{\star}$ assign most of the red optical slopes and prominent Balmer breaks to opacity by gas, not dust \citep[e.g.,][]{2025A&A...701A.168D}.

The fourth, and final, point to take with a grain of salt is the extrapolation of properties derived for some LRDs whose spectroscopy has been analyzed in a number of papers \citep{2023ApJ...954L...4K,2024A&A...691A..52K,2024ApJ...969L..13W,2024ApJ...964...39G,2025arXiv251121820D} to a more general and numerous number of sources identified as LRDs with NIRCam photometry \citep{2024ApJ...968...38K,2025ApJ...986..126K,2024ApJ...963..128B}. In this regard, the limited overlap between LRD samples, whose selection (and consequently, definition) is not standardized, is concerning \citep{2025ApJ...979..138H,2025A&A...702A..57H,2025arXiv251215853B,2026arXiv260407138R}. LRDs appear to be a heterogeneous group of sources in terms of their SED and spectral properties, and 
it is possible that their nature may not all be attributable to a single mechanism/scenario. 

In this paper, we address these four debated topics by analyzing a complete sample of photometrically selected LRDs and gathering all the available spectroscopy provided by JWST, as well as MIRI imaging covering the rest-frame near- and mid-infrared (IR). The goal is to understand the nature of LRDs by, (1) investigating whether any heterogeneity in their (photometric, spectral) properties is linked to the different selection techniques, and (2) carefully understanding whether the LRDs with currently available  spectroscopy are representative of the more general photometric sample. Our analysis technique aims to escape from the degeneracies introduced by the studies based on photometry alone \citep[e.g.,][]{2024ApJ...968....4P} or bright emission line profiles and ratios alone \citep{2023ApJ...954L...4K,2025ApJ...986..126K,2025ApJ...989L...7T,2025ApJ...992...71R}, by predominantly using the continuum emission, jointly with the faint emission lines, in a wide spectral range covering the UV, optical, and near-IR.

This paper is organized as follows. Section~\ref{sec:selection} presents the photometric and spectroscopic samples used in this paper, describing the NIRCam, NIRSpec and MIRI datasets gathered to analyze a statistically representative sample of LRDs. Section~\ref{sec:stacks} introduces a stack of spectra for nearly 250 LRDs and discusses some interesting features in the UV, optical, and near-IR. Section~\ref{sec:subtypes} divides the full sample of LRDs in several subtypes and analyzes the heterogeneity of properties within the LRD population. Section~\ref{sec:continuum_properties} further analyzes the different subtypes of LRDs based on SED fitting performed on the NIRSpec and MIRI stacks. Section~\ref{sec:discussion} discusses all our findings in comparison with the literature. Finally, Section~\ref{sec:conclusions} presents a summary of our work and conclusions.

In this paper, we assume a flat $\Lambda$CDM cosmology with $\mathrm{\Omega_M\, =\, 0.3,\, \Omega_{\Lambda}\, =\, 0.7}$, and a Hubble constant $\mathrm{H_0\, =\, 70\, km\,s^{-1} Mpc^{-1}}$. We use AB magnitudes \citep{1983ApJ...266..713O}. We presume a universal \citet{2003PASP..115..763C} initial mass function (IMF). We refer to all emission lines via their vacuum wavelengths.

\begin{figure*}%[htp!]%[ht!]
\centering
%\fbox{
\includegraphics[clip, trim=0.4cm 0.8cm 1.9cm 2.0cm,width=8.5cm,angle=0]{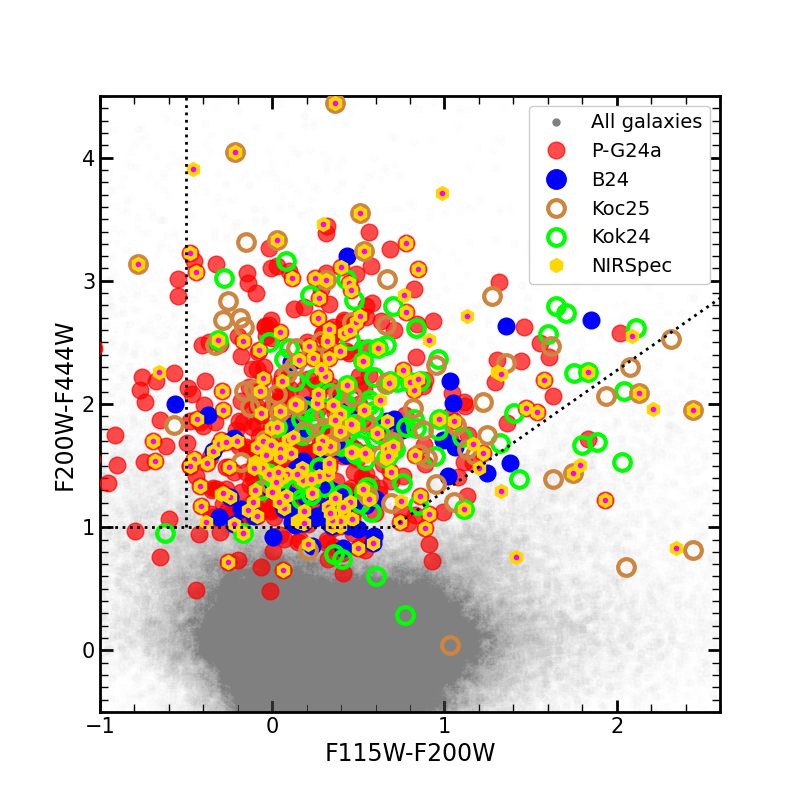}
\includegraphics[clip, trim=0.4cm 0.8cm 1.9cm 2.0cm,width=8.5cm,angle=0]{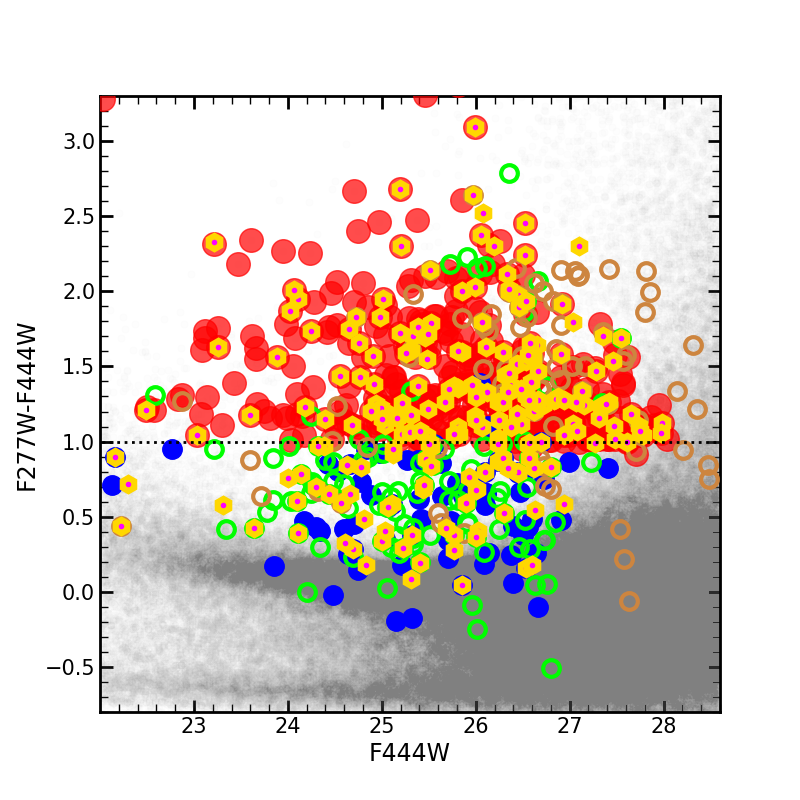}
%}
%\fbox{
%}
%\fbox{
\includegraphics[clip, trim=0.4cm 0.0cm 1.9cm 1.0cm,width=8.cm,angle=0]{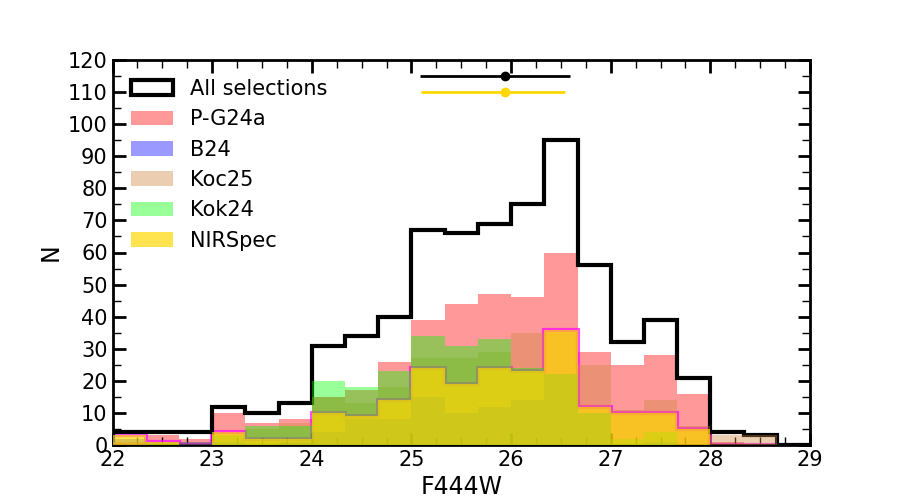}
\includegraphics[clip, trim=0.4cm 0.0cm 1.9cm 1.0cm,width=8.cm,angle=0]{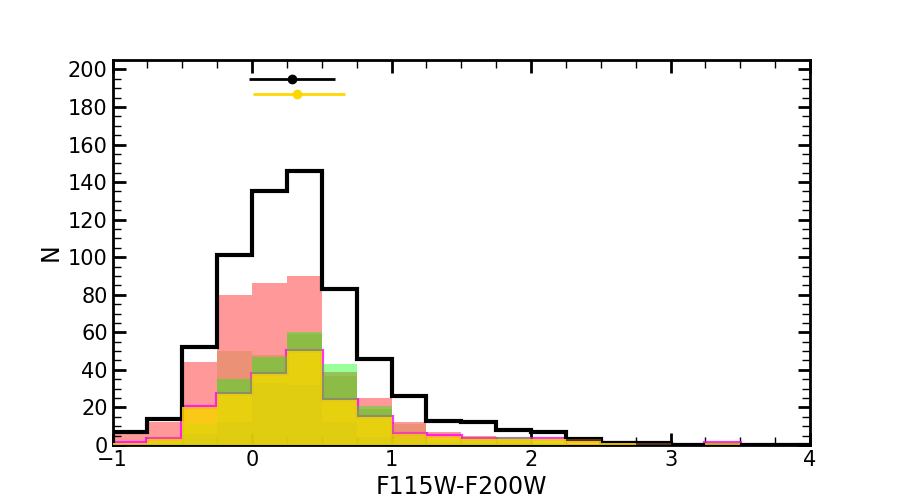}
\includegraphics[clip, trim=0.4cm 0.0cm 1.9cm 0.0cm,width=8.cm,angle=0]{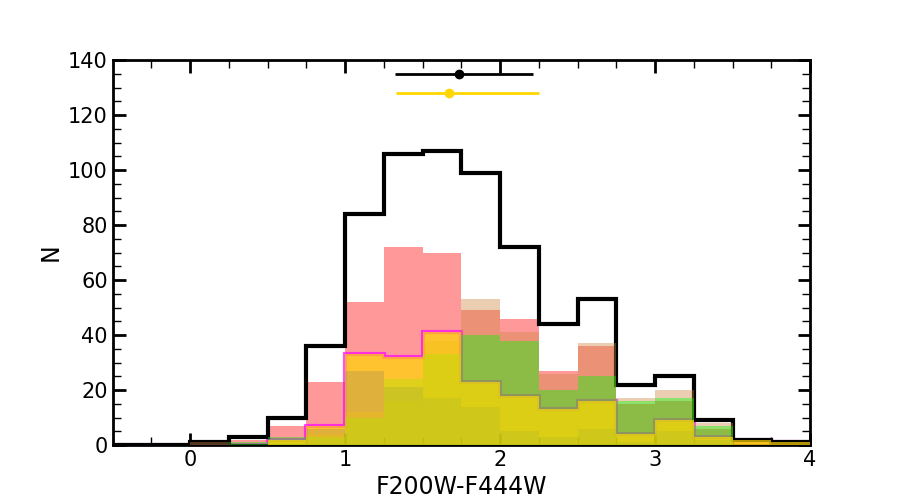}
\includegraphics[clip, trim=0.4cm 0.0cm 1.9cm 1.0cm,width=8.cm,angle=0]{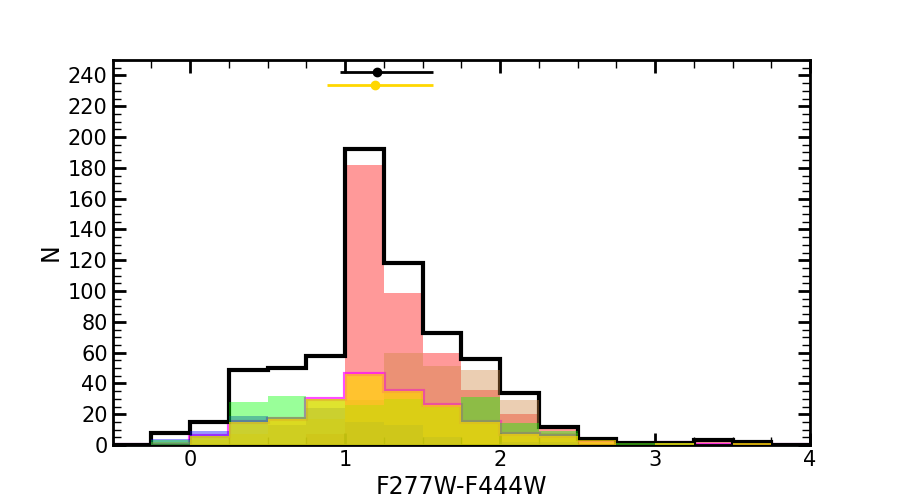}
%}
\caption{\label{fig:selection}Photometric selection of LRDs, and the spectroscopic sample used in this work. {\it Top left:} Color-color diagram introduced by \citet{2024arXiv241201887B}, with the selection boundary represented by the dotted lines. The LRD selection criterion used in \citet{2024ApJ...968....4P} based on F277W-F444W colors is shown with red symbols, LRDs identified in \citet{2025ApJ...986..126K} and \citet{2024ApJ...968...38K} are shown as orange and lime symbols, respectively, while blue symbols depict the LRDs identified solely by \citet{2024arXiv241201887B}. Our spectroscopic sample (with NIRSpec prism observations) is depicted with yellow symbols. Gray symbols show the general NIRCam sample (drawn from the ASTRODEEP–JWST catalogs presented in \citealt{2024A&A...691A.240M}). {\it Top right:} Color-magnitude plot with LRD selection criterion used in \citet{2024ApJ...968....4P} based on F277W-F444W colors (marked with a dotted line). We use the same symbols described for the previous panel. {\it Middle left:} F444W magnitude distribution of the different samples of LRDs. {\it Middle right:} F115W-F200W color distribution of LRDs. {\it Bottom left:} F200W-F444W color distribution of LRDs. {\it Bottom right:} F277W-F444W color distribution of LRDs. In all the panels with histograms, the median and quartiles for the photometric (black) and spectroscopic (gold) samples are shown.}
\end{figure*}

\section{Data and Little Red Dot selections}
\label{sec:selection}

\subsection{Photometric selection}

Using the ASTRODEEP–JWST catalogs presented in
\citet{2024A&A...691A.240M}, we construct a sample of 787 photometrically selected LRDs across six JWST deep fields (also used in \citealt{2025arXiv251215853B}): GOODS-S and GOODS-N from JADES \citep{2020IAUS..352..337R,2020IAUS..352..342B,2023arXiv230602465E,2025ApJS..281...50E}, EGS from CEERS \citep{2025ApJ...983L...4F}, COSMOS and UDS from PRIMER (Magee et al., in prep.), and A2744 from GLASS-JWST \citep{2022ApJ...935..110T}
and UNCOVER \citep{2024ApJ...974...92B}. The ASTRODEEP-JWST catalog provides NIRCam photometry measured in small apertures in PSF-matched images and then corrected to total \citet{1980ApJS...43..305K} apertures.

Our LRD selection is designed to be as complete as possible by combining multiple, complementary photometric criteria. The full selection technique is illustrated in Figure~\ref{fig:selection}. The core of the selection follows the color–color approach introduced by \citet{2024arXiv241201887B}, using the F115W–F200W versus F200W–F444W diagram to identify sources with blue rest-frame ultraviolet continua and red rest-frame optical slopes, i.e., utilizing the main distinctive feature of LRDs, the V-shaped SED. This method efficiently recovers the bulk of the LRD population over $z\sim3$–7 and captures a wide
range of intrinsic UV–to–optical colors. To minimize contamination from brown dwarfs, we apply an additional photometric cut of F115W–F200W $> -0.5$ \citep{2024ApJ...964...39G,2024ApJ...964...66H}. A small number of spectroscopically confirmed LRDs can scatter blueward of this threshold (see four galaxies with spectroscopy --i.e., in yellow-- in the top left panel of Figure~\ref{fig:selection}), but this cut effectively removes the bulk of stellar contaminants without significantly impacting completeness. At the other extreme of UV colors, the color-color diagram introduced by \citet{2024arXiv241201887B} misses some LRDs with red F200W–F444W and F115W–F200W colors, or noisy colors in the short wavelength (SW) channels probing the UV slope. As noted by \citet[][Section~3.1]{2024arXiv241201887B}, this is the case for peculiar objects such as The Cliff \citep{2025A&A...701A.168D}, and all the LRDs (marked with colored points) to the right of the diagonal dotted line in the upper left panel of Figure~\ref{fig:selection}. This region is heavily contaminated by dusty galaxies (gray symbols in the figure) with red colors in both the optical and the UV, i.e., not presenting the typical V-shaped SED of LRDs. However, this is also the locus of LRDs with very strong Balmer breaks, which explain the red F115W-F200W, while some other colors such as F115W-F150W reveal a relatively flat UV spectral slope (and confirm the V-shape SED). These sources were added manually to the selection in \citet{2024arXiv241201887B}, and are nevertheless well recovered by the other selection methods we use in this paper (see below).

To improve completeness, particularly at higher redshifts, we additionally apply the F277W–F444W versus F444W color–magnitude selection used in \citet{2024ApJ...963..128B} and \citet{2024ApJ...968....4P}, selecting LRDs with a color F277W–F444W\,$>1$~mag. This criterion
is effective at recovering LRDs that may fall outside the color–color selection, especially at $z\gtrsim8$, where (i) the F150W–F200W color
begins to trace the Lyman break; (ii) the bands redder than 4~$\mu$m can be heavily contaminated by [OIII]$+$H$\beta$ emission, and  (iii) the magnitudes become fainter, all resulting on the selection becoming increasingly uncertain, unreliable, and/or incomplete.

Finally, we supplement the photometrically selected sample with a small number ($\lesssim$5\%) of additional LRDs drawn from the catalogs of \citet{2024ApJ...968...38K} and
\citet{2025ApJ...986..126K}. These sources are not recovered by the primary color selections but are included to ensure maximal completeness of the LRD population. The combined selection (adding sources selected with any of the methods above) therefore
recovers nearly all previously reported LRDs while extending the earlier, more restrictive definitions to bluer UV–to–optical colors.

Morphological compactness in the rest-frame optical is enforced by requiring a small-aperture flux ratio of F[F444W(${d=0\farcs5}$)]/F[F444W(${d=0\farcs2}$)] $< 1.5$,
ensuring that selected sources are unresolved or marginally resolved
in the rest-frame optical \citep[see, e.g.,][]{2024ApJ...964...39G,2024ApJ...977L..13B}.

The use of four different methods in the selection of LRDs is meant to provide the highest completeness possible. In that regard, we remark that the methods have distinct efficiencies for different redshifts. Indeed, the \citet{2024ApJ...968....4P} selection reaches a sample with median redshift $z=7.1$, similarly to the \citet{2025ApJ...986..126K} selection, median redshift $z=7.1$ too. The \citet{2024arXiv241201887B} and \citet{2024ApJ...968...38K} probe lower redshifts, median $z=5.7$ and $z=6.1$, respectively. Alternatively, we can characterize the redshift sensitivity of the different methods by quantifying the fraction of sources in the final sample recovered by each selection for the lowest ($z<6$) and highest ($z>=6$) redshifts. While for $z<6$, the \citet{2024arXiv241201887B} selection recovers $\sim65$\% of the full sample, the other methods select 30\% \citep{2024ApJ...968....4P} to 45\% \citep{2024ApJ...968...38K,2025ApJ...986..126K}. At $z>6$, the most effective method is the one in \citet{2024ApJ...968....4P} (75\% of sources), with the rest of methods all around 50\% contribution.

In summary, our work compiles the LRDs in GOODS-S and GOODS-N (from JADES), UDS and COSMOS (from PRIMER), EGS (from CEERS), and the A2744 cluster field (from UNCOVER), probing an approximate total area of 690~arcmin$^2$. A total of 787 LRDs are included in our photometric sample. This number was obtained after removing sources from the parent ASTRODEEP–JWST catalog which had measured fluxes but were found by visual inspection to lie in the edge of the NIRcam mosaics, i.e., with only partial coverage in some bands(s). Low redshift (photometric redshift $z\lesssim1$) galaxies (which passed the compact criterion) were also removed from the sample, since the V-shape for these sources was linked to a different origin compared to LRDs (hot/warm dust and 3~$\mu$m PAH emission and optical/near-infrared stellar emission), based on the photo-z template best-fitting the photometry (no spectroscopy is available for any of these sources). Field to field variations of the number density are found at a 25\% level. The average and standard deviation of the densities for all fields is $1.1\pm0.3$~sources\,arcmin$^{-2}$. 

\subsection{Spectroscopic observations of LRDs}
\label{sec:spectra}

Our parent sample of LRDs was matched with the list of spectroscopically observed sources in all fields. In this paper, we concentrate our analysis on the NIRSpec prism data only, due to our focus on the continuum and faint lines. We found a total of 249 observations with the NIRSpec prism carried out by a variety of programs. The program identifications and number of sources per field are given in Table~\ref{tab:obs}.

We downloaded fully calibrated 2D spectra from MAST. The calibration was carried out with the  {\sc jwst} python package, pipeline version 1.20.2, reference files 1464.pmap.  We extracted 1D spectra for all objects with a custom aperture that ranged from 3 to 5 pixels (for a few very bright objects we used 7 pixels) in the spatial direction (0.1 arcsec/pixel), determined to maximize the median signal-to-noise ratio of the final spectrum. The spectrum for each object was normalized to match its NIRCam broad-band fluxes, as extracted from the ASTRODEEP catalog, by applying a single multiplicative factor calculated as a weighted average of the ratios between the fluxes measured in photometry and in the spectra for the NIRCam passbands. 

Virtually all of the spectra showed the emission lines [O\,III]$\lambda\lambda4960,5008$ and,  for sources at  $z\lesssim7$, H$\alpha$. Several additional lines (Ly$\alpha$, C\,III], [S\,II], [O\,II], H$\beta$, Paschen lines) were detected in many spectra. All were used to measure accurate (typical error $<0.005$) spectroscopic redshifts.

\subsection{MIRI data}

All MIRI data publicly available in November 2025 for the six fields covered in this paper were reduced with the JWST Rainbow pipeline following the methodology described in \citet[][see also \citealt{2024ApJ...969L..10P,2025A&A...696A..57O}]{2024ApJ...968....4P}. The observations come from different programs, PIDs provided in Table~\ref{tab:obs}. Mosaics were constructed for filters F560W, F770W, F1000W, F1280W, F1500W, F1800W, F2100W, and F2550W. Details about this reduction will be presented in P\'erez-Gonz\'alez et al. (in prep.), while here we briefly describe  the methodology. The JWST Rainbow reduction scheme uses the official JWST pipeline adding some extra bespoke reduction steps to deal with the background homogenization of the MIRI images. For this paper, the reduction was carried out with the {\sc jwst} code, version v1.19.41, reference files in jwst\_1413.pmap. 
A key component of this workflow is the `superbackground' strategy, in which the background for each observation is flattened by constructing a template that combines all other images taken with the same filter within a 3-month  period. Known sources are masked (after applying a dilation) to avoid background overestimates. This approach yields a highly homogeneous background in terms of both level and noise. This methodology has been demonstrated to improve depth by up to 0.8 mag in the final mosaics compared to archive data. We refer the reader to Appendix A in \citet{2024ApJ...968....4P} for further details.

Photometry was performed by assuming LRDs are point-like sources and measuring in small apertures with diameters equal to the FWHM of the PSF for each filter. Correlated noise was taken into account in the estimation of uncertainties. See \citet{2024ApJ...968....4P,2024ApJ...969L..10P} for more details. The MIRI fluxes for sources with $\mathrm{SNR}>3$ were added to the analysis in the rest of the paper by measuring a F444W-MIRI color in the aperture mentioned before, and applying it to the ASTRODEEP-JWST SED (which was used as a reference for the spectra).

The depth of the observations varies significantly in all filters, from field to field and within each one of them. Typical (median for all fields of the medians for each filter) depths ($5\sigma$) in AB mag are 26.2 for F560W, 25.7 for F770W, 24.3 for F1000W, 24.1 for F1280W, 23.5 for F1500W, 22.8 for F1800W, 22.0 for F2100W, and 20.2 for F2550W. 

The coverage of the MIRI observations for our spectroscopic sample is limited. The fraction of LRDs with MIRI data varies from 0\% in F2550W to 60\% in F770W, but the typical percentage is 30-40\%,  except for F560W, where only 12\% of spectroscopic LRDs have data. The MIRI detection fraction among those LRDs in our spectroscopic sample with available data is high for wavelengths shorter than 10~$\mu$m (50\% detected in F1000W, 80\% in F560W and F770W). The detection fraction is much smaller for the reddest wavelengths, ranging from 30\% in F1280W to 10\% in F2100W, 40\% and 20\% for F1500W and F1800W, respectively.

Very similar numbers are obtained for the photometric sample, with only a small bias in the reddest bands, in the sense that there is a lack of spectroscopy for photometrically selected LRDs detected in F1800W and F2100W.

%PIDs 1181, 1264, 2926, 4530, 4586, 4762, and 5407 for GOODS-N; 1180, 1207, 1283, 2516, 3954, 4498, 5279, 5407, and 6511 for GOODS-S; 1345, 1448, 3794, and 4586 for EGS; 1657, 1837, and 7814 for UDS, 1284, 1727, 1762, 1837, 2417, 2775, 3954, 5893, and 6595 for COSMOS, and 4530 and 6743 for A2744.

\subsection{Classification of LRDs}
\label{sec:classes}

The full photometric and spectroscopic samples of LRDs were divided in different types. Given that the main distinctive spectral feature of LRDs is their V-shaped SED (i.e., blue or flat UV and red optical slopes), we decided to sort them by spectral slope. We use the optical-to-UV luminosity ratio as a proxy for the optical slope (as in, e.g., \citealt{2025ApJ...986..126K} and \citealt{2025ApJ...979..138H}), $\mathrm{L_{5100}/L_{2500}}$ (defined as luminosity densities in erg~s$^{-1}$~Hz$^{-1}$). We note that the use of $\mathrm{L_{5100}/L_{2500}}$ as an empirical classifier can be linked to both AGN and stellar population diagnostics, reflecting differences in, for example, dust geometry, stellar age, and/or viewing angle. We refer the reader to Section~\ref{sec:subtypes} for more details.

Based on the optical-to-UV luminosity ratio, and using the photometric sample as a reference to avoid possible biases of the spectroscopic sample, we divided the list of LRDs with spectroscopy into four subsamples: 
\begin{itemize}
\item {{\it extreme} LRDs, $\mathrm{^xLRD}$s, defined as extremely red sources that are 1$\sigma$ above the median ($\mathrm{L_{5100}/L_{2500}}>6.3$). Our spectroscopic sample has 29 $\mathrm{^xLRD}$s (12\% of the whole sample).
%In some of the following sections, we will also refer to very extreme LRDs, $^\mathrm{xx}$LRDs, defined as $\mathrm{L_{5100}/L_{2500}}>10$. We have 9 $^\mathrm{xx}$LRDs.
}
\item {LRDs slightly above the median luminosity ratio, $\mathrm{^+LRD}$s, which are within the median and $+$1$\sigma$ ($3.1<\mathrm{L_{5100}/L_{2500}}<6.3$). Our sample has 76 $\mathrm{^+LRD}$s (31\% of the whole sample).}
\item LRDs slightly below the median luminosity ratio, $\mathrm{^-LRD}$s, which are between the median and --1$\sigma$ ($1.8<\mathrm{L_{5100}/L_{2500}}<3.1$). Our sample has 77 $\mathrm{^-LRD}$s (31\% of the whole sample).
\item {\it blue} LRDs, $\mathrm{^bLRD}$s, defined as those with $\mathrm{L_{5100}/L_{2500}}<1.8$. Our sample has 67 $\mathrm{^bLRD}$s (27\% of the whole sample).
\end{itemize}

\begin{figure*}%[hb!]%[htp!]%[ht!]
\begin{center}
\includegraphics[clip, trim=3.5cm 0.3cm 3.7cm 2.cm,width=17.cm,angle=0]{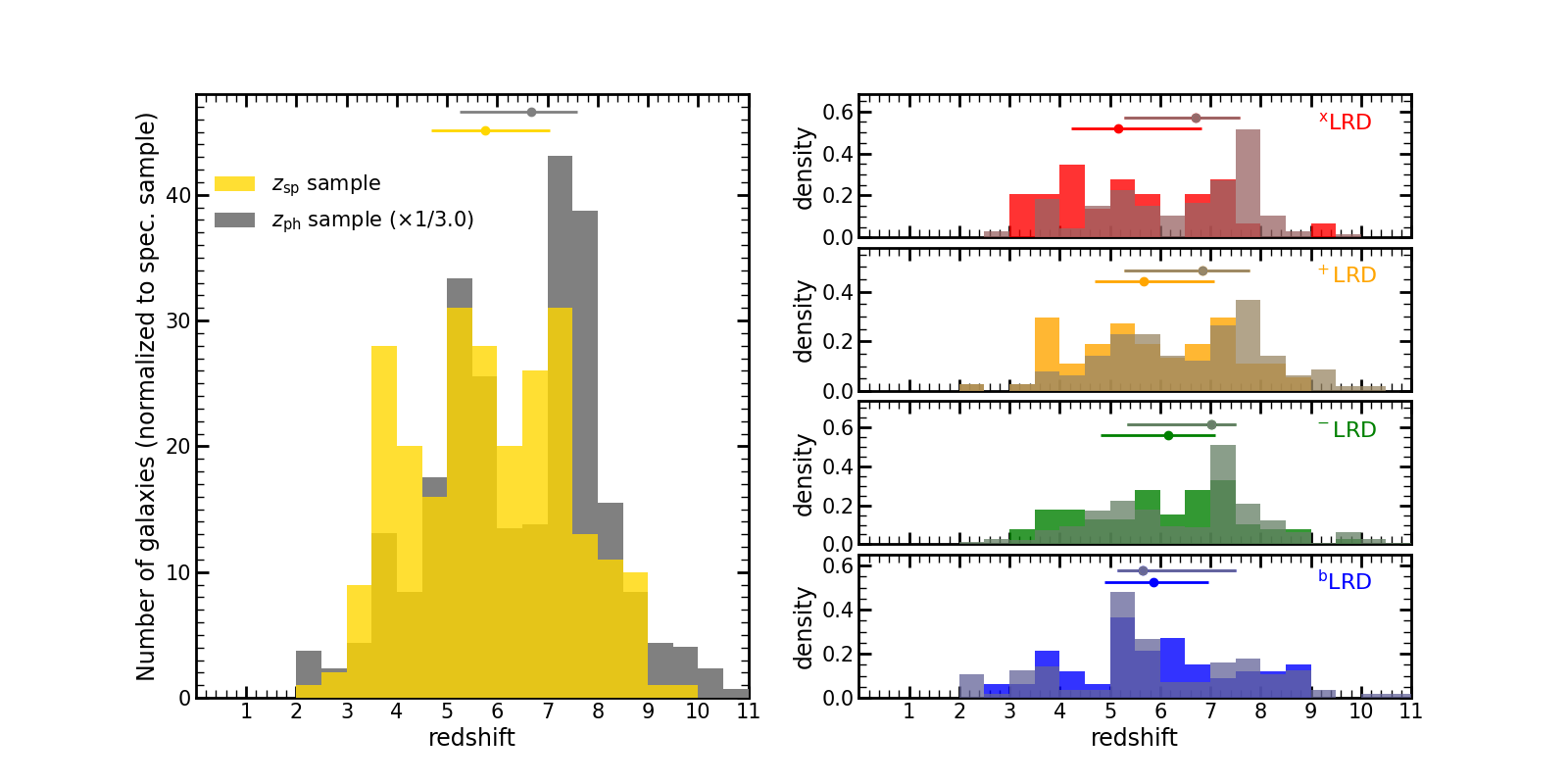}
\caption{\label{fig:redshifts}{\it Left:} Redshift distribution for the spectroscopic sample of LRDs characterized in this paper (gold histogram) and the general sample of photometrically selected LRDs presented in Figure~\ref{fig:selection} (gray histogram, scaled down by a factor of $\sim4$ to make the comparison easier). {\it Right:} Same but for the different LRD subtypes defined in Section~\ref{sec:classes}, based on optical-to-UV luminosity ratios, running from the reddest LRDs (top) to the bluest (bottom). The photometric sample plotted again in gray, and this time using normalized histograms. In all panels, the median and quartiles of the distributions are shown with dots and segments, the gray histograms refer to the photometric sample and the colored ones to the spectroscopic subsamples.}
\end{center}
\end{figure*}

Our LRD spectroscopic sample was cross-correlated with the list of LRDs observed with higher spectral resolution data ($R=1000$, mainly), published in \citet{2025ApJS..277....4D}, \citet{2025A&A...702A..57H}, \citet{2025arXiv251007376J},  and \citet{2026MNRAS.546ag086J}. Nearly one third (27\%) of our galaxies were found in those catalogs. Out of them, virtually all (97\%) have broad ($>900$~km~s$^{-1}$) H$\alpha$ emission, the rest are two sources from Dark Horse (\citealt{2026MNRAS.549ag824D}) which have no measurements. Concerning our LRD sub-types, $\sim20$\% the bluest LRDs ($\mathrm{^-LRD}$s and $\mathrm{^bLRD}$s) have high-resolution NIRSpec data, the fraction increases to 31\% for $\mathrm{^+LRD}$s and 45\% for $\mathrm{^xLRD}$s. The two sources with no measurements are classified as a $\mathrm{^+LRD}$ and a $\mathrm{^xLRD}$. We conclude that there is a very high incidence of broad emission lines among LRDs, with no significant differences among the sub-types.

\subsection{Similarity of the spectroscopic and photometric  samples}

The observational properties of the spectroscopic sample compiled in this paper compared to photometrically selected LRDs are presented in Figure~\ref{fig:selection}. The color-magnitude \citep{2023Natur.616..266L,2023ApJ...946L..16P,2024ApJ...963..128B,2024ApJ...968....4P} and color-color \citep{2024arXiv241201887B} diagrams used in the photometric selection of LRDs (see also \citealt{2025ApJ...986..126K}) show that our sample of LRDs with prism spectroscopy is representative of the whole photometric LRD selection down to F444W magnitude around 27 (with a few sources up to one magnitude fainter). In particular, by using the comprehensive selection technique presented in \citet{2024arXiv241201887B}, our spectroscopy compilation covers the selections of \citet{2024ApJ...968...38K} and \citet{2025ApJ...986..126K}, which are biased towards blue and red LRDs, respectively.

The main photometric properties of the parent sample of photometrically selected LRDs and the sample with spectroscopy compiled for this paper are given in Table~\ref{tab:stats_photobs}. Summarizing the properties of these samples, in this paper we are analyzing spectroscopic LRDs with F444W magnitude around 26, median and quartiles $\mathrm{F444W}=25.9^{26.5}_{25.0}$~mag\footnote{In the rest of the paper, we distinguish in the nomenclature between median and scatter (represented by quartiles), given with sub- and super-script format and not showing any sign, and uncertainties, given as an upper and lower error with a plus or minus sign.}, consistent within 0.1~mag with the values for the photometric sample. Concerning colors, the spectroscopic sample of LRDs presents a F277W-F444W color around 1.2~mag, and a F115W-F200W color around 0.3~mag (consistent with the V-shape SED selection); almost identical values are obtained for the photometric and spectroscopic samples. We conclude that, globally, the spectroscopic sample is representative of the photometric sample. Table~\ref{tab:stats_photobs} also provides statistical information for subsamples drawn from the LRD spectroscopic sample defined in Section~\ref{sec:classes}, to be discussed in Section~\ref{sec:stacks}, and for which the similarity to the photometric sources is weaker.

In Figure~\ref{fig:redshifts}, we present the redshift distribution of the different samples. The median redshift of our spectroscopic sample is slightly lower than that of the parent photometric sample. Median and quartiles are $z=5.8_{4.7}^{7.0}$ for the NIRSpec sample, and $z=6.5_{5.2}^{7.6}$ for the NIRCam one, both pointing to the fall of the LRD population at $z\sim3$ (\citealt{2025arXiv250315323E,2025ApJ...986..126K,2025ApJ...988L..22I,2026ApJ..1000...59M}; although some selection effects might partially explain this decline). We do not interpret this bias as a photometric redshift problem, but it would be rather linked mainly to the low completeness of the spectroscopic sample at faint magnitudes (F444W magnitude fainter than 27). Indeed, if we cut the photometric sample to $\mathrm{F444W}<27$~mag, the redshift statistics, $z=6.2_{4.8}^{7.5}$, become more similar to the spectroscopic sample. We conclude that the spectroscopic sample of LRDs is biased towards the lower redshift end of the photometrically selected population. 

Figure~\ref{fig:redshifts} also shows redshift histograms for four different types of LRDs based on optical-to-UV luminosity ratios, from the reddest (top) to the bluest (bottom), presented in the previous subsection (see also Section~\ref{sec:subtypes}). These histograms show that, generally, the spectroscopic samples are biased towards lower redshifts compared to photometric samples, the median redshifts being around $z=6$ for the former and $z=7$ for the latter in all subtypes except the bluest. %For the bluest LRDs, the photometric sample distribution extends to lower values compared to the sources with spectroscopic observations.
Remarkably, the bias in the spectroscopic samples is largest for the reddest LRDs, median redshifts being $z\sim6.9$ and $z\sim5.2$ for the photometric and spectroscopic  samples, respectively. Overall, the redshift distributions of the photometric subsamples are asymmetric, with the distance from the median to the 25\% quartile being larger than to the 75\% quartile. This might indicate a selection effect against sources beyond $z\sim8$, i.e., LRDs at higher redshifts than currently known \citep{2025ApJ...989L...7T},  which will only be revealed  with deep MIRI data \citep[see, e.g.,][]{2024ApJ...969L..10P,2025ApJ...989..160I,2025ApJ...995...21T}.

% In addition, these blue LRDs are more sparsely sampled with spectroscopy. 

Compared to the sample of LRDs with spectroscopy discussed in \citet{2025arXiv251121820D}, our sample extends to higher redshifts and fainter magnitudes. The median and quantiles for their paper are $z=5.0^{4.2}_{6.0}$ and F444W $=25.4^{26.2}_{24.7}$~mag (after applying a 0.6~mag correction to their quoted values in Table~1), while we cover a median redshift of $z\sim6$ and a median magnitude fainter by 0.5~mag. A similar result is obtained when comparing with the spectroscopic samples in \citet{2025arXiv251215853B} and \citet{2026arXiv260120929S}.

% applied 0.6 mag correction to de Graaff.

\section{Median properties of Little Red Dots based on a stacked spectrum}
\label{sec:stacks}

\begin{figure*}[ht!]%[ht!]
\centering
%\fbox{
\includegraphics[clip, trim=1.8cm 0.3cm 2.7cm 2.0cm,width=12.0cm,angle=0]
{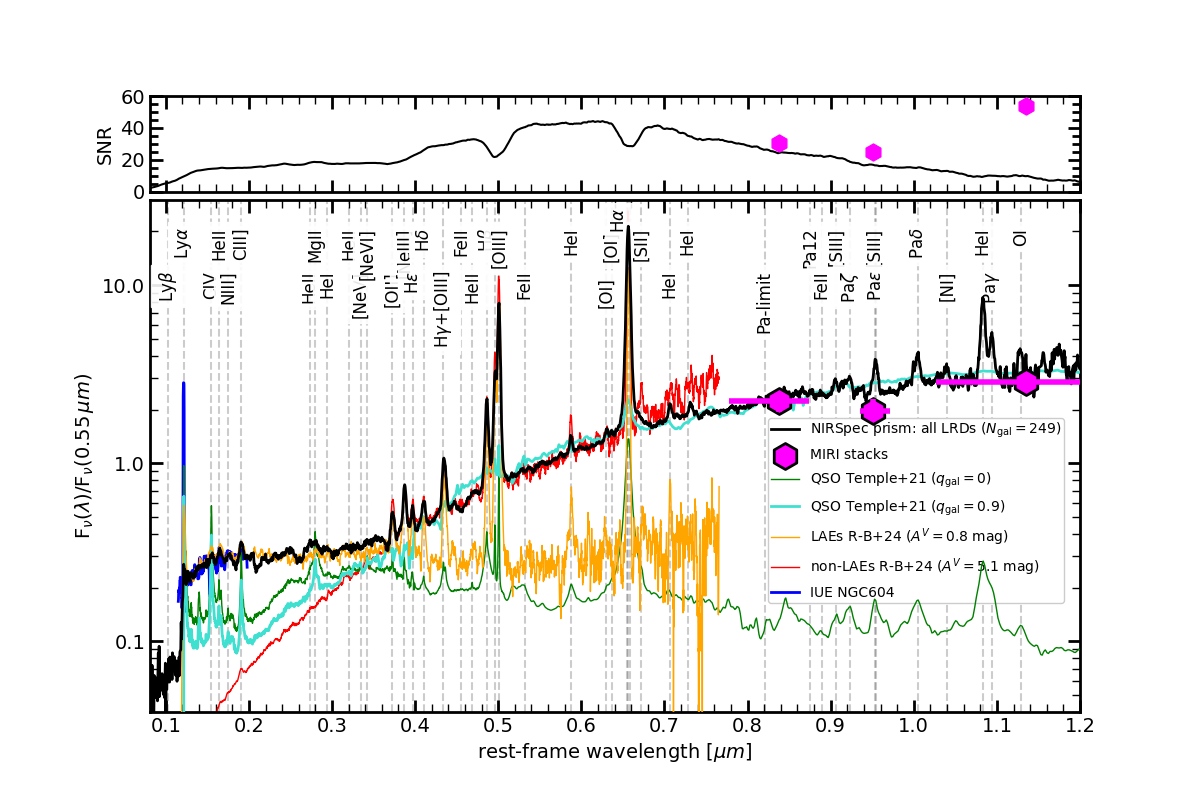}
\includegraphics[clip, trim=0.6cm 0.cm 2.7cm 1.8cm,width=8.24cm,angle=0]{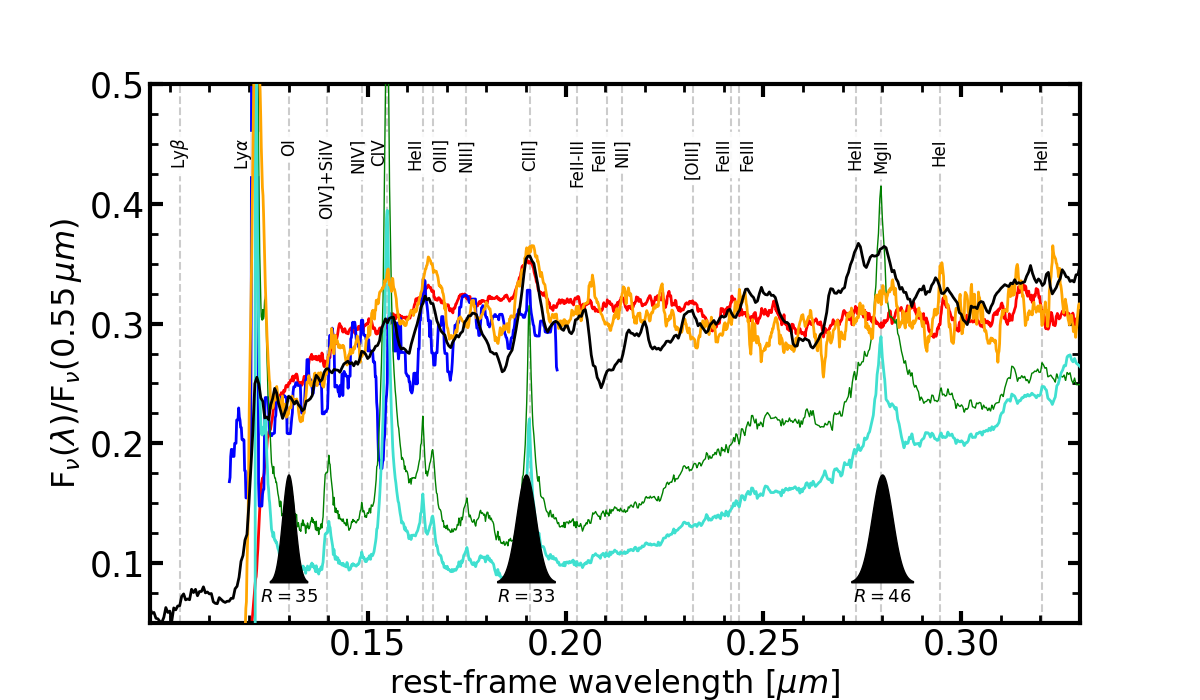}
\includegraphics[clip, trim=0.6cm 0.cm 2.7cm 1.8cm,width=8.24cm,angle=0]{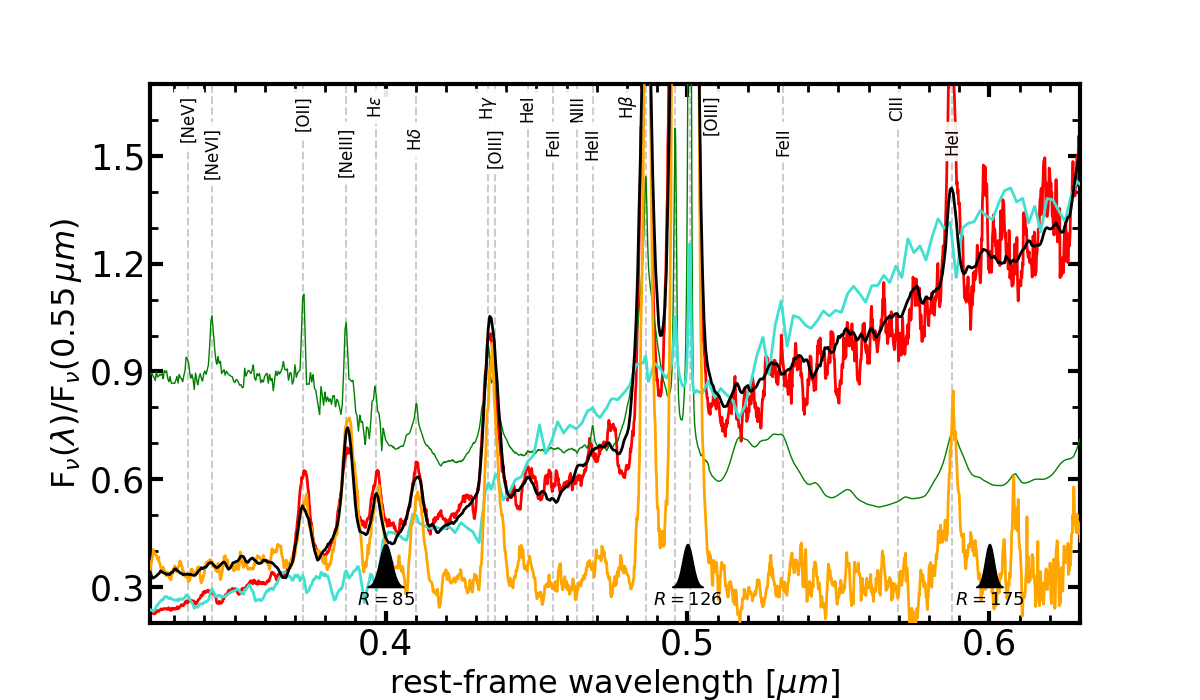}
\includegraphics[clip, trim=0.6cm 0.cm 2.7cm 1.8cm,width=8.24cm,angle=0]{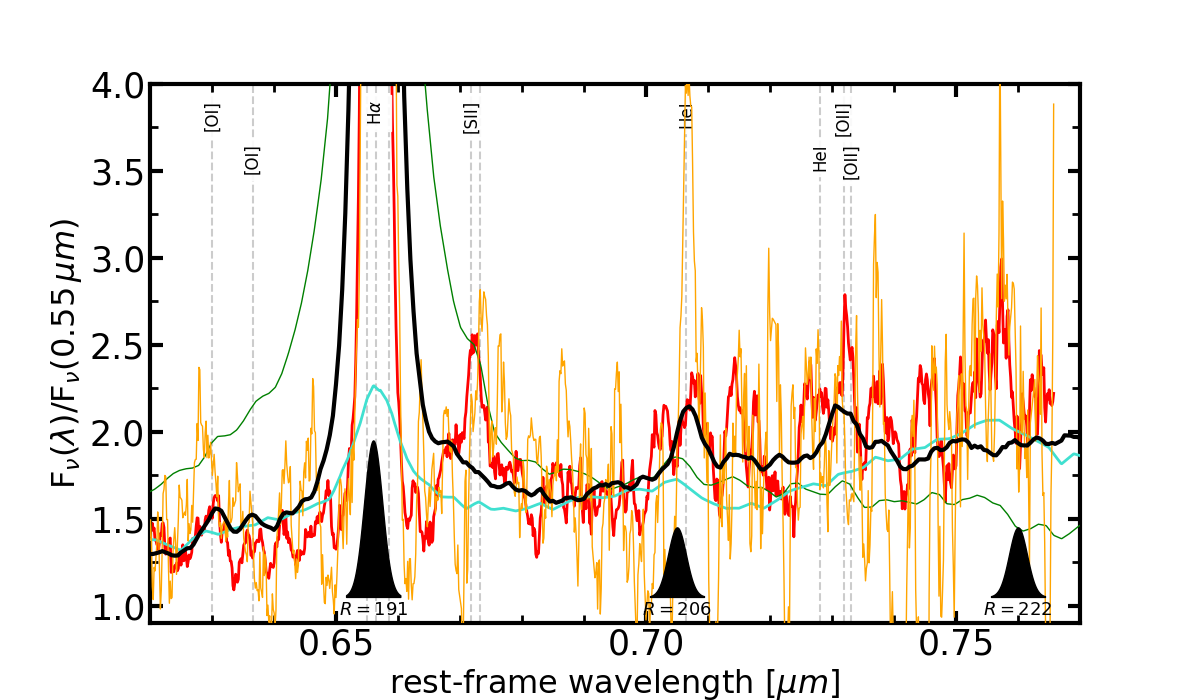}
\includegraphics[clip, trim=0.6cm 0.cm 2.7cm 1.8cm,width=8.24cm,angle=0]{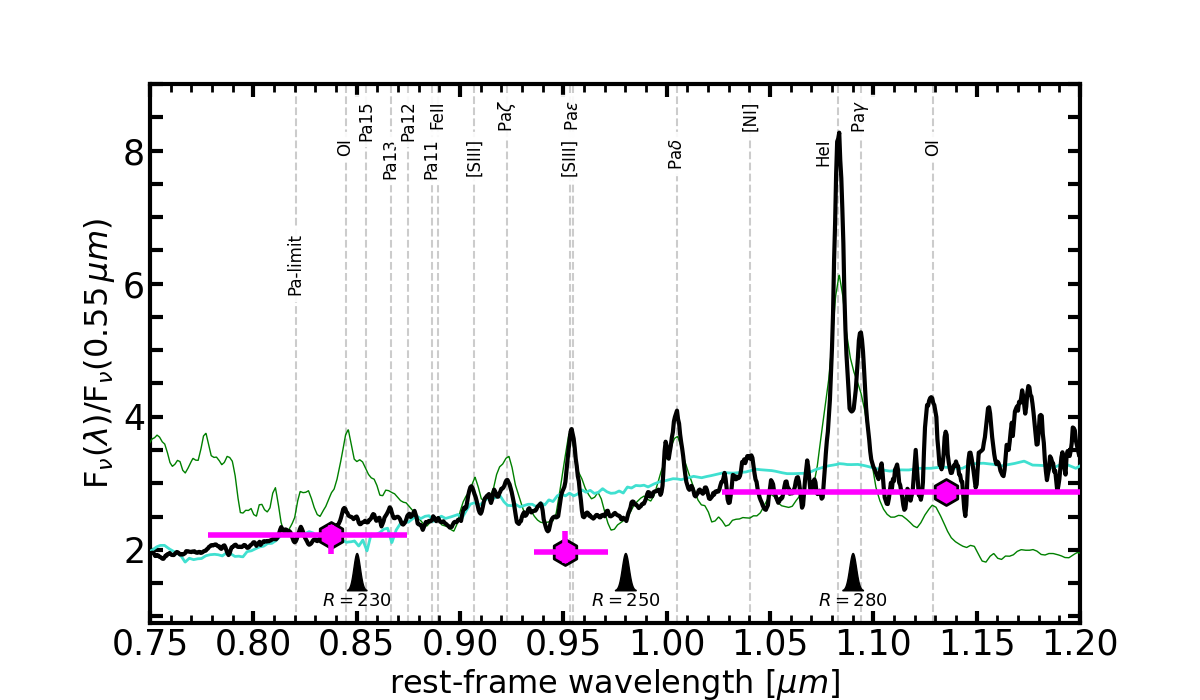}
%}
\caption{\label{fig:stacked_all}The {\it top panel} shows the stacked spectrum of the 249 LRDs with NIRSpec prism spectroscopy selected in this paper (black line). The MIRI stacks in the same rest-frame wavelength range are shown (including a horizontal error bar enclosing the wavelength range for 68\% of the MIRI points entering the stack), to demonstrate that they agree with the spectroscopic stack (see next figure for redder MIRI stacks). The spectrum is compared with QSO models from \citet{2021MNRAS.508..737T} for a pure AGN ($q_\mathrm{gal}=0$, green line) and 90\% contribution from a host ($q_\mathrm{gal}=90$\%, cyan line).
In orange and red, we show the stacked spectra for LAEs and non-LAEs presented in \citet{2024ApJ...976..193R}, after applying $A^V=0.8$ and 5.1~mag attenuation, respectively,  following a \citet{2000ApJ...533..682C} law. The QSO templates are normalized to the emission around the MgII$\lambda2798$ line, the LAE template is normalized to the emission around 0.28~$\mu$m, and the non-LAE template and median LRD SED are normalized at 0.55~$\mu$m. The panels below the main one show zoomed-in versions (in linear scale for the flux axis) in the UV ({\it middle left} panel), optical around [OIII]+H$\beta$ and H$\alpha$ regions ({\it middle right} and {\it bottom left}, respectively), and the near-IR ({\it bottom right}). In these panels, the comparison spectra are normalized to the median in the spectral window covered by the plot. The Gaussians at the bottom of the panels show the spectral resolution at different wavelengths.}
\end{figure*}
%The median SED of LRDs constructed with NIRCam and MIRI photometry and presented in \citet{2024ApJ...968....4P} is shown in magenta (median SED and scatter), and the median model given in \citet{2025ApJ...980L..29A} is shown in purple. 

%with $3\sigma$ errors shown with a gray shade

To characterize the typical properties of LRDs, we stacked all the spectra gathered in this work. We used two methods to stack, obtaining almost identical results on the analysis presented in the following sections.

For the first method, we built a rest-frame wavelength grid from 0.07 to 1.2~$\mu$m with a pixel spectral scale given by the resolution of the NIRSpec prism \citep{2022A&A...661A..80J} for a $z=6$ source (the median redshift of our sample). To calculate the sampling bins, we took into account two facts. First, we are stacking (almost) point-like sources, which lie at different redshifts and thus probe the rest-frame at a variety of nodes. Second, the spectral resolution of NIRSpec prism observations 
has been proven to be better than pre-launch expectations by up to 50\% \citep{2025A&A...702L..12S}. This improvement can be explained because the spectral resolution was measured for a uniformly illuminated shutter, compared to which point-like sources (such as LRDs) will have higher resolution  values. With this in mind, and after some testing, we used a wavelength array with 0.5 times smaller bins compared to the fiducial grid (fiducial meaning the default pixel sampling of the 1D NIRSpec spectra output by the pipeline).

For the second method, we used a constant pixel spectral size, set to 5~\AA, found to be the optimal taking into account the considerations about spectral resolution presented before.

All the analysis presented in the remainder of this paper was performed in the stacks built with the two methods, obtaining almost identical results. We choose the second method (constant pixel size) for the plots in the rest of the paper.

Since we are combining data for sources at different redshifts, we degraded all the individual spectra to a common resolution, which varies significantly with wavelength for the prism. For this common resolution, we tested using the minimum and the median values of the distribution of resolutions of the individual sources entering the stack at a given wavelength. The test showed minor differences in the final spectra, with completely negligible effects on the results.

In summary, we produce stacks with the median of all normalized spectra of individual LRDs. These spectra were rebinned to a rest-frame wavelength array typically with constant spectral pixel size (5~\AA), normalized to the flux at rest-frame 0.55~$\mu$m (averaged between 0.51 and 0.59~$\mu$m, a spectral region free of bright emission lines), and then stacked, calculating the 16\%, 50\% (i.e., we produced median stacks), and 84\% quantiles.  Following the method presented in \citet{2024ApJ...976..193R}, the difference between the 16\% and 84\% quantiles, divided by the square root of the number of objects entering the stack at each wavelength, was used as the estimated error of the median. We note that this calculation is linked to two effects: the intrinsic diversity of the SEDs of LRDs, and the noise in the individual spectra. The stacks were scaled in flux to match the median optical luminosity density (at 5100~\AA) of the different samples used in the analysis assuming $z=7$ (approximate median redshift of the photometric sample).

The resulting stacked spectrum combines data for more than 200 LRDs spanning from Ly$\alpha$ to H$\alpha$ (both included), and more than 60 LRDs up to Pa$\gamma$. Beyond that wavelength, only 10 LRD spectra reached Pa$\beta$, so we cut the spectroscopic stack at 1.2~$\mu$m. However, we complement the stack with MIRI photometric data to probe the behavior of the median LRD up to 4~$\mu$m rest-frame wavelength. For the MIRI extension, we proceeded in the same way as for the spectra: we normalized the fluxes, and we calculated medians and uncertainties in bins. In the imaging case, we constructed these bins to match the SNR of the spectral stack up to the longest wavelength possible, maximizing the spectral information. This was achieved by using at least  ten sources per bin, which provided SNR around 10 up to 2~$\mu$m and above 5 up to 2~$\mu$m. This strategy implies that each bin has a different width.

The stacked spectrum of LRDs is presented in Figure~\ref{fig:stacked_all}. The SNR of the stack, defined as the median flux divided by its error, is larger than 10 (for 5~\AA\ spectral bins) for the full range between Ly$\alpha$ to Pa$\gamma$. We note that the error of the median should be interpreted both in terms of the statistical uncertainty in the determination of the median spectrum as well as in terms of the intrinsic dispersion of the LRD population  (i.e., their diversity). It is also affected by the normalization performed before stacking and the number and quality of spectra included at each wavelength. These factors explain the structure seen in the SNR curve shown in the top panel of Figure~\ref{fig:stacked_all}. In particular, we interpret the local minima in the brightest emission lines in terms of a larger real dispersion of the nebular emission in the LRD population, given the other factors should be very similar for the narrow wavelength ranges around the line peaks.

The LRD stack is very similar to the median SED of LRDs constructed in \citet{2024ApJ...968....4P} with NIRCam and MIRI data (see also the imaging stack in \citealt{2025ApJ...991...37A}). The median LRD presents an optical-to-UV luminosity ratio, $\mathrm{L_{5100}/L_{2500}}$, (defined as the ratio between $\mathrm{L}_\nu$ values at 0.51 and 0.25~$\mu$m) of 2.7, which for the typical redshift of LRDs, $z\sim6$, would translate to $\mathrm{F200W-F356W}\sim1$~mag and an optical slope $\beta^\mathrm{opt}_\nu\sim2.5$ (as defined in \citealt{2025arXiv251215853B}). In the UV, the spectrum is quite flat, with a slope $\beta^\mathrm{UV}_\nu\sim0.2$. The median spectrum of LRDs presents a flattening and slight curvature in the near-infrared, at wavelengths beyond $\sim0.8$~$\mu$m, as found in previous analyses of MIRI data \citep{2024ApJ...968....4P,2024ApJ...968...34W}. The SED rises slowly at least up to 4~$\mu$m (see the discussion in the next subsection, pointing to Figure~\ref{fig:stacked_all_miri}). 

\subsection{Comparison of the median LRD spectrum with other types of sources}

The median LRD spectrum presented in Figure~\ref{fig:stacked_all} is now compared to some relevant templates of different types of sources. We first compare with QSO templates from \citet{2021MNRAS.508..737T}. The colors of the pure QSO template are dominated by the accretion disk and broad line region (BLR) in the UV, and do not present  the typical V-shaped SED of LRDs. If we normalize the LRD stack and the QSO template to the UV emission (at $\sim0.28$~$\mu$m), we see that the QSO spectrum is much redder than LRDs in the UV and bluer in the optical and near-IR. A QSO template with 90\% contribution from the host galaxy, which is still dominated by the AGN in the UV (the SED is very similar in both \citealt{2021MNRAS.508..737T} templates bluewards of 0.3~$\mu$m), would fit the continuum of LRDs quite nicely at all wavelengths redwards of 0.28~$\mu$m, but would not reproduce the UV either. Some relevant lines are present both in QSOs and in LRDs, for example, N\,IV]$\lambda$1486, and very remarkably, redwards of 0.2~$\mu$m, Fe\,III emission complexes at wavelengths around 2030, 2105, or 2418~\AA, or Fe\,II around 3150 and 3280~\AA\ \citep{1997ApJ...489..656L,2001ApJS..134....1V}.

We also compare the LRD median spectrum with the stacks of star-forming galaxies presented in \citet{2024ApJ...976..193R} for both Ly$\alpha$ emitters (LAE) and non-LAEs. The general shape of the LAE template does not fit the LRD. The intrinsic UV slope of LAE is $\beta_\lambda^\mathrm{UV}\sim-2.5$, which is bluer than the typical $\beta_\lambda^\mathrm{UV}\sim-2$ for LRDs and star-forming galaxies at $z<10$ \citep{2022ApJ...941..153T,2024MNRAS.531..997C,2025ApJ...994..212M}. Thus, in Figure~\ref{fig:stacked_all}, we apply some dust attenuation to the \citet{2024ApJ...976..193R} stacks based on the \citet{2000ApJ...533..682C} law ($A^V=0.8$~mag). Apart from the general shape, the spectral features in the UV for LRDs are very similar to those from the LAE stack, as seen in the zoomed version of the spectrum comparison in the middle-left panel of Figure~\ref{fig:stacked_all}.  Both the LRD and LAE spectra present some Ly$\alpha$ emission, fainter in the case of LRDs. Other relevant emission lines seen in both spectra are C\,III]$\lambda$1908, He\,II$\lambda$1640 and Mg\,II$\lambda$2798. Although not covered by the LAE stack, we remark on the non-zero emission bluewards of Ly$\alpha$.

%, with LRDs showing some Ly$\beta$ emission (blended with OVI$\lambda$1035, if present, at our resolution). 

The LRD composite is also remarkably similar to the IUE far-UV spectrum of NGC604, a famous H\,II region in the Triangulum galaxy \citep{2000MNRAS.317...64G,2003AJ....125.3082B}, shown here to represent a pure star-forming source (in fact, an OB association) with no contamination from an AGN. The spectrum of NGC604 covers both the central young (3~Myr old) cluster as well as the HII region, showing features from both nebular and wind resonance ultraviolet stellar lines. Remarkably, the continuum emissions near Ly$\alpha$ for LRDs and the giant HII region are very similar, potentially due to two-photon emission although, especially for LRDs, damped Ly$\alpha$ features might be relevant.

The comparison of the LRD and H\,II region spectra presents several distinct differences, e.g., a stronger Ly$\alpha$ emission and also stronger emission bluewards of it in NGC604  (which can be related to differences in the IGM absorption at low and high redshift), and a P-Cygni C\,IV absorption in NGC604 (typically linked to stellar winds and ISM absorption). Nevertheless, this absorption feature would be undetectable in our low spectral resolution spectrum and, interestingly, the peak emission around the C\,IV line (redshifted from the rest-frame value) coincides very well with the local H\,II region and the median LRD spectrum. 

All these comparisons would point to a significant contribution of star formation to the continuum emission of the typical LRDs in the UV (0.1--0.4~$\mu$m), especially in the far-UV (bluewards of $\sim0.2\,\mu$m). Consequently, the UV emission lines (at least those bluewards of 0.35~$\mu$m) could have a more even contribution between star formation and an AGN.

While the LAE stacked spectrum in  \citet{2024ApJ...976..193R} is remarkably similar to the UV spectrum of LRDs (after applying some attenuation, $A^V=0.8$~mag) and even reproduces the emission lines bluewards of H$\beta$ and [O\,III]$\lambda\lambda4960,5008$ (see middle right panel in Figure~\ref{fig:stacked_all}), it would not be able to explain the continuum redwards of $\sim0.4$~$\mu$m. Some other component must contribute to these wavelengths, accounting for more than 70\% of the flux at 0.6~$\mu$m, and even higher fractions at $0.75$~$\mu$m (the limit of the LAE template). The spectra of LRDs in the optical and near-infrared present less structure than in the UV. The only prominent features (at the resolution of the prism) are the hydrogen Balmer lines, [OII]$\lambda3728$,  [O\,III]$\lambda4960,5008$ and [O\,III]$\lambda4363$, [NeIII]$\lambda3870$, the HeI lines at 5877, 7067, and 10833~\AA, [NI]$\lambda$10400 (seen in low ionization AGN, \citealt{2006A&A...457...61R}). The whole optical spectrum is very similar to the non-LAE stack in \citet{2024ApJ...976..193R}, after applying a significant amount of reddening to reproduce the continuum ($A^V\sim5$~mag using a \citealt{2000ApJ...533..682C} law). However, this large attenuation would produce a spectrum which is too red to fit the LRD beyond H$\alpha$. The spectral slope for LRDs there would need a significantly lower attenuation to be reproduced by the non-LAE spectrum, around $A^V\sim1.5$~mag. 

We conclude that the UV part of the spectrum of LRDs seems dominated by star formation, more clearly in the far-UV, given the similarity with the H\,II region spectrum. The UV spectrum does not show the typical features and profile of QSOs. We cannot exclude {\it a priori} emission from accretion disk and BLR in the near-UV (around 2600-3000~\AA). An AGN with a hotter accretion disk could be brighter than the \citet{2021MNRAS.508..737T} template in the far-UV and reproduce the LRD UV slope more closely. The nature of the optical and  the near-infrared emission is elusive, we cannot find a convincing agreement with empirical or theoretical single-component templates. 

In the following sections, we analyze each spectral range separately, looking for clues about the nature of the typical (median) LRD.

\subsection{UV range}

%The anchor points to estimate the continuum are marked with red points, the spline fitting them is shown with a red line.

As remarked in the previous section, the most prominent emission lines detected in the UV (i.e., the ones with higher equivalent width, EW), are C\,III]$\lambda1908$,  Mg\,II$\lambda2798$, Ly$\alpha\lambda1216$, or He\,II$\lambda1640$. Some fainter peaks might be identified with He\,II$\lambda2733$, or C\,IV$\lambda$1549.

\begin{figure}[htp!]%[htp!]%[ht!]
\includegraphics[clip, trim=1.0cm 0.5cm 2.0cm 2.cm,width=8.5cm,angle=0]{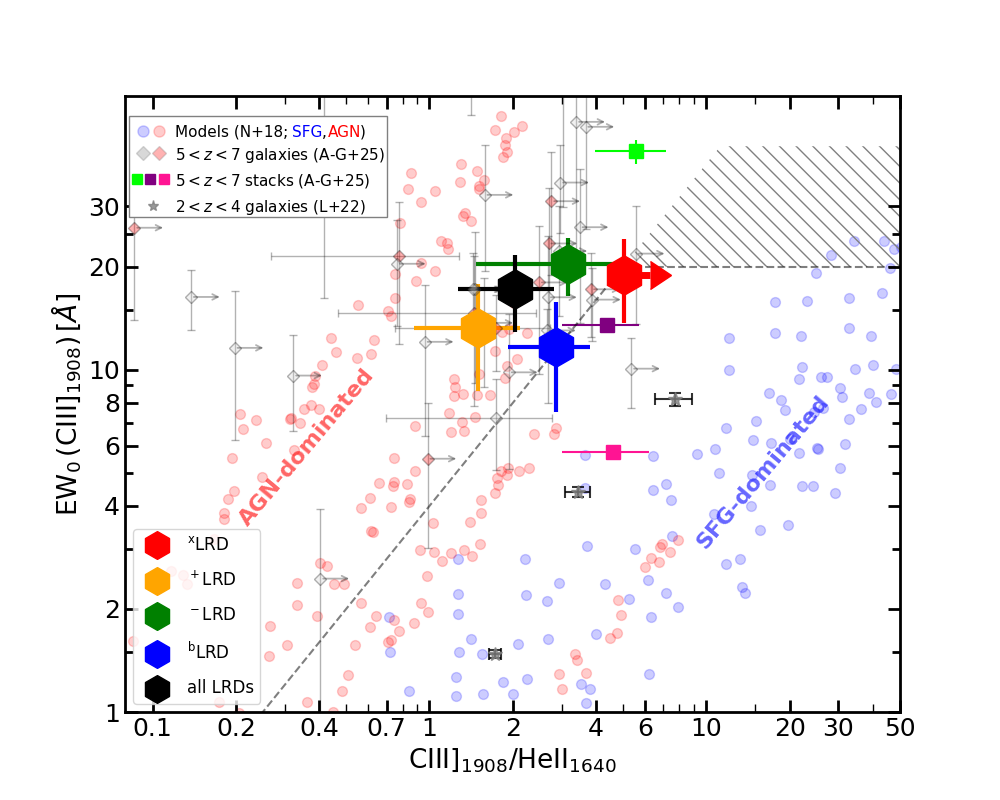}
\caption{\label{fig:ciii_heii}Position of the LRD median (black hexagon) in a UV line diagnostic plot comparing C\,III] EW $vs.$ C\,III]/He\,II intensity ratio. We also show the measurements for different subtypes of LRDs (colored hexagons) based on optical-to-UV luminosity ratios, $\mathrm{L_{5100}/L_{2500}}$, as defined in Section~\ref{sec:classes}. We compare with a compilation of measurements for $z>2$ galaxies, extracted from \citet{2025arXiv251216365A}. Gray diamonds depict individual sources at $5<z<7$ (presented in that same paper), with those showing broad H$\alpha$ emission filled in red; stars show sources at $2<z<4$ (from \citealt{2022A&A...659A..16L}). In lime, magenta and deep pink, stacks for strong, weak and non-detection C\,III] emitters are shown with squares \citep{2025arXiv251216365A}. Models from \citet{2022MNRAS.513.5134N} of AGN-dominated and star formation-dominated sources are shown with orange and blue dots, also depicting the separation lines between them, as well as the hybrid region (dashed).}
\end{figure}

In Figure~\ref{fig:ciii_heii} we show a diagnostic diagram involving the most prominent of these UV lines, typical of both AGN and star-forming galaxies \citep{2001AJ....122..549V}. The fluxes and equivalent widths of the lines in this plot were measured with the {\sc lime} software \citep{2024A&A...688A..69F}, first determining the continuum with a spline fitting a set of anchor points constructed with the 5\% lowest flux values in the 1200 to 2400~\AA\ region and using Gaussians with widths larger than those corresponding to the spectral resolution of the stack at each specific wavelength (see Figure~\ref{fig:stacked_all}). For the He\,II measurement, a OIII]$\lambda\lambda$1661,1666 component was also considered, which would be unresolved in our stacks given that the nominal spectral resolution is $R\sim35$, $R\sim50$ considering the improvement for point-like sources, corresponding to a FWHM of 33~\AA. The measurements include uncertainties in the determination of the continuum, which in fact are the main contributor given that the UV spectrum of LRDs (especially the reddest) presents significant structure.

The point for the stack of all LRDs (and also for the subtypes) lies near the boundary region between star-formation and AGN dominated galaxies according to the models presented in \citet{2018A&A...612A..94N}. For the reddest LRDs (red hexagon), He\,II is not detected, so we provide a lower limit for the C\,III]/He\,II ratio. In general, all LRD types have very similar C\,III] and He\,II line emission as  super star clusters at Cosmic Noon \citep{2014MNRAS.445.3200S,2017ApJ...842...47V}. 

%If we combine the CIII]/He\,II ratios shown in Figure~\ref{fig:ciii_heii}, jointly with He\,II EW (around 1--2~\AA, not shown in plot), the result would be consistent with chemically evolved systems according to \citet{2022MNRAS.513.5134N}. Very similar results are obtained when considering the CIV line compared to He\,II.

%although the actual positions of the peaks are offset, probably indicating that several weaker lines contribute to the signal at those wavelengths (or some more complex effects introduced by the limited spectral resolution)

\subsection{Optical range}
\label{sec:optical}

The optical range exhibits strong emission lines of the hydrogen Balmer series, [Ne\,III] at 3870 and 3969~\AA, [O\,II] at 3728~\AA\ (the doublet is not resolved), and [O\,III] at 4364, 4960, and 5008~\AA, as well as He\,I at 5877~\AA, and [O\,I] at 6302 and 6366~\AA. There is also a very peculiar feature at 4550~\AA, which we describe in the second subsection below.

\subsubsection{Prominent emission in the WR blue bump}
\label{sec:wr_full}

\begin{figure*}[htp!]%[htp!]%[ht!]
\includegraphics[clip, trim=2.6cm 0.cm 3.5cm 0.cm,width=8.8cm,angle=0]{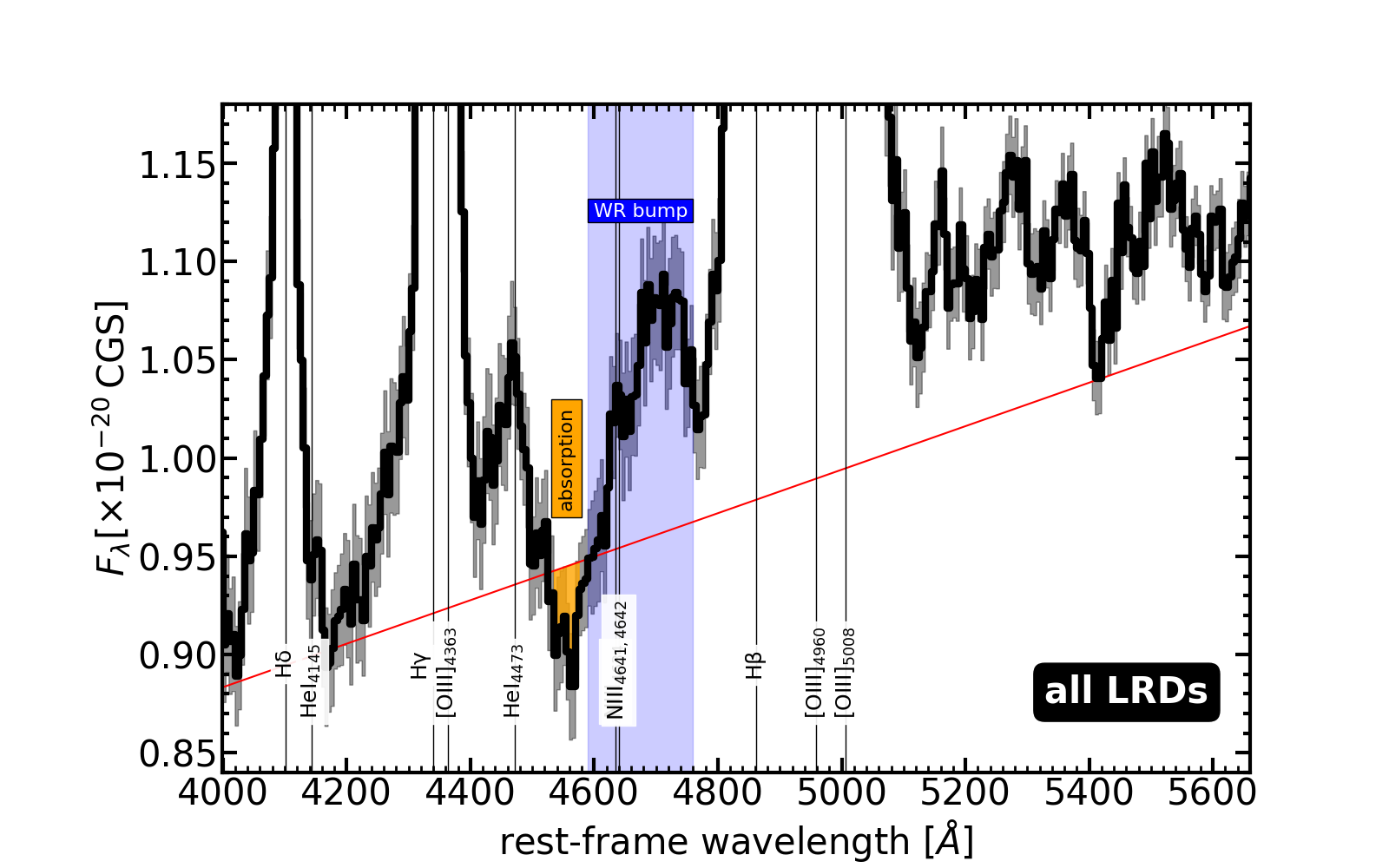}
\hspace{0.2cm}
\includegraphics[clip, trim=2.cm 0.cm 3.9cm 0.cm,width=8.8cm,angle=0]{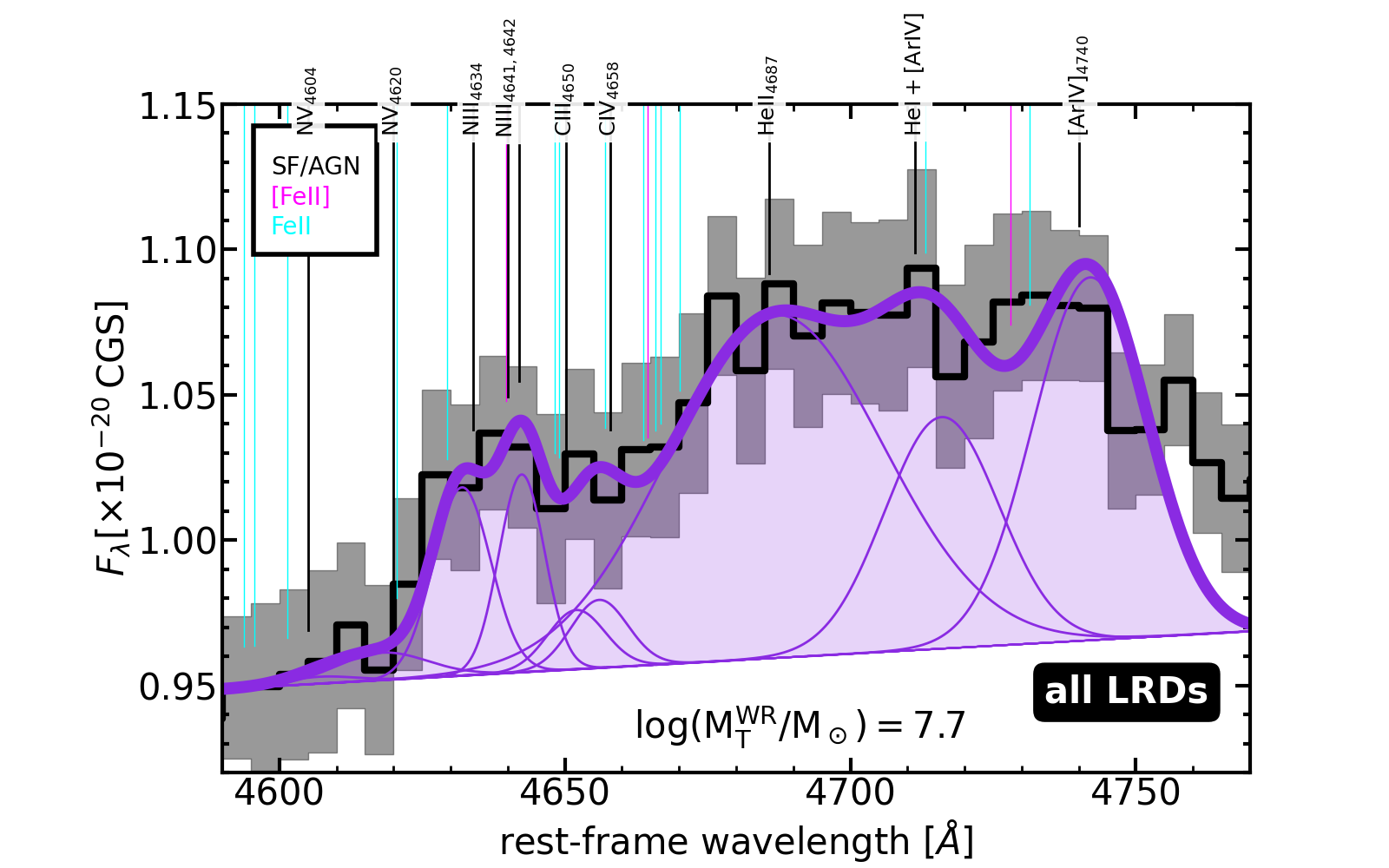}
\caption{\label{fig:wr}{\it Left}: Plot showing the estimation of the continuum emission around the WR blue bump (marked with a blue shade) for the stack of all LRDs, scaled in rest-frame flux to match the optical luminosity density at 5100~\AA of the full spectroscopic sample for $z=7$ (approximate median redshift of the photometric sample). The estimated continuum is shown with a red line. Several relevant spectral features are marked, including an absorption at 4550~\AA\ (see Section~\ref{sec:absorption} for details). {\it Right}: Spectral region around the WR blue bump. The stack of spectra for all LRDs (black line with errors represented by the gray shading) is fitted with Gaussians placed at the wavelengths of emission lines typical of this kind of massive stars and from AGN (marked with black vertical lines in the plot). The He\,II$\lambda4687$ line is allowed to have a narrow and a wide component (the latter defined as $FWHM>2000$~km~s$^{-1}$), and the intrinsic widths of the remaining lines are free. The continuum is measured as shown in the left panel. In cyan, we also depict the wavelengths for the Fe\,II line forest, and in pink the [Fe\,II] lines. The plot gives the calculated total mass of a starburst explaining the He\,II emission in terms of WR stars (WN subtype).}
\end{figure*}

Apart from the bright lines, our stack shows some fainter lines in key spectral ranges. Figure~\ref{fig:wr} shows the region around the blue bump typical of Wolf Rayet (WR) stars. Several papers have claimed the detection of WR features at $2.5\lesssim z<6$ with JWST \citep{2024A&A...690A.269R,2025ApJ...979...87M,2025arXiv251113591B}. Our spectrum shows peaks at wavelengths consistent with N\,III$\lambda\lambda4634,4640$, N\,V$\lambda\lambda4605,4620$, and  N\,III$\lambda$4511 (also  N\,IV]$\lambda1486$, not shown in Figure~\ref{fig:wr}, but see Figure~\ref{fig:stacked_all}), as well as a strong bump which would correspond to broad He\,II$\lambda4687$ emission. The detection of this line for LRDs has also been reported in \citet{2025arXiv250818358W},  where they discuss a possible coexistence of an AGN and nuclear starbursts, a scenario supported and quantified in this section. Three even fainter lines compatible with Fe\,II$\lambda4590$ and [Ar\,IV]$\lambda\lambda4711,4740$ are observed in the same region, although higher-resolution spectroscopy is required to confirm the identity of each individual feature in the prism blend. Some indication of faint carbon lines in this bump are visible, but are weaker than the nitrogen features. The potential presence of carbon could be confirmed by studying the WR red bump at around 5800~\AA, produced by carbon emission, but our stack is inconclusive, due to the contamination of He\,I$\lambda5877$ in our low spectral resolution data (see Figure~\ref{fig:stacked_all}).

%, but can be deblended from He\,II$\lambda$4471, see Figure~\ref{fig:stacked_all_miri}

Tentatively interpreting the spectral region around 4650~\AA\ as a WR signature (i.e., the typical WR blue bump), and using all the lines mentioned above, we fitted the spectrum shown in Figure~\ref{fig:wr} with the {\sc lime} software \citetext{\citealp{2024A&A...688A..69F}; we also checked the results obtained with the {\sc fantasy} code, \citealp{2023ApJS..267...19I}}. The continuum was measured in two ways. First, we used a power law between 4550 and 4770~\AA. We repeated the power-law fit, this time using the 5\% lowest flux values in the 4000 to 5650~\AA\ range, given that there were indications of an absorption at $\sim4550$~\AA (see next section), and also because the red end of the WR bump is could be affected by the H$\beta$ broad component. We show and discuss results with the second method, pointing that line fluxes change by $\sim15$\% when using the first continuum determination. We warn the reader that, even though the continuum in the optical shows less structure than in the UV, the determination of the continuum is the main uncertainty contributor to the line flux errors.

After the determination of the continuum, Gaussians were placed at the wavelengths for emission lines in this spectral range. The widths of the lines were fitted for each line separately, allowing the He\,II line to have a wide and narrow component, while for the remaining lines only one narrow component was used. Fixing the width of all the narrow lines to be the same reached almost identical results. Based on the nominal NIRSpec spectral resolution, the instrumental $\sigma$ we used is 16~\AA, FWHM 38~\AA, i.e., $R\sim120$ (which is the median of all sources  entering the stack at these wavelengths, the scatter being 30). However, the GTO Team is typically applying a 1./0.7 $\sim$ 1.4 multiplicative correction to the nominal NIRSpec spectral resolution (i.e., the one corresponding to the uniformly illuminated shutter), consistent with the findings by \citet{2025A&A...702L..12S}. Assuming the spectral resolution is indeed better than nominal (which should be the case for our point-like sources), the instrumental $\sigma$ would be 12~\AA\ (resolution $R\sim170$). 

As a reference, for the [OIII]$\lambda\lambda4960,5008$ lines, for which we have very good signal-to-noise ratio (SNR), we measure $\sigma$ around 13~\AA, which gives $R\sim160$. The nominal resolution for the stack at 0.5~$\mu$m is $R=130$, 1/0.9 with respect to the 160 value given before. If we consider the 0.7 correction in \citet{2025A&A...702L..12S}, those narrow lines would be slightly resolved (or close to resolved). In any case, we conclude that the calculations about the resolution used for Figure~\ref{fig:wr} (and throughout the rest of the paper) are robust, running on the conservative side.

The fits shown in Figure~\ref{fig:wr} would favor the presence of nitrogen-rich Wolf-Rayet (WN) stars rather than carbon-rich WC stars \citep{2007ARA&A..45..177C}, which could be linked to the high nitrogen abundances seen in many high redshift galaxies \citep{2023A&A...677A..88B,2023MNRAS.523.3516C,2024A&A...681A..30M,2024A&A...687L..11S,2024MNRAS.535..881J,2025arXiv251113591B}, also in the LRD in \citet{2026MNRAS.545f2110J}. This characteristic has been claimed to be linked to spatial compactness \citep{2024MNRAS.529.3301T}. Our fits indicate $\sigma=1130\pm310$~km~s$^{-1}$ (intrinsic) for the He\,II line, which is a typical value in nearby and intermediate redshift WR galaxies \citep{2008A&A...485..657B,2010A&A...516A.104L,2010A&A...514A..24H}. This value is 25\% smaller than the broad component we measure for H$\beta$, compatible within errors, thus we cannot rule out that the He\,II line is linked to an AGN (if the Balmer lines are dominated by nuclear activity).

The rest-frame equivalent width of the He\,II line is $\sim6$~\AA. The width is comparable to some of the highest values measured for WR galaxies in the local Universe by \citet{1992ApJ...401..543V}; the galaxy with the most extreme line in this paper has twice the EW of our stack. We also compare with  \citet{2008A&A...485..657B}, who studied WR galaxies based on the "blue bump" EW, adding all lines together; in our case we measure an EW of ~13~\AA for the whole "blue bump" (although we also include He\,I and Ar\,IV lines in the calculation, so not directly comparable), comparable to the most extreme WR galaxies in this work. Finally, our measurements can also be compared with \citet{1998ApJ...497..618S}, who find that our "blue bump" EW could only be reached in a very narrow  range of ages (around 4~Myr) for 0.4\,Z$_\odot$ galaxies and for more extended periods at higher metallicities.

Translating the flux of this broad emission line to stellar content by using the typical luminosity of local low metallicity late-type Wolf-Rayet nitrogen stars (WNL, \citealt{2006A&A...449..711C}), assuming the WNL to O-type stellar ratio in \citep{1998ApJ...497..618S}, and a Chabrier IMF to calculate the total mass of the starburst responsible for this emission, we would obtain a total mass of $\mathrm{M_\bigstar}=10^{7.7\pm0.3}$~M$_\odot$. This stellar mass is consistent with the values obtained from clustering analysis in \citet{2025ApJ...988..246M}, $10^{7.7\pm0.3}$~M$_\odot$, as well as with dynamical mass constraints \citep{2024ApJ...969L..13W,2024MNRAS.535..853J,2025MNRAS.tmp.1995D,2025arXiv251000101D}.

We note that the WR interpretation would need a detection of broad He\,II$\lambda1640$, with a typical Case-B ratio with respect to He\,II$\lambda4687$ of 7.8 \citep{2019AJ....158..192L}. The He\,II$\lambda1640$ is well detected in our stack ($\mathrm{SNR}\sim10$), but the spectral resolution of the prism is low in the rest-frame UV, so we cannot study the line profile. Assuming the entire line flux comes from a broad component with the same origin as the blue bump emission (i.e., massive stars), we obtain $F(\mathrm{He\,II_{1640}})/F(\mathrm{He\,II_{4687}})=3.5\pm0.4$, which would imply 0.9~mag attenuation between the two wavelengths, translating to $A^V\sim1$~mag with the \citet{2000ApJ...533..682C} dust law. This value is similar to that applied to the LAE stacks discussed in Figure~\ref{fig:stacked_all}, and the SED fitting exercise described in  Section~\ref{sec:sedfit}. This level of dust obscuration would imply infrared luminosities smaller than $10^{11}$~L$_\odot$ (see Section~\ref{sec:timescales} for more details).

% 2006A&A...449..711C: 2.2e35 WNL
%

\subsubsection{Unknown absorption feature at 4550~\AA}
\label{sec:absorption}

Our stack and spectral analysis (including the fit of the continuum in the 3000-6000~\AA\ range) also highlight the presence of a strong absorption feature near 4550~\AA\ (Figure~\ref{fig:wr}, left).
This feature can be observed in most LRDs with adequate SNR \citep{2024arXiv241204557L,2025ApJ...994L...6T}, but was initially interpreted as lack of emission infill from the broad permitted Fe\,II \citep[see e.g. the models presented in ][]{2024arXiv241204557L}.
However, deep higher-resolution spectroscopy confirmed the surprising absorption nature of this dip \citep{2026MNRAS.545f2235J,2025arXiv251000101D}, even though both works did not report a clear identification.

While LRDs are known to display strong hydrogen and helium absorption \citep[e.g.,][]{2024ApJ...963..129M,Wang_BLAGN,2024MNRAS.535..853J,BigredDot,2025arXiv251000101D}, no known H or He feature 
matches 4550~\AA.
Metal absorption features have been observed in many individual LRDs, both in the rest-frame optical
starting from Na\,I and Ca\,II at $z\sim2$ \citep{2024MNRAS.535..853J}, to Ca\,I, Fe\,I, Fe\,II, Ba\,II, K\,I at $z\sim0.1$ \citep{2025arXiv250710659L,2026MNRAS.545f2235J}, and Ca\,II and Na\,I at $z\sim7$ \citep{2025arXiv251000101D};
as well as in the rest-frame UV \citetext{Fe\,II UV1--UV3, \citealp{2025arXiv251000101D,2025arXiv251000103T}}.

The widespread observation of metal absorption in the spectra of LRDs suggests that the 4550-\AA\ dip may also be related to metal absorption. Given the common presence of neutral and low-ionization species, an identification with one or more of Fe\,II, Ba\,II, Fe\,I may be credible, even though it remains to be seen whether these transitions are consistent with what is seen (or absent) in other regions of the optical spectrum.

Among the possibilities, the most plausible alternative is a link to Fe\,II emission. The region around 4550~\AA\ is covered by the Fe\,II F-band \citep[e.g.,][]{2025A&A...694A.289K}, which in fact is used (compared to H$\beta$ emission) to estimate Eddington ratios \citep{1992ApJS...80..109B}. The Fe\,II emission at these wavelengths have been claimed to be very weak and possibly a sign of  metal-poor BLRs at high redshift \citep{2025A&A...700A.203T}. However, we identify with Fe\,II some strong emission in the UV and in the 5300~\AA region (the G band in \citealt{2025A&A...694A.289K}), thus the absorption would be rather linked to high gas densities, low ionization parameters \citep{2010ApJS..189...15K}, or depletion of iron into dust grains \citep{2010ApJ...721.1835S}. Weak F-band emission compared to the G-band is common in AGN with high Balmer decrements \citep{2010ApJS..189...15K}, as is observed for LRDs.

An intriguing alternative is represented by molecular absorption, since at least two studies reported the
presence of G-band absorption at 4300~\AA\ \citep{2026ApJ...997..364L,2026MNRAS.545f2235J}, a feature typical of low-temperature stars and attributed to CH. However, we found no suitable match with equally abundant diatomic molecules, NH, OH, and H$_2$. Strong absorption in this region is also found in supergiants (see, e.g., \citealt{2015A&A...581A..70K}).

%and in some Wolf Rayet stars in binary and multiple systems \citep{}. 

\begin{figure*}[htp]%[htp!]%[ht!]
\includegraphics[clip, trim=0.3cm 0.7cm 1.5cm 2.0cm,width=8.5cm,angle=0]{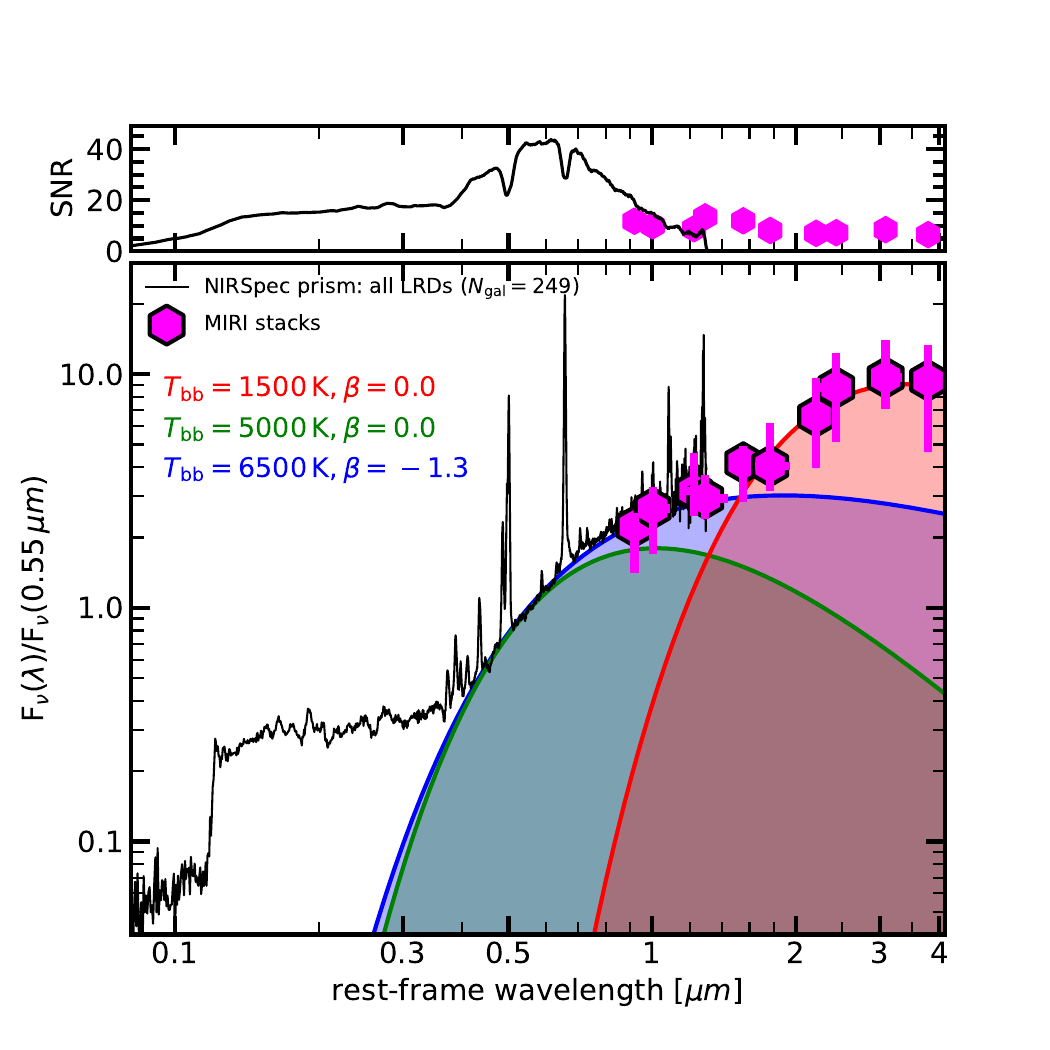}
\includegraphics[clip, trim=0.3cm 0.7cm 1.5cm 2.0cm,width=9.6cm,angle=0]{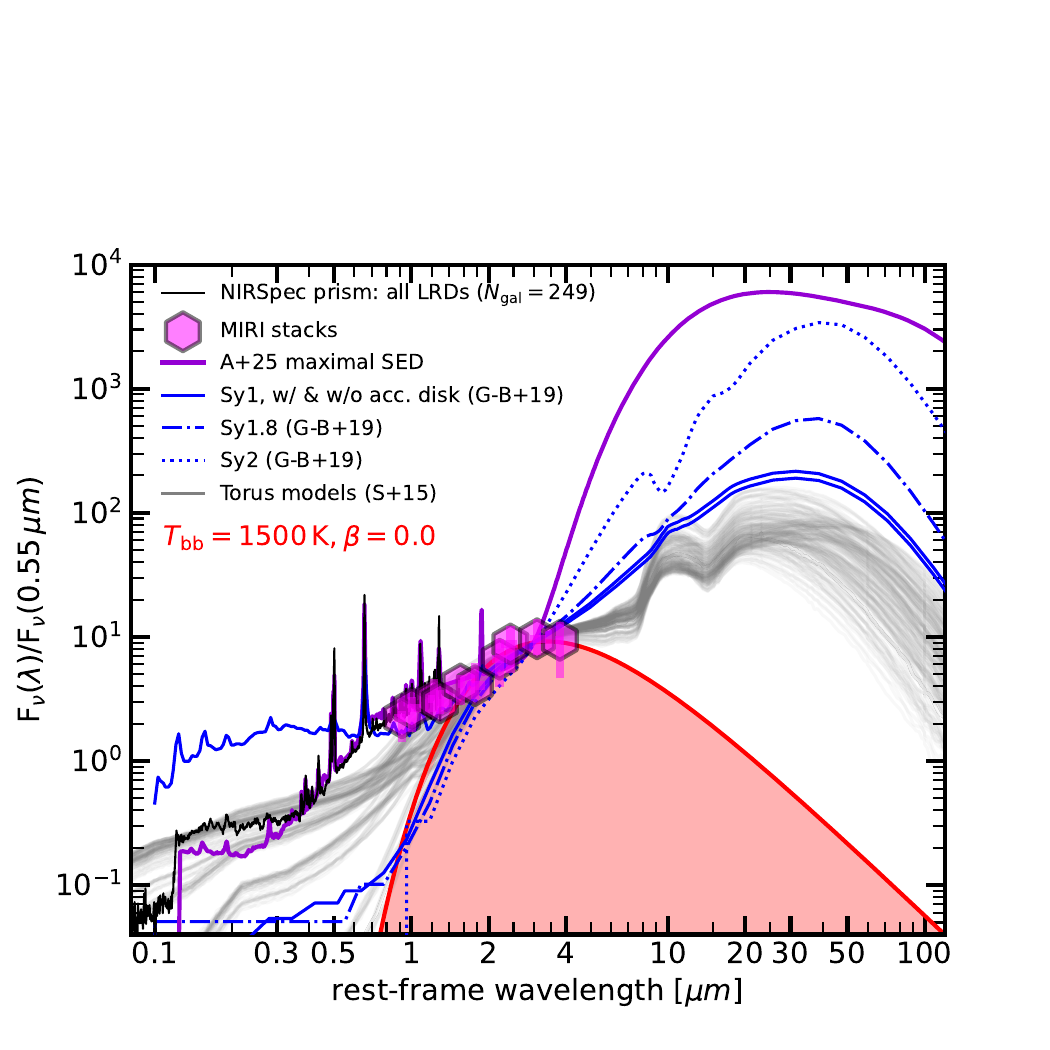}
\caption{\label{fig:stacked_all_miri}{\it Left:} Same as Figure~\ref{fig:stacked_all}, showing the LRD stacked spectrum extended to $\sim4\,\mu$m using MIRI data (magenta points and error bars, showing the median and scatter). %The magenta and purple lines show the median LRD SEDs from \citet{2024ApJ...968....4P} and \citet{2025ApJ...980L..29A}. 
The shaded blue and green regions illustrate two MBBs fitting the rest-frame optical part of the spectrum, as in \citet{2025arXiv251121820D} and \citet{2025arXiv251204208U}; we use representative parameters for the variety of LRDs, from the hot (6500~K) to the cold end (5000~K). Neither can reproduce the MIRI photometry at $\lambda >1.5\mu$m, suggesting the presence of hot dust emission. This spectral region covered by MIRI is fit to another MBB, at 1500~K (in red). {\it Right:} LRD stack spectrum and near-infrared data (from MIRI) compared to several templates and models, all normalized to the SED level of LRDs at 3~$\mu$m. In red, we plot the same MBB shown in the left panel. In blue, we show the stacked SEDs of Sy1, Sy2, and intermediate Sy subtypes (Sy1.8 in legend) from \citet{2019MNRAS.486.4917G}. In gray, we show the torus models from \citet{2015A&A...583A.120S} which present color in the 2--4~$\mu$m spectral regio similar (within 1$\sigma$) to the LRD stack. In purple, we show the maximal SED model for LRDs presented in \citet{2025ApJ...991...37A}.}
\end{figure*}

Currently, the main hurdle preventing a secure identification is the trade-off between high-SNR, low-resolution prism data (as seen in our stack)
and high-resolution, low-SNR grating data. Deep, higher resolution spectroscopy is required to securely identify this feature. In any case, its omnipresent appearance over most of the LRD range $z=0.1\text{--}7$ strongly supports the presence of metals, as well as a warm or cool gas medium capable of supporting low-ionization species, and maybe even molecules (e.g., water, see \citealt{2026arXiv260206024W}).

\subsection{Near-infrared range}
\label{sec:nir}

Our NIRSpec stack is significantly noisier in the near-infrared compared to bluer wavelengths (given that fewer LRDs contribute to the stacks). However, significant peaks are detected in several Paschen lines, in He\,I at 1.083~$\mu$m, and [S\,III] at 9071 and 9533~\AA. %The  slope of the continuum is relatively flat, with a flux ratio of $\sim1.5$ between 0.8 and 1.1~$\mu$m, the ratio increasing to $\sim4$ from 0.8 to 4~$\mu$m.

Using MIRI imaging, we extend the stacked SED of the full LRD sample to rest-frame wavelengths of $\sim$4~$\mu$m, as shown in Figure~\ref{fig:stacked_all_miri} (magenta points). The stacked SED with the MIRI extension shows a clear flattening beyond  $\sim$2~$\mu$m, in excellent agreement with, and smoothly extending, the NIRSpec continuum (see also Figure~\ref{fig:stacked_all} for a zoom-in to $\sim$1.2~$\mu$m). This trend confirms previous photometric results indicating that LRDs exhibit moderate near-IR colors (e.g., \citealt{2024ApJ...968....4P, 2024ApJ...968...34W,  2025ApJ...980L..29A, 2024ApJ...975L...4C,2025arXiv250820177R}), in contrast to the steeply rising continua of obscured type-1 QSOs. The trend is now verified in a sample of spectroscopically confirmed LRDs.

The MIRI detections reveal a mild upturn beyond $\sim2$~$\mu$m, producing a rest-frame color of $[1-3~\mu m]\sim1.3$~mag, comparable to the median $[1-3~\mu m]=1.5$~mag measured for the F1800W-detected LRDs in \citet{2024arXiv241201887B}. While these colors are not extremely red, they suggest the presence of hot dust, with characteristic temperatures of $T\sim1500$~K. As discussed in \citet[][see also \citealt{2025ApJ...992...26L} and \citealt{2025arXiv250820177R}]{2024arXiv241201887B}, in the absence of dust, the near-infrared continuum of either a stellar population or a standard AGN accretion disk would be significantly bluer, with $[1-3~\mu m]\sim-0.5$~mag. The emission of a dust torus, represented in Figure~\ref{fig:stacked_all_miri}  by a $T\sim1500$~K blackbody, peaks around 3~$\mu$m (in $f_\nu$, 2~$\mu$m in $f_\lambda$), making the SED redder. The observed colors therefore imply a substantial hot-dust contribution. In particular, using the simple AGN-based models shown in Figure~8 of \citet{2024arXiv241201887B}, in which a hot-dust modified blackbody component is added to an obscured accretion-disk continuum, colors in the range observed for the bulk of the F1800W-detected LRDs, $[1-3~\mu m]\sim1$--2.25~mag, correspond to hot-dust fractions of $f^{\rm HD}{(3~\mu{\rm m})}\sim0.5$--0.8. Here $f^{\rm HD}{(3~\mu{\rm m})}$ denotes the fractional contribution of the hot-dust component to the total rest-frame 3~$\mu$m flux.

%corresponding to a fraction of the total emission of $f^{\rm HD}_{3 \mu{\rm m}}\sim50$–80\% (see Figure~8 of \citealt{2024arXiv241201887B}).

In the context of more recent black hole star (BH$^{\star}$) or black-hole-envelope (BHE) models \citep[][see also \citealt{2025ApJ...980L..27I} that argue in favor of the presence of dense gas in LRDs, but not necessarily as a BHE or BH$^{\star}$]{2025arXiv250316596N,2025MNRAS.544.3407K,2025arXiv251121820D}, which mathematically parameterize the continuum using modified blackbodies (MBBs), the requirement for hot dust becomes even stronger.
Note that MBBs are generally employed to model dust emission, where dust-grain emissivity in the Rayleigh-Jeans tail behaves like a powerlaw \citep{draine2006}, while -- in the context of LRDs and
optical emission -- a MBB has to be considered as an effective shape function, given that for
$T\sim 5000$~K, thermal emission is dominated by gas, not dust.
The MBB SED with temperatures $T\sim5000$~K declines rapidly toward the near-infrared, predicting much bluer $[1-3~\mu m]$ colors. This is illustrated in Figure~\ref{fig:stacked_all_miri} by two MBB models: first, a true blackbody model with $T=5000$~K, $\beta=0$ (green shaded region), representative of the best-fit results in \citet{2025arXiv251121820D} for spectroscopic LRDs at $z<4.5$; and second, a MBB $T=6500$~K, $\beta=-1.3$ model (purple shaded region). These MBBs correspond to our best fits to the stacked NIRSpec spectrum using the optical regions described in \citet{2025arXiv251121820D}, i.e., without using the MIRI photometry. While the latter reproduces the flattening at $\lambda\sim1$–2~$\mu$m, it predicts a near-zero $[1-3~\mu m]$ color, significantly bluer than observed. This discrepancy demonstrates that even within MBB-based BH$^{\star}$ frameworks, an additional hot-dust emission component (or another emitting source with cooler temperatures) is required to match the MIRI constraints \citep[see, e.g.,][]{2024arXiv241201887B, 2024MNRAS.535..853J,2025A&A...704A.313D,2026MNRAS.545f2235J}.

We illustrate this effect in Figure~\ref{fig:stacked_all_miri} by plotting a $T\sim1500$~K, $\beta=0$ MBB (in fact, a pure BB) fitting the MIRI data. This dust emission with small $\beta$ is typical of very large grains, commonly found in dense environments such as AGN tori given that they can survive in harsh environments \citep[e.g.][]{2001A&A...365...28M}, as well as iron bearing grains \citep{2016ApJ...825..136D}. %This type of dust typically produces gray attenuation laws, which have been claimed to be needed to reproduce the (UV) SED of LRDs \citep[e.g.,][]{2024ApJ...963..128B,2024ApJ...968....4P}.
See the Discussion section for more details about the possible presence of dust in LRDs.

On the right panel of Figure~\ref{fig:stacked_all_miri}, we show a comparison of the near-infrared emission of LRDs with several templates and models for AGN and tori. The SED of LRDs in the 2--4~$\mu$m spectral region resembles that of nearby Sy~1 galaxies (which have bolometric luminosities similar to LRDs), represented with blue lines showing the stacks published in \citet{2019MNRAS.486.4917G}. Later types (intermediate Sy and Sy~2 galaxies) are significantly redder. The emission of the torus in these AGN gets fainter at wavelengths bluer than $\sim1.5$~$\mu$m, following the shape of the 1500~K BB (the rough limit for dust sublimation).

To explore what might happen redwards of the wavelength (currently) covered by the MIRI data, we plot in the right panel of Figure~\ref{fig:stacked_all_miri} a subset of the torus models in \citet{2015A&A...583A.120S}. The \citet{2015A&A...583A.120S} models are characterized by five parameters: the viewing angle of the central engine, the inner radius of the torus, the cloud volume filling factor, and the optical depths of individual clouds in the torus and the disk midplane. We chose those models which have similar colors (within a factor of 1.5) to the LRD stack in the 2-4~$\mu$m spectral region. We also cut the models to those with UV emission  lying below the LRD spectrum (within a 1.5 factor). This UV emission comes from scattered light and the figure demonstrates that it can be relevant to understand the V-shape SED of LRDs. The dust emission for these subset of models peak at around 30--50~$\mu$m, exactly the same behavior observed for Sy~1 galaxies. The maximum emission for the models are 1.5--2~dex fainter than the maximal flux reported by \citet{2025ApJ...991...37A} based on MIPS, Herschel, and sub-mm data (purple curve in the Figure~\ref{fig:stacked_all_miri}). The torus models which 
more closely resemble the colors probed by MIRI present lower peak fluxes compared to Sy~1 galaxies (by up to a factor of 2). These models are typically characterized by large viewing angles ($\sim65^o$ from the polar axis), small internal radii of the torus ($\sim8\times10^{17}$~cm), no particular value of the filling factor, relatively low optical depths in individual clouds ($\tau_V\sim5$) and in the disk midplane ($\tau_V\sim30$).
%The purple and magenta curves in Figure~\ref{fig:stacked_all_miri} show comparisons with the median MIRI-detected LRD stacks from \citet{2024ApJ...968....4P} and \citet{2025ApJ...980L..29A}. The \citet{2025ApJ...980L..29A} stack differs slightly in the UV, showing a higher $\mathrm{L_{5100}/L_{2500}}$ ratio, as expected given their selection bias toward the reddest LRDs ($F277W-F444W>1.5$~mag).

\section{The diversity of Little Red Dots: stacks of different subtypes}
\label{sec:subtypes}

\begin{figure}[ht]%[htp!]%[ht!]
\includegraphics[clip, trim=1.0cm 0.5cm 2.0cm 1.1cm,width=8.3cm,angle=0]{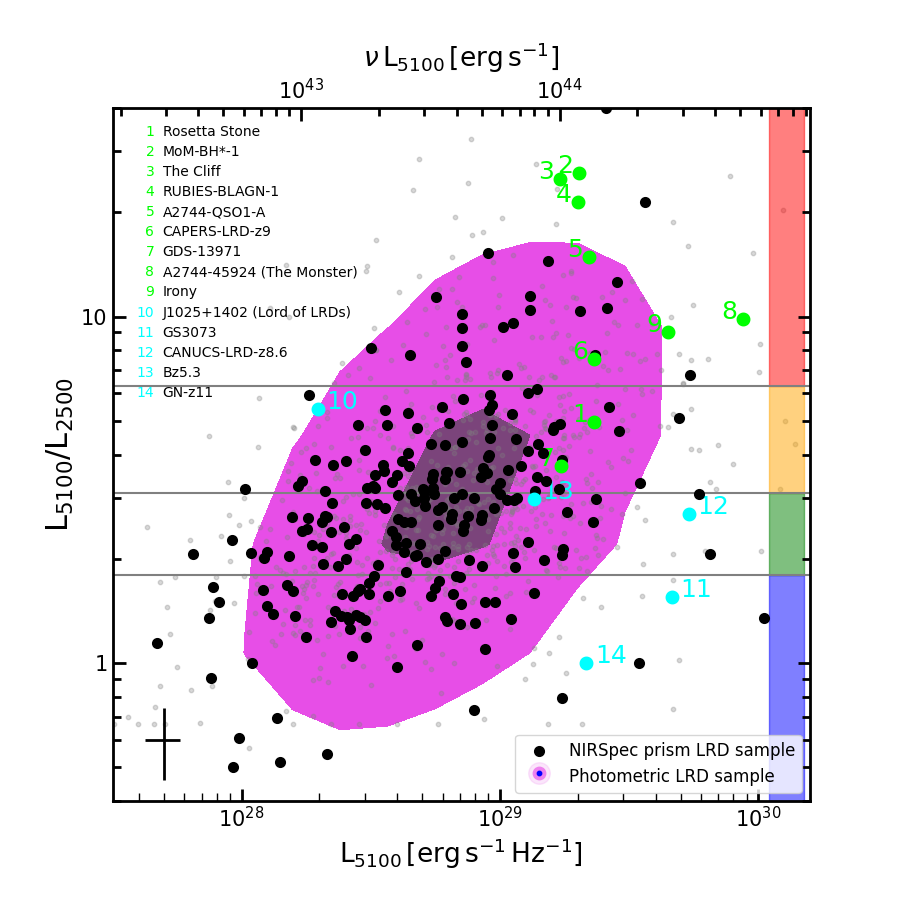}
\caption{\label{fig:lumratio}Optical-to-UV luminosity density ratio {\it vs.} optical luminosity density for the parent sample of photometrically selected LRDs (gray dots, purple regions showing 1,2$\sigma$ loci) and the spectroscopic sample of LRDs characterized in this paper (black and lime points). A scale in energy units $\nu\,L_\nu$ for the x-axis is  given. The gray lines separate the sample as a function of optical-to-UV luminosity ratio in 4 subtypes, which are represented by vertical shaded rectangles: extreme $\mathrm{L_{5100}/L_{2500}}$ LRDs ($\mathrm{^xLRD}$s, red shade), above-median $\mathrm{L_{5100}/L_{2500}}$ LRDs ($\mathrm{^+LRD}$, orange), sub-median $\mathrm{L_{5100}/L_{2500}}$ LRDs ($\mathrm{^-LRD}$s, green), and blue $\mathrm{L_{5100}/L_{2500}}$ LRDs ($\mathrm{^bLRD}$, blue). %Median and quartiles for each subsample are provided in the plot. 
Some interesting LRDs presented and characterized in individual papers are marked in lime (for sources included in our sample) and cyan (for sources not in our sample because they are not in our fields or do not have MSA data -but IFU observations-), identified with a number. The Rosetta Stone LRD comes from \citet{2024MNRAS.535..853J}, MoM-BH$^{\star}$-1 from \citet{2025arXiv250316596N}, The Cliff from \citet{2025A&A...701A.168D}, RUBIES-BLAGN-1 from \citet{2024ApJ...969L..13W}, A2744-QSO1-A from \citet[][see also \citealt{2025MNRAS.tmp.1770J}]{2023ApJ...952..142F}, CAPERS-LRD-z9 from \citet{2025ApJ...989L...7T}, GDS-13971 from \citet{2024ApJ...963..129M}, A2744-45924 from \citet{2024arXiv241204557L}, Irony from \citet[][also in \citealt{2024ApJ...969L..13W,2025arXiv250909607L,2025ApJ...991..217T,2025arXiv250820177R,2025arXiv250818358W}]{2025arXiv251000101D}, J1025+1402 (also known as Lord of LRDs) from \citet{2026MNRAS.545f2235J} and \citet{2026ApJ...997..364L}, GS3073 from \citet[][see also \citealt{2024MNRAS.535..881J} and \citealt{2025arXiv251209996V}]{2023A&A...677A.145U}, CANUCS-LRD-z8.6 \citep{2025NatCo..16.9830T}, Bz5.3 \citep{2025ApJ...994L...6T}, and GN-z11 \citep[included here due to its red optical to near-infrared continuum, compatible with LRDs;][]{2016ApJ...819..129O,2023A&A...677A..88B,2026A&A...706A..46C}.}
\end{figure}

To advance our understanding of LRDs, in this section we analyze stacked spectra of the several subsamples presented in Section~\ref{sec:classes} constructed in terms of the optical-to-UV luminosity density ratio, $\mathrm{L_{5100}/L_{2500}}$. Figure~\ref{fig:lumratio} shows this ratio compared to the optical luminosity density, $\mathrm{L_{5100}}$. The median value and 1$\sigma$ scatter of $\mathrm{L_{5100}/L_{2500}}$ for the whole spectroscopic sample is $3.1^{6.3}_{1.8}$. The figure clearly shows that there is a trend for more luminous LRDs to be redder. It is also clear that there exists a bias for the LRDs discussed individually in different papers (see caption for references), all of them being at the bright end of the distribution of optical luminosities.

\begin{figure*}[htp!]%[ht!]
\centering
%\fbox{
\includegraphics[clip, trim=1.1cm 0.3cm 2.7cm 2.0cm,width=15.cm,angle=0]{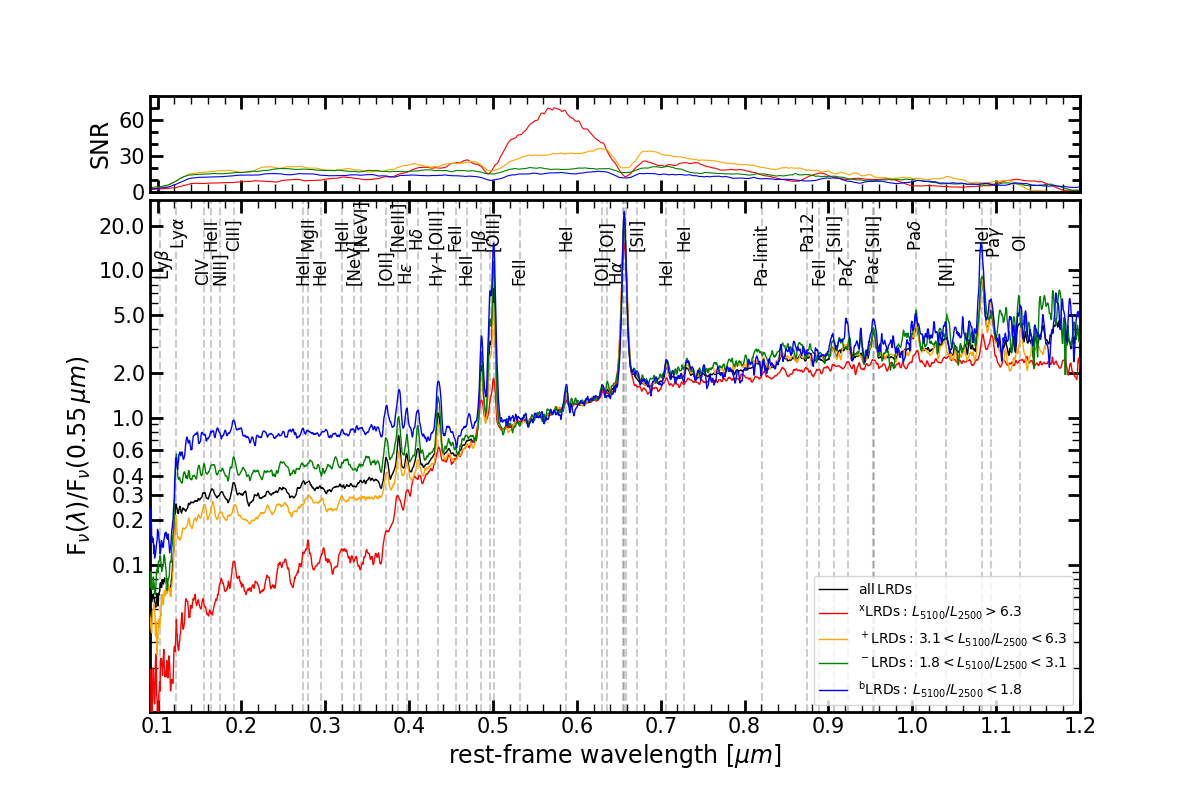}
\includegraphics[clip, trim=0.5cm 0.cm 2.7cm 1.8cm,width=8.2cm,angle=0]{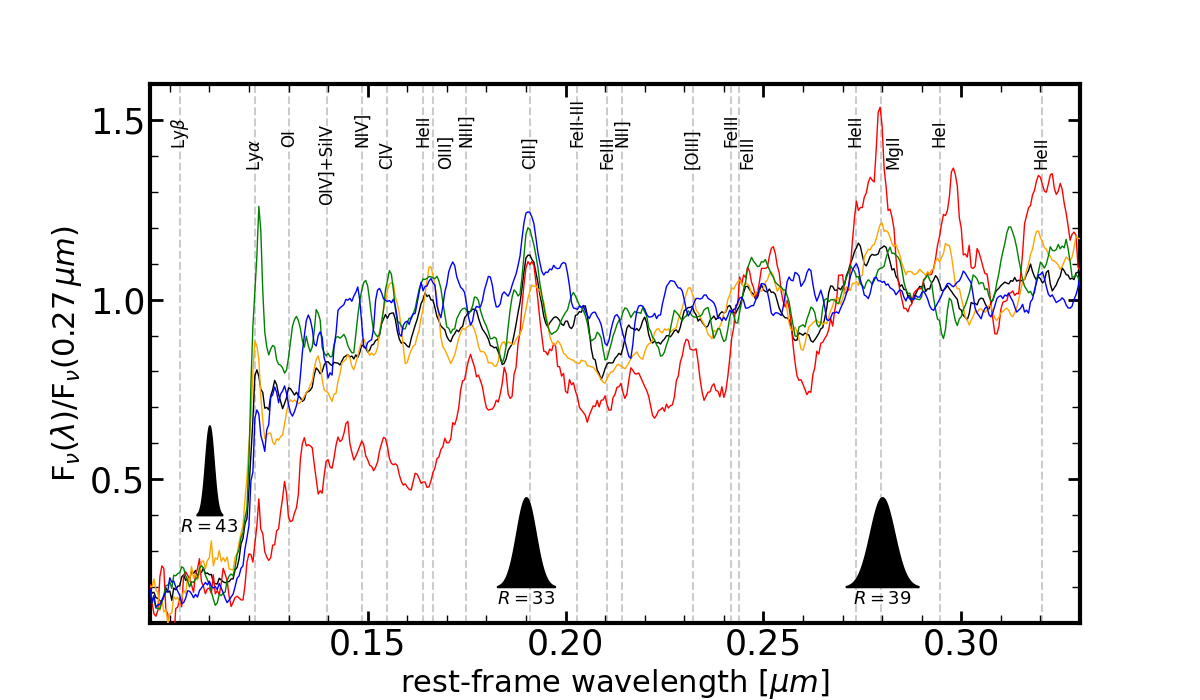}
\includegraphics[clip, trim=0.5cm 0.cm 2.7cm 1.8cm,width=8.2cm,angle=0]{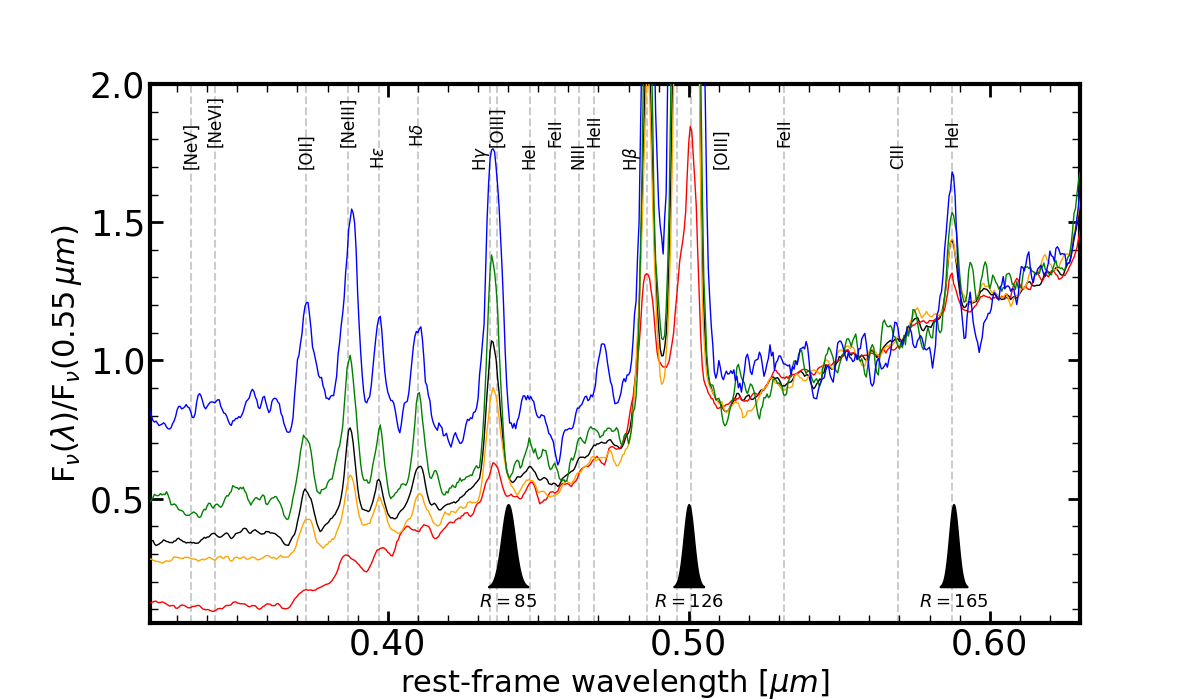}
\includegraphics[clip, trim=0.5cm 0.cm 2.7cm 1.8cm,width=8.2cm,angle=0]{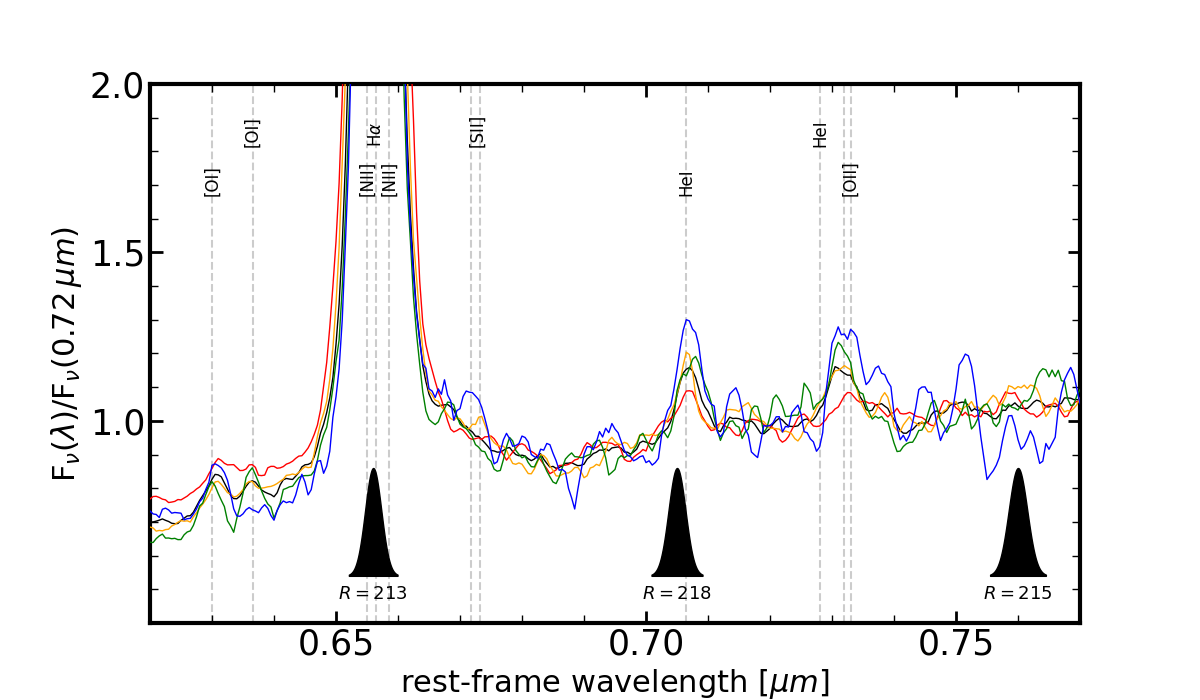}
\includegraphics[clip, trim=0.5cm 0.cm 2.7cm 1.8cm,width=8.2cm,angle=0]{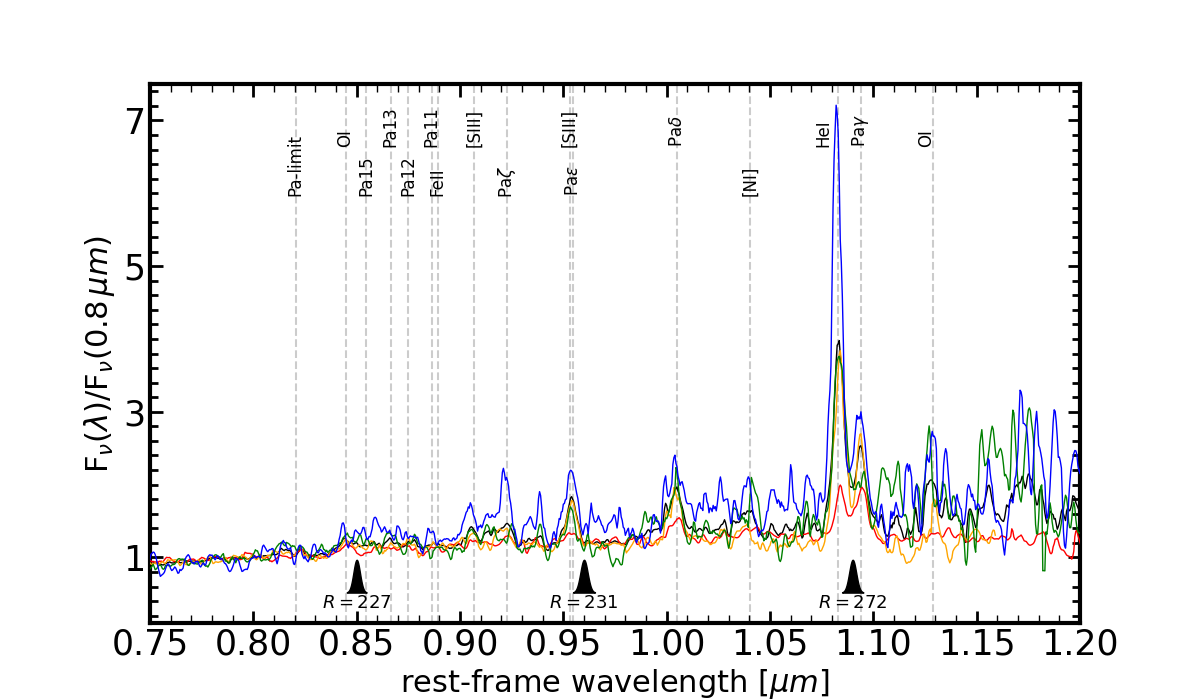}
%}
\caption{\label{fig:stacked_lumratio}Stacks of LRDs divided by their optical-to-UV luminosity ratio, $\mathrm{L_{5100}/L_{2500}}$. $\mathrm{^xLRD}$s, ($\mathrm{L_{5100}/L_{2500}}>6.3$), are shown in red; $\mathrm{^+LRD}$s, $2.7<\mathrm{L_{5100}/L_{2500}}<6.3$, are shown in orange; $\mathrm{^-LRD}$, $1.4<\mathrm{L_{5100}/L_{2500}}<2.7$, are shown in green; and $\mathrm{^bLRD}$s, $\mathrm{L_{5100}/L_{2500}}<1.4$, are shown in blue. In the main panel, all spectra are normalized to 0.55~$\mu$m. In the bottom panels, we zoom into several spectral ranges. To make the comparison easier, we normalize the four spectra to different wavelengths in each panel (specific value shown in the y-axis label). The varying spectral resolution is shown with Gaussians in each zoomed-in panel, providing the average of the four stacks at each wavelength.}
\end{figure*}

%For clarity, the stacks shown in this plot have been smoothed using a 3 spectral pixels weighted average filter. The SNR shown in the top panel takes into account this smoothing.

The stacked spectra for the four subsamples defined in Section~\ref{sec:classes}, $\mathrm{^xLRD}$, $\mathrm{^+LRD}$, $\mathrm{^-LRD}$, and $\mathrm{^bLRD}$, are shown in Figure~\ref{fig:stacked_lumratio}.

%\subsection{Spectroscopic diagnostics of the LRD subtypes}
%\label{sec:specprops}

In the following subsections, we quantify and discuss some relevant spectroscopic properties of LRDs based on our general stack and the stacks for the different subtypes. We present an analysis of some interesting spectral regions. The main measurements used in the rest of the paper are compiled in Table~\ref{tab:stats_specobs}. In Appendix~\ref{app:sedexamples}, we provide more details about the stacks for LRD subtypes, focusing on the scatter of SED shapes.

The spectra shown in Figures~\ref{fig:ironbytype} to \ref{fig:wrall}  were fitted to nebular emission from hydrogen, helium, and metals  with the {\sc fantasy} code \citep{2023ApJS..267...19I}. Among those metals, we emphasize that {\sc fantasy} considers iron line complexes, which are very relevant in the analysis of AGN. The fitting procedure follows. First, we subtracted a continuum calculated from anchor points identified among the 5\% lowest fluxes in the spectral region of interest. Then, the fits were performed by considering a narrow component for hydrogen, helium, and all metals except iron (e.g., oxygen for Figure~\ref{fig:ironbytype}), an independent broad component for hydrogen and helium, and a final independent (with their own width) component for Fe\,II lines, all sharing the same velocity shift.

\subsection{Ubiquitous iron emission}
\label{subsec:ironlines}

The ubiquity of singly ionized iron, Fe\,II, lines in AGN in the optical was discovered in the very first QSOs and prototypical AGNs such as I~Zw~1 \citep{1967ApJ...148..695W,1968ApJ...152L..31S,1985ApJ...288...94W}. There are several other spectral regions where these emission complexes are very prominent in type-1 AGN, including the UV, red optical, and near-infrared.

\begin{figure}[htp!]%[htp!]%[ht!]
\includegraphics[clip, trim=1.2cm 3.5cm 3.5cm 3.2cm,width=8cm,angle=0]{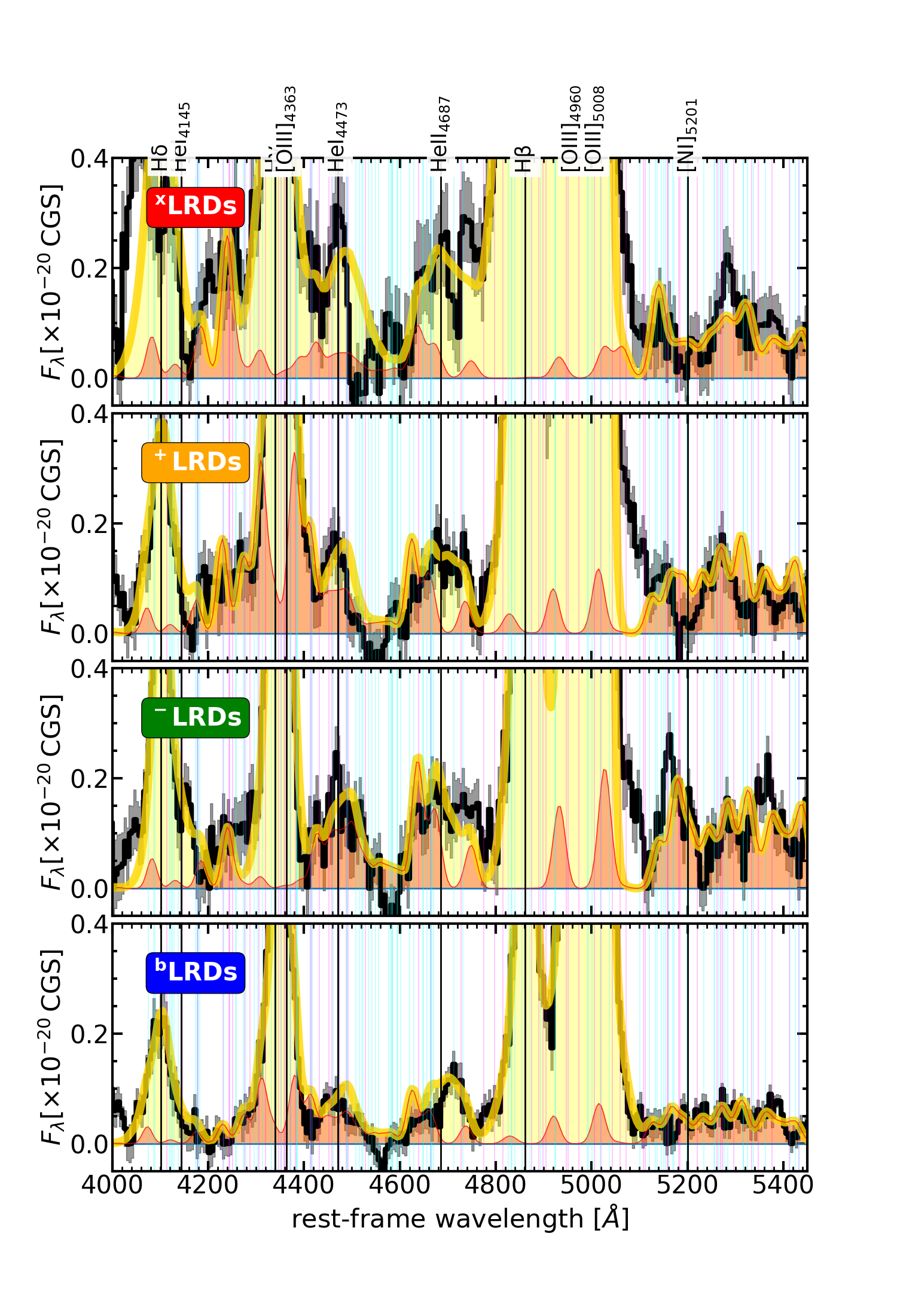}
\caption{\label{fig:ironbytype} Spectra of the different types of LRDs in the optical Fe\,II region. The continuum has been subtracted by fitting a spline to a set of anchor points constructed with the 5\% lowest flux values in the region. The spectra have been expanded in intensity to emphasize the faint lines, mainly forming a pseudo-continuum created by Fe\,II emission. %As the legend on the top right indicates, 
The main hydrogen, helium and oxygen lines (typical of SFG and AGN) are marked in black, and the Fe\,II lines (consistent, inconsistent, semi-forbidden, and forbidden, see \citealt{2025A&A...694A.289K}) are marked in cyan and magenta. The fit to the spectra performed with the {\sc fantasy} code \citep{2023ApJS..267...19I} is shown in gold (all components) and red (Fe contribution).}
\end{figure}

Figure~\ref{fig:ironbytype} shows the spectra of the different types of LRDs in the region around the H$\beta$ emission line, covered by the ``optical Fe\,II lines''\footnote{We note that this spectral region includes the WR blue bump treated in Section~\ref{sec:optical}, which will be discussed for LRD subtypes in the next section.}. This spectral range shows a distinctive signal from multiple Fe\,II lines forming a pseudo-continuum in AGN \citep[e.g.][]{2010ApJS..189...15K}. The lines (marked with vertical cyan lines in the plot) are organized in this region in three groups, one for each of the lower transition levels of Fe\,II: the F lines around 4550~\AA\ (also found in AGN at high redshift observed with JWST; \citealt{2025A&A...700A.203T}), the S lines around 5000~\AA, and the G lines around 5200~\AA. Further Fe\,II emission extends bluewards down to 4100~\AA, in a group that \citet{2025A&A...694A.289K} qualifies as inconsistent Fe\,II lines based on the fact that they are clearly seen in spectra of AGN despite their low transition probabilities \citep{1983ApJ...275..445N,1999ApJS..120..101V,2008ApJ...675...83B}, and also jointly with some semi-forbidden and forbidden lines of Fe arising from higher levels of excitation. In our low spectral resolution data, the consistent lines around 5300~\AA\ are the most visible, given that the spectral region from 4000 to 5000~\AA\ is dominated by strong nebular lines from hydrogen and oxygen and the identification of iron lines is more challenging.

\subsection{Fe~II and Eddington Ratios}
\label{sec:eddington}

\begin{figure}[htp!]%[htp!]%[ht!]
\includegraphics[clip, trim=1.2cm 3.5cm 2.5cm 3.0cm,width=8.5cm,angle=0]{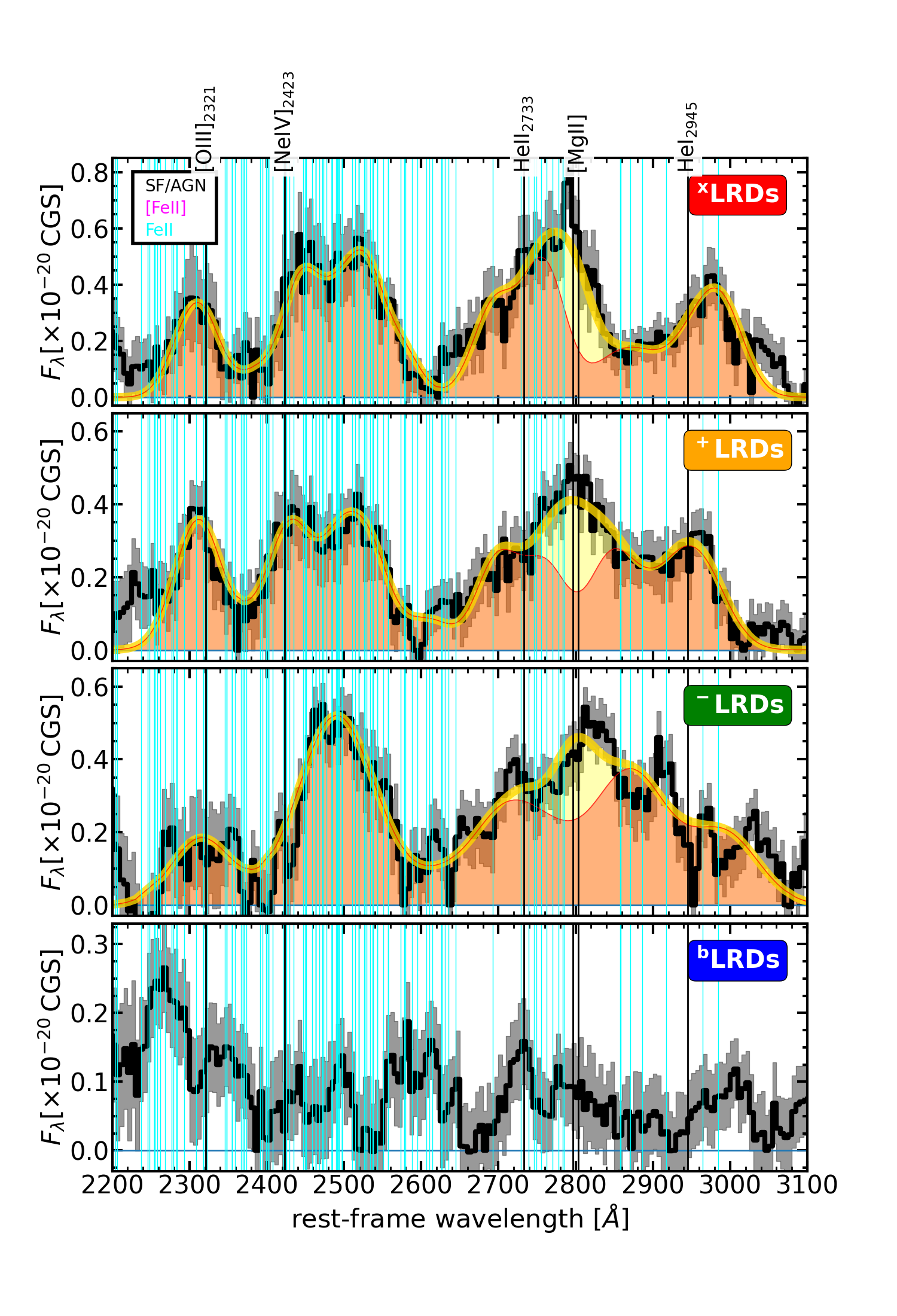}
\caption{\label{fig:uvironbytype} Spectra of the different types of LRDs in the UV Fe\,II spectral region. The continuum has been subtracted by fitting a spline to a set of anchor points constructed with the 5\% lowest flux values in the region. The main helium and magnesium lines are marked in black. The Fe\,II lines are marked in cyan. The spectra for $\mathrm{^xLRD}$s and $\mathrm{^+LRD}$s in the region between 2200 and 3100~\AA\ are fitted with the {\sc fantasy} code (in gold for the full fit, in red for the Fe component; see text for details). We do not attempt any fit for blues LRDs, given that a significant contribution from stellar emission is found in our SED analysis for the UV spectral region (Section~\ref{sec:sedfit}).}
\end{figure}

All different LRD types present Fe\,II emission, which is more prominent in the reddest sources. The Fe\,II emission has been shown to correlate with the Eddington ratios in nearby AGN and QSOs \citep[see, e.g.,][]{2014Natur.513..210S,2023ApJS..267...19I}. 
For the reddest LRDs, information about Eddington ratios can be extracted from UV line ratios involving Fe\,II and Mg\,II, as shown in Figure~\ref{fig:uvironbytype}. For the $\mathrm{^xLRD}$s, we measure a line ratio $\mathrm{Fe\,II/Mg~II}=11.3\pm2.8$\footnote{Following \citet{2011ApJ...736...86D}, the Fe\,II emission is integrated from 2200 to 3090~\AA.}. This value is comparable to the most luminous QSOs \citep[e.g.,][]{2003ApJ...596L.155M,2003ApJ...596..817D}. Using the compilation of different types of AGN presented in \citet[][see also \citealt{2019ApJ...874...22S,2021ApJ...917..107S}]{2011ApJ...736...86D}, that line ratio would translate to $\lambda_\mathrm{Edd}=0.6\pm0.2$. A similar Eddington ratio is obtained for the $\mathrm{^+LRD}$s and $\mathrm{^-LRD}$s, for which we measure $\mathrm{Fe\,II/Mg~II}=10.2\pm2.7$ and $12.0\pm3.0$, translating to $\lambda_\mathrm{Edd}=0.5\pm0.2$ and $0.6\pm0.2$, respectively. We conclude that LRDs are likely to have relatively high accretion ratios, but not super-Eddington (see \citealt{2026arXiv260212548P}).

%We note that a correlation between Eddington ratio and emission line ratios is more meaningful than between Eddington ratio and a direct luminosity.

Another estimation of the Eddington ratio can be performed using the results in \citet{2025A&A...694A.289K}, who present a correlation between the Eddington ratio and the ratio of the Fe\,II intensities in the UV and the optical. We find a optical-to-UV intensity ratio (using only the consistent lines in the optical, and the group-60 Fe\,II lines in the UV, as done in the paper referenced above) of $3.8\pm1.5$ for $\mathrm{^xLRD}$s (this is a direct measurement without any correction for dust attenuation). This would translate to $\lambda_\mathrm{Edd}\sim0.6$ based on Figure~10 in \citet{2025A&A...694A.289K}, given that our value would be in between group 4 and 5 in that plot. Similar calculations  place $\mathrm{^+LRD}$s and $\mathrm{^-LRD}$s in groups 3-4 in the figure in \citet{2025A&A...694A.289K}, which implies $\lambda_\mathrm{Edd}=0.2-0.6$. 

%.For $\mathrm{^xLRD}$, and assuming the same dust correction (we do not estimate it with our SED fitting analysis), we find $3.2\pm1.2$, translating to the same $\lambda_\mathrm{Edd}=0.1-0.2$. As we will show in Section~\ref{sec:sedfit}, the UV spectral region for the other subtypes of LRDs ($\mathrm{^-LRD}$ and $\mathrm{^bLRD}$) would have a non-negligible ($\sim40$\%) contribution from stellar emission, and we are not able to measure the UV-Fe\,II emission in our prism spectra.

%considering a dust attenuation of $A^V_{BH}=0.6$~mag following a \citet{2000ApJ...533..682C} law (see SED fitting results in Section~\ref{sec:sedfit})

LRDs are very weak in the X-ray \citep[e.g.,][]{2024ApJ...974L..26Y,2025ApJ...986..126K,2026arXiv260109778H}. This property is shared with high-redshift moderate-luminosity AGNs in general \citep[e.g.,][see also \citealt{2019A&A...630A.118V}]{2024A&A...691A.145M}, for which a median Eddington ratio of $\sim$ 0.4 has been found \citep[e.g.,][]{2023ApJ...959...39H,2023ApJ...957L...7K,2023ApJ...953L..29L,2024A&A...691A.145M}. This value is consistent with the ones we have deduced for LRDs from their Fe\,II emission.

As an alternative,  \citet{2024ApJ...976...96P} have suggested that the weak X-rays may indicate mildly super-Eddington emission with Eddington (mass) ratios of 1.4 - 4\footnote{We note that \citet{2024ApJ...976...96P} do not discuss Eddington ratios in terms of luminosity but in terms of mass accretion rates. This quantity can significantly increase at high accretion rates, while $\lambda_\mathrm{Edd}$ does not linearly increase accordingly, since at high accretion rates, the disk becomes radiatively inefficient.}. Their conclusion requires very specific properties for the radiating black holes, i.e.,  zero spin and viewed at angles $ > 30^{\rm o}$ from their 
poles. It seems unlikely that these conditions can hold for {\it nearly all} LRDs and high-redshift AGNs, so alternative possibilities for the weak X-rays are favored, such as high levels of absorption \citep{2025MNRAS.544..726I,2025ApJ...989L..30S}.

We conclude that LRDs present high but sub-Eddington accretion rates. Indeed, LRDs have quite different properties compared to clear super-Eddington systems at high redshift (e.g., narrower Fe\,II emission and higher column density of neutral gas; see, e.g., \citealt{2026arXiv260212325F}).  

\begin{figure}[htp!]%[htp!]%[ht!]
\includegraphics[clip, trim=1.2cm 3.5cm 3.5cm 2.5cm,width=8cm,angle=0]{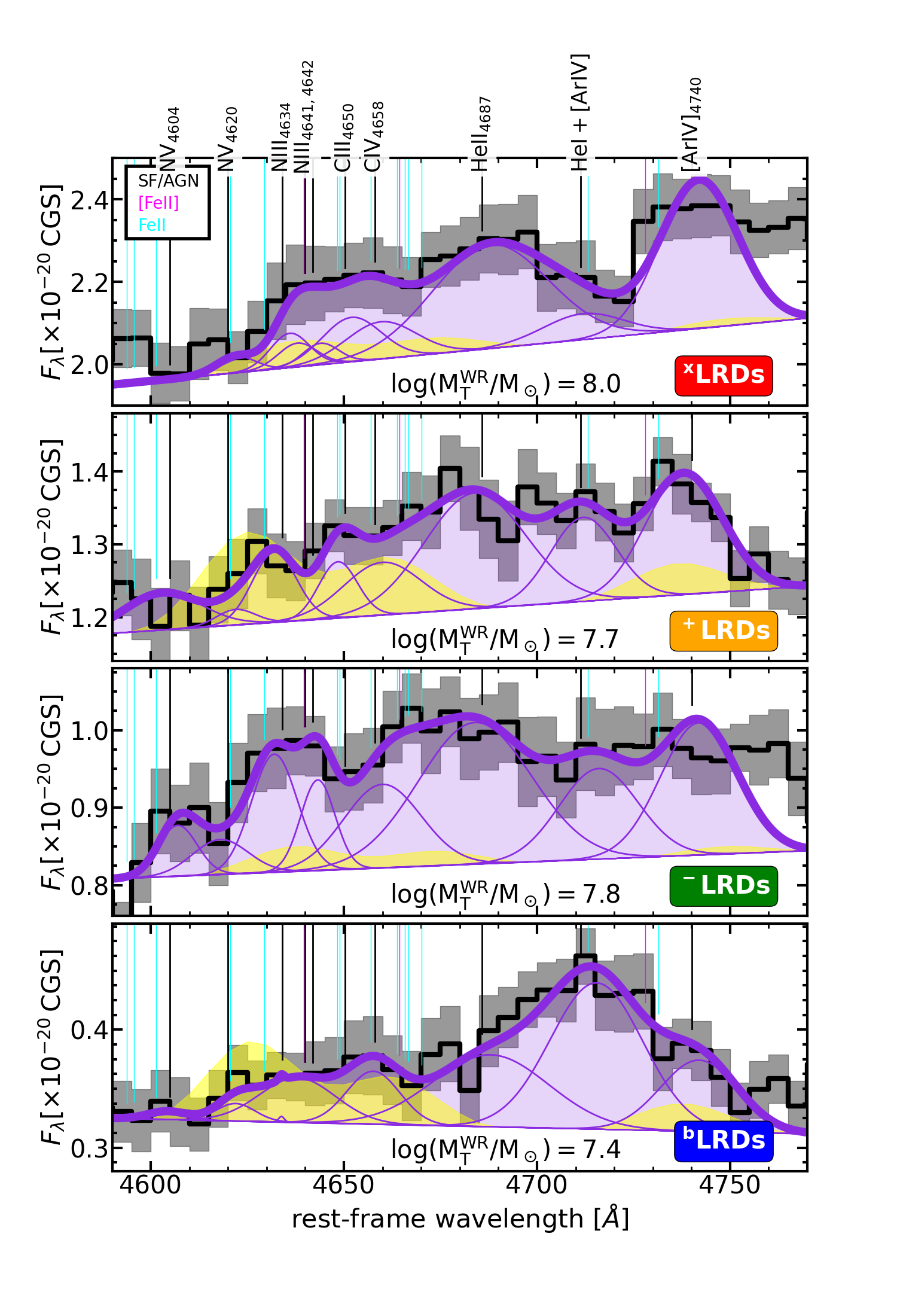}
\caption{\label{fig:wrall} Spectra of the different types of LRDs in the WR blue bump region. The fit with the {\sc lime} code \citep{2024A&A...688A..69F} is shown in purple (thin lines for individual lines, the total fit shown with a thick line). In yellow, we show the contribution of Fe\,II lines as measured with the {\sc fantasy} code \citep{2023ApJS..267...19I}. }
\end{figure}

\subsection{WR features in LRD subtypes}

In Section~\ref{sec:optical}, we discussed the spectral region around the WR blue bump for the median LRD spectrum. In Figure~\ref{fig:wrall}, we show the spectra in the same region for the different LRD subtypes. The same fitting approach explained in Section~\ref{sec:optical} has been used, taking into account the absorption around 4550~\AA\ which was presented in Section~\ref{sec:absorption}. The widths and intensity of the He\,II$\lambda4687$ line are similar in all the LRD subtypes except for the bluest. If interpreted as a WR signature, all LRD subtypes would host young stars, the stellar masses of the recent burst would range from $10^{7.4}$ to $10^{8.0}$~M$_\odot$. The rest-frame equivalent widths of the He\,II line decrease as we move to bluer LRDs, ranging from 10~\AA\, for $\mathrm{^xLRD}$ to 2~\AA\, for $\mathrm{^bLRD}$. Even in the extreme case that all the He\, II emission is linked to WR stars but $\sim$50\% of the continuum is linked a different source (see Section~\ref{sec:sedfit}), the EWs would be compatible with the measurements in nearby WR galaxies (see Section~\ref{sec:wr_full} for references).

\subsection{Line ratios and Balmer break}

\begin{figure}[htp!]%[htp!]%[ht!]
\includegraphics[clip, trim=0.7cm 3.5cm 1.0cm 16.9cm,width=8.1cm,angle=0]{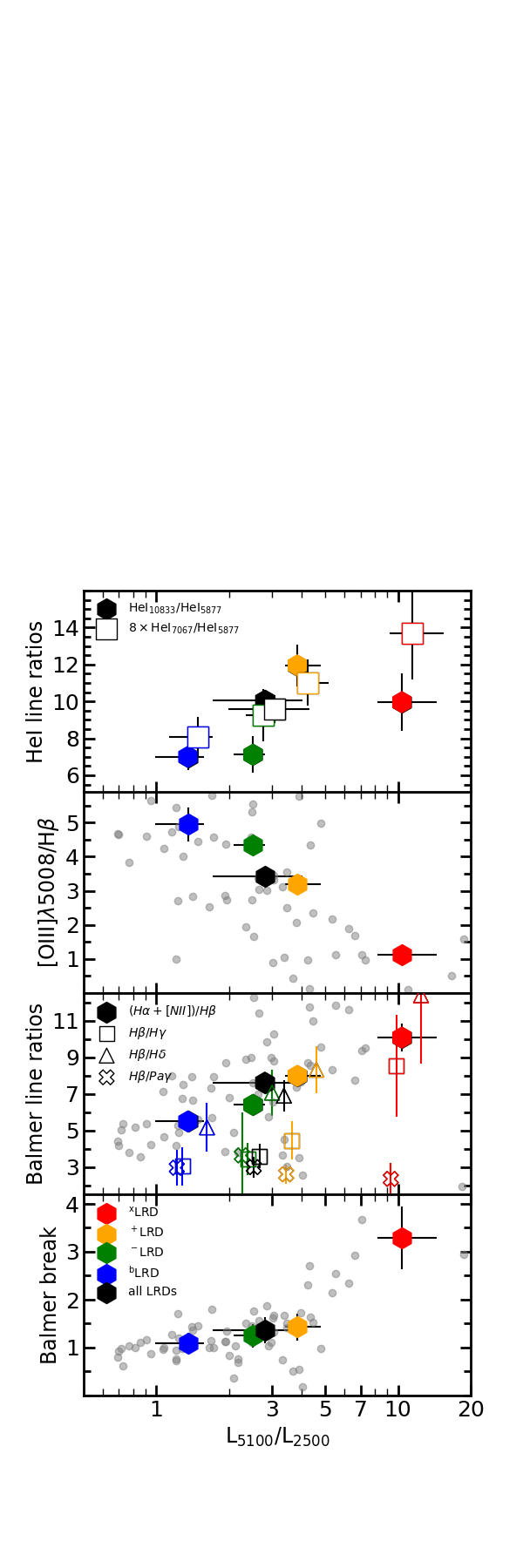}
\caption{\label{fig:ratios}Correlation of spectroscopic features with the optical-to-UV continuum luminosity ratios. $\mathrm{^xLRD}$s are shown in red, $\mathrm{^+LRD}$s in orange, $\mathrm{^-LRD}$s in green, and $\mathrm{^bLRD}$s in blue. In the panels showing several point sets, each one of them is offset in the x-axis for clarity. Top panel shows the ratio between several He\,I lines in the optical and infrared (multiplied by a factor to plot them in the same scale). The panel below depicts the ratio between [OIII]$\lambda5008$ and the integrated flux of H$\beta$ (including broad and narrow components). The middle panel shows the Balmer decrement for the full flux of the hydrogen lines and for the broad component alone (symbols with thick black edge, artificially offset in the x-axis for clarity). Errors in the y-axis depict measurement uncertainties, in the x-axis they refer to scatter of the luminosity ratio. The bottom panel shows the Balmer break. In all panels, measurements for individual LRDs based on DJA catalogs and presented in \citet{2025arXiv251215853B} are shown as grey dots.}
\end{figure}

Figure~\ref{fig:ratios} shows how the optical-to-UV luminosity ratios used in our paper to distinguish different types of LRDs correlate with spectral features.

The most extreme LRDs (around 15\% of the population) are characterized by large Balmer breaks (not compatible with stellar properties, see \citealt{2025A&A...701A.168D}, \citealt{2025arXiv250316596N} or \citealt{2025ApJ...989L...7T}). The rest show more moderate values of the Balmer break amplitude, around 1.0-1.5 with a slight trend for redder sources to present larger Balmer breaks; all the values around 1.0-1.5 could in principle be explained with stars.

The hydrogen line ratios also correlate with $\mathrm{L_{5100}/L_{2500}}$. Interpreted as dust attenuation, the Balmer decrements would mean an increasing dust content as we move to redder LRDs, with $A^V$ values changing from 0.2 to nearly 3~mag. However, in the scenario invoking dense gas enshrouding a SMBH to explain LRDs (see, e.g., \citealt{2025ApJ...980L..27I}), the Balmer line ratios are significantly affected by radiative transfer processes such as Thomson, Raman, and resonance scattering \citep{2026MNRAS.545f2131C}, so this dust content estimate should be taken with caution. Indeed, in a few cases where accurate narrow-broad line decomposition has been performed, the observed line ratios seem inconsistent with standard dust attenuation \citep{2025arXiv251000101D,2025arXiv251006362N}, favoring instead collisional excitation.

%Nevertheless, it is interesting that the Balmer decrements for the broad component of the H$\alpha$ and H$\beta$ lines do not change as dramatically as do the ratios for integrated fluxes. Again, interpreting this ratio as the effect of dust, the BLRs seem to be affected by only a moderate amount of dust obscuration, with  $A^V$ values changing from 0.2 to 0.5~mag. 

In our stacks, the behavior of the hydrogen line ratios is not-standard for case-B and dust interpretations. As noted by several authors \citep{2025arXiv251121820D,2025arXiv251215853B,2025ApJ...986..177B}, the H$\alpha$/H$\beta$ ratios are larger for the reddest LRDs, reaching values as high as 10 and above. We measure the same increase for H$\beta$/H$\gamma$ and H$\beta$/H$\gamma$. In our stacks, we are able to measure the H$\beta$/Pa$\gamma$ ratio, obtaining values around 2--3, which are very different from the $\sim11$ value expected for case-B recombination. Taking the results from \citep{2026MNRAS.545f2131C}, we infer that the small values we measure for H$\beta$/Pa$\gamma$ would be consistent with column densities of hydrogen atoms in the 2s state of nearly $10^{14}$~cm$^{-2}$, consistent with the high Balmer decrements.

%A further interesting measurement in our stacks linked to radiative transfer effects is the relationship between H$\alpha$ and H$\beta$ widths. For $\mathrm{^xLRD}$s, we H$\beta$ is broader than H$\alpha$ by 20\%, a very similar value to that observed for QSOs and classical AGN \citep{2004ApJ...606..749K,2005ApJ...630..122G,2017ApJ...851...21G}. However, for the bulk of the LRD population, $\mathrm{^+LRD}$s and $\mathrm{^-LRD}$s, we measure larger widths for H$\alpha$, the average ratio being $\mathrm{FWHM(H\beta)/FWHM(H\alpha)}=0.6\pm0.1$. As discussed in \citet{2026MNRAS.545f2131C}, this level of discrepancy in the width of the two brightest Balmer lines is expected in Raman scattering \citep[see][]{2025ApJ...995...24K}. We, however, warn the reader that disentangling and interpreting the different line components in a stacked spectrum is complex, given that objects with different line widths and contributions from narrow and broad components are mixed together, and some other effects such as self-absorption might be present for some sources entering the stack and affect the measurements in low spectral resolution data. 

Figure~\ref{fig:ratios} also shows the correlation between [O\,III]$\lambda5008$/H$\beta$ and $\mathrm{L_{5100}/L_{2500}}$, revealing a clear trend in which redder LRDs exhibit systematically lower [O\,III]/H$\beta$ ratios. At face value, this behavior could be interpreted in terms of metallicity variations or density effects, particularly if the gas density exceeds the critical density of [O\,III] ($\sim10^{5}$~cm$^{-3}$), leading to collisional suppression of the line. Variations in ionization state and in the relative contributions of broad and narrow H$\beta$ components may also contribute. %In addition, very dense gas surrounding the ionizing source might suppress significantly the higher ionization lines.

An alternative, and likely complementary, explanation is that the observed line-ratio trend reflects the changing relative contribution of the galaxy host and the LRD core along the color sequence. In this picture, [O III]/H$\beta$ increases toward bluer LRDs as the stellar host contributes more strongly to both the continuum and the narrow-line emission. As shown in \citet{2025arXiv251121820D}. \citet{2025arXiv251215853B}, and \citet{2026arXiv260120929S}, this interpretation is supported by the systematic rise in the equivalent width of [O III] toward bluer systems, consistent with a growing prominence of host-galaxy emission relative to the LRD core. Within this framework, the low [O III]/H$\beta$ ratios observed in the reddest LRDs arise primarily from strong broad H$\beta$ emission combined with weak or suppressed narrow-line emission from the host, rather than requiring extreme metallicity or density conditions alone.

The effect of dense gas is also noticeable in the He\,I line ratios plotted in the top panel of Figure~\ref{fig:ratios}. Our stacks show high-SNR detections of three He\,I lines in the optical and near-IR, namely He\,I$\lambda$5877, He\,I$\lambda$7067 and He\,I$\lambda$10833. These lines, especially the latter, but also He\,I$\lambda$7067, are highly sensitive to electron density and temperature, as well as optical depth \citep[see][and also \citealt{2012MNRAS.425L..28P}]{2026ApJ...996...68B}. The LRD subtypes differ significantly in the He\,I line ratios, which are, in general, relatively high. Qualitatively (a quantitative approach is delayed to a future publication), they indicate high electron densities, $n_e>10^5$~cm$^{-3}$, optically thick material for He\,I lines, and temperatures around 5000-7000~K.

\section{Modeling the BH and stellar continuum properties of the LRDs}
\label{sec:continuum_properties}

In this section, we model the spectral stacks to infer stellar and black hole masses, as well as monochromatic and bolometric luminosities, and examine the correlations among these properties. We use two approaches, described in the next two subsections.

\begin{figure*}[htp!]%[htp!]%[ht!]
\includegraphics[clip, trim=0.0cm 0.2cm 1.0cm 1.7cm,width=9cm,angle=0]{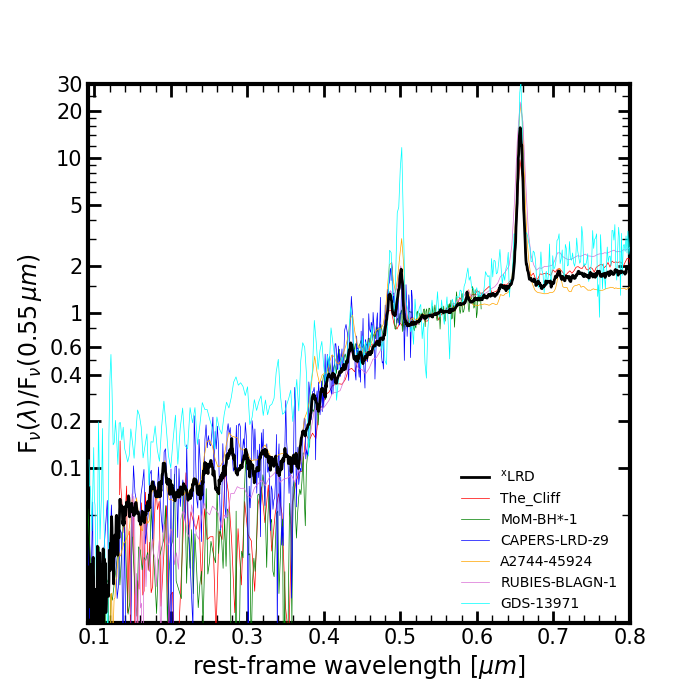}
\includegraphics[clip, trim=0.0cm 0.2cm 1.0cm 1.7cm,width=9cm,angle=0]{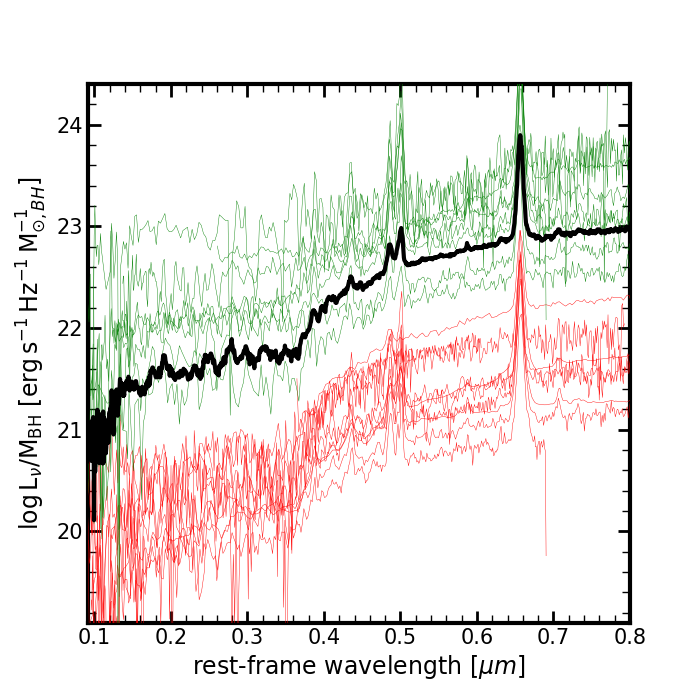}
\caption{\label{fig:xLRDs} {\it Left:} Spectrum of the extreme LRDs in our paper (in black), compared to individual sources (see caption of Figure~\ref{fig:lumratio} for references), all normalized to 0.55~$\mu$m. {\it Right:} Same spectra but normalized to the BH mass. All spectra for individual sources shown in the left panel, also including the galaxies in both the \citet{2025arXiv251007376J} paper and our sample, are now plotted in red. Their BH masses come from  single-epoch estimations based on H$\alpha$ or H$\beta$ lines. The BH masses for our stacks come from bolometric luminosities (Equation~\ref{eq:bhmass}). In green, we show the ten sources in common between our sample and the sources for which the BH mass has been calculated by \citet{2025arXiv250316595R} based on high spectral resolution data (mainly $R=1000$) and a using single-epoch BH-mass estimator applied to the intrinsic narrow component, which yields up to an order of magnitude smaller SMBH masses compared to using the observed line width.}
\end{figure*}

%We also show a stack of our sample for extremely high optical-to-UV luminosity ratios, $\mathrm{L_{5100}/L_{2500}}>10$ (in magenta), more similar to individual BH$^{\star}$ sources such as The Cliff, MoM-BH*-1 or CAPERS-LRD-z9.

%\begin{figure*}[htp!]%[htp!]%[ht!]
%\includegraphics[clip, trim=0.0cm 0.cm 0.0cm %0.0cm,width=18cm,angle=0]{Figures/bestfit_stacks.pdf}
%\caption{\label{fig:linefits}.}
%\end{figure*}

\subsection{The two-component LRD-core plus galaxy-host SED model}
\label{sec:sedfit}

We model the SEDs of LRD subtypes using a two-component framework consisting of a stellar host and an ``LRD core''. The latter captures the red, steep optical continuum, the strong Balmer break, and the characteristic V-shaped UV–optical slope (i.e., an inflection in the SED, \citealt{2025ApJ...995..118S}) observed in LRDs. This Balmer-break component is likely associated with a dense gas envelope that helps reddening the intrinsic emission, even with little or no dust extinction. The detailed nature of this LRD core remains uncertain and may be linked to a BH$^{\star}$ or BHE, or very compact and intense gas-rich starburst with the possible presence of a super-massive star (or several). In our analysis, we remain agnostic about its physical nature, and use a template for this component (see below).

The primary motivation for this approach is that previous two-component SED models combining stellar hosts with standard AGN templates (such as those from, e.g., \citealt{2021MNRAS.508..737T}) failed to reproduce
the strong (large amplitude) yet smooth Balmer break observed in LRDs \citep[e.g.;][]{ma2025b}. In fact, stellar atmospheres have either smooth but weak breaks \citep[with microturbulence up to 10--15 km\,$s^{-1}$;][]{smith+1998}, or strong but sharp breaks \citetext{typical of A-type stars; see \citealp{2025MNRAS.tmp.1995D} for a discussion in the context of LRDs}. In contrast, models where the Balmer break is due to dense gas \citep{2025ApJ...980L..27I} can reproduce both the observed break strength in LRDs, as well as its smoothness, provided the break itself is smoothed by high turbulence \citep[$v\sim 100~\mathrm{km\,s^{-1}}$;][]{2025MNRAS.tmp.1770J,2025arXiv250316596N,2025A&A...701A.168D}.
Models combining a stellar host with an LRD core exhibiting a pronounced break successfully reproduce the composite spectra \citep[see, e.g.,][]{2025arXiv250316596N,2025arXiv251204208U,2025arXiv251215853B}.

We use an empirical LRD-core template based on the $\mathrm{^xLRD}$ spectrum, which exhibits the most pronounced V-shaped continuum (Figure~\ref{fig:xLRDs}; black line). To allow for possible variations in the optical continuum shape among LRDs, we include a dust attenuation parameter (following a \citealt{2000ApJ...533..682C} law) that may reflect the presence of a developing torus, an external dust screen, or a higher gas column density relative to the fiducial LRD core. We note that indeed the most extreme of the $\mathrm{^xLRDs}$, some individual examples are shown in Figure~\ref{fig:xLRDs}, are well represented by this approach when adding an extra opacity of $A^V\sim1$~mag. Jointly with the LRD core, we consider a galaxy host with a composite stellar population.

Consistent with the inclusion of dust, we also model the associated mid-IR re-emission using a compact torus viewed at $45^\circ$–$60^\circ$, following \citet{2015A&A...583A.120S}. While rest-frame mid-IR wavelengths are only loosely constrained by the NIRSpec spectra, they are effectively probed by the available MIRI photometry (see Section~\ref{sec:nir} and Figure~\ref{fig:stacked_all_miri}). We do not apply any energy-balance algorithm to the LRD core reddening (which, as we remarked in the previous paragraph, can be linked to dust or differences in the column density for distinct LRD subtypes) and dust emission models.

The stellar host component is modeled using {\sc synthesizer} \citep{2003MNRAS.338..508P,2008ApJ...675..234P,2024ApJ...968....4P}, following a similar approach to that adopted in \citet{2024ApJ...968....4P}. We allow for two stellar populations: a very young component ($<10$~Myr) to account for emission lines and WR features (see Section~\ref{sec:optical}), and an older population with a free age and independent dust attenuation, representing either an underlying stellar population or a deeply embedded burst. The star-formation history is parameterized as a delayed-exponential burst, with metallicities ranging from 2 to 20\% solar, a \citet{2003PASP..115..763C} IMF with mass limits of 0.1–100~M$_\odot$, and dust attenuation described by a \citet{2000ApJ...533..682C} law. We use \citet{2003MNRAS.344.1000B} models.

The combined host + LRD-core SED model depends on a total of 12 free parameters. Three parameters describe the LRD core: its normalization, attenuation, and the normalization of the mid-IR torus component. The remaining nine parameters describe the stellar populations, including the timescale, age, metallicity, and attenuation of each stellar component, as well as the relative burst strength, defined as the fraction of the total stellar mass associated with the youngest population. The main results of this SED-fitting analysis are compiled in Table~\ref{tab:stats_specobs}.

\subsection{The two-component modified blackbody plus galaxy-host SED model}

An alternative SED modeling approach for LRDs has recently been presented by \citet{2025arXiv251121820D} and \citet{2025arXiv251204208U}, in which the LRD core is described using a parametric MBB model rather than an empirical template. We refer the reader to Section~\ref{sec:nir} for caveats about the MBB model in the context of $T\sim 5000$~K gas.
Nevertheless, given its relevance for comparison and interpretation, we apply this approach to our stacked spectra as well.

Specifically, we model the LRD core using the MBB formulation of \citet{2025arXiv251121820D}, consisting of a Planck function modified by a power-law slope, $\beta$, relative to a pivot wavelength at $\lambda=5500$~\AA. This parameterization introduces three free parameters: the temperature $T$, the slope $\beta$, which primarily affects the Rayleigh–Jeans tail, and a normalization. Following \citet{2025arXiv251204208U}, we additionally include a sharp drop at the Balmer limit, which cannot be reproduced by a pure blackbody. Rather than a Heaviside function, we adopt a sigmoid profile centered at the Balmer edge with a width of 0.05~$\mu$m to allow for a smoother transition.

For consistency with our default SED modeling, we combine the MBB LRD-core component with a stellar host contribution, resulting in a two-component model analogous to that described in the previous subsection. This differs from the single-component MBB approach adopted in \citet{2025arXiv251121820D}. The impact of including an explicit host-galaxy component is discussed in the following sections.

\subsection{BH and stellar masses and luminosities of LRDs}

Under the assumption that the LRD core traces a black-hole–powered component, whether in the form of a BH$^{\star}$ or a BHE, the best-fit SED models provide estimates of both stellar and black hole masses and luminosities. For the stellar component, the stellar mass $\mathrm{M}_\bigstar$ is obtained directly from the population synthesis modeling, while the infrared luminosity associated with stars, $\mathrm{L}_\bigstar^\mathrm{IR}$, is inferred from the fitted attenuation $\mathrm{A}^V_\bigstar$, assuming that the absorbed energy is re-emitted by dust in the infrared.

For the LRD core, the bolometric luminosity $\mathrm{L}_\mathrm{BH}^\mathrm{bol}$  is obtained by integrating the core SED. The black hole mass $\mathrm{M}_\mathrm{BH}$ is not a direct fit parameter, but is instead inferred from $\mathrm{L}_\mathrm{BH}^\mathrm{bol}$, also including an Eddington ratio, $\lambda_\mathrm{Edd}$, through the relation:

\begin{equation}
\label{eq:bhmass}
\mathrm{M}_\mathrm{BH}/\mathrm{M}\odot = 0.81 \times 10^5 \, \lambda_\mathrm{Edd}^{-1}
\left( \frac{\mathrm{L}_\mathrm{BH}^\mathrm{bol}}{10^{43}\,\mathrm{erg\,s^{-1}}} \right),
\end{equation}

As discussed in Section~\ref{sec:eddington}, we adopt Eddington ratios of $\lambda_\mathrm{Edd}=0.6$ for all LRDs. We note that this method is similar to that presented in \citet{2025arXiv251204208U}, but they obtain the bolometric emission with a constant correction applied to optical luminosities, and we use SED fitting, resulting in different bolometric corrections for each subtype.

A different (but not completely independent) BH mass estimation is obtained by using the normalized $\mathrm{^xLRD}$ spectrum presented in Figure~\ref{fig:xLRDs} (right panel, see next section for details). The factor that should be applied to that spectrum to fit the SED of the different subtypes of LRDs, after correcting for attenuation $\mathrm{A}^V_\mathrm{BH}$, provides an estimation of $\mathrm{M}_\mathrm{BH}$.

\subsection{Discrepancies between the emission line width and bolometric luminosity-based \texorpdfstring{$\mathrm{M}_\mathrm{BH}$}~  estimates}
\label{sec:discrepancies}

Black hole masses inferred from our bolometric-luminosity–based approach, assuming a gas-enshrouded SMBH and accretion disk, are systematically smaller than the median values obtained from single-epoch estimates based on emission-line widths and luminosities \citep[e.g.,][]{2025arXiv251007376J}, as also noted by \citet{2025arXiv251204208U}. This difference is illustrated in the right panel of Figure~\ref{fig:xLRDs}, where the $\mathrm{^xLRD}$ stack is plotted in luminosity-density units normalized by the inferred black hole mass (using the estimations based on bolometric luminosity). Individual sources with black hole masses derived from H$\alpha$ line widths using classical single-epoch calibrations \citep[e.g.,][]{2013ApJ...775..116R} are shown in red. We find that the mass-normalized LRD stacks differ from the individual single-epoch estimates by $\sim$1–2~dex, with a scatter of $\sim$1~dex among individual sources.

\begin{figure*}[htp!]%[htp!]%[ht!]
\includegraphics[clip, trim=1.5cm 1.5cm 2.0cm 0.0cm,width=9cm,angle=0]{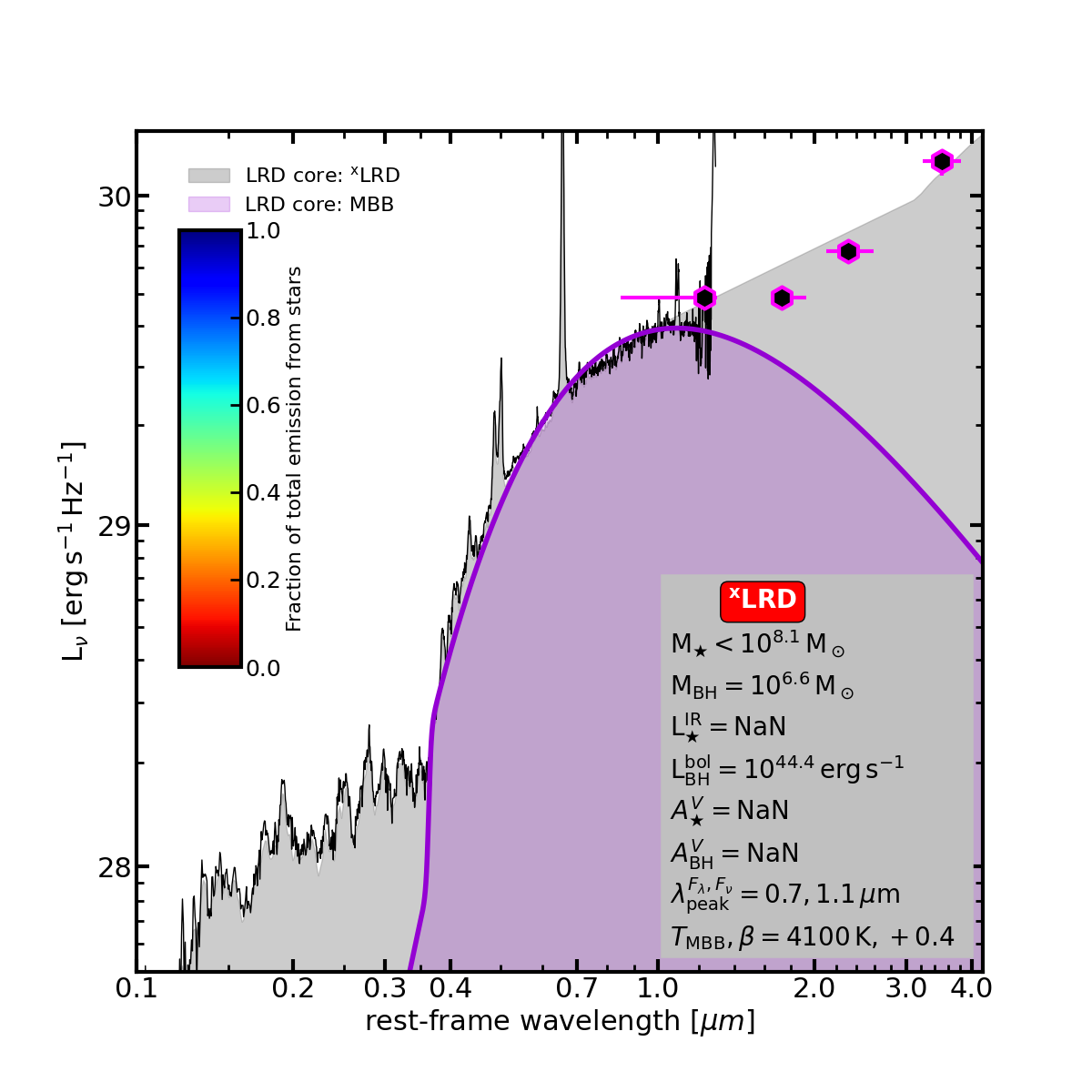}
\includegraphics[clip, trim=1.5cm 1.5cm 2.0cm 0.0cm,width=9cm,angle=0]{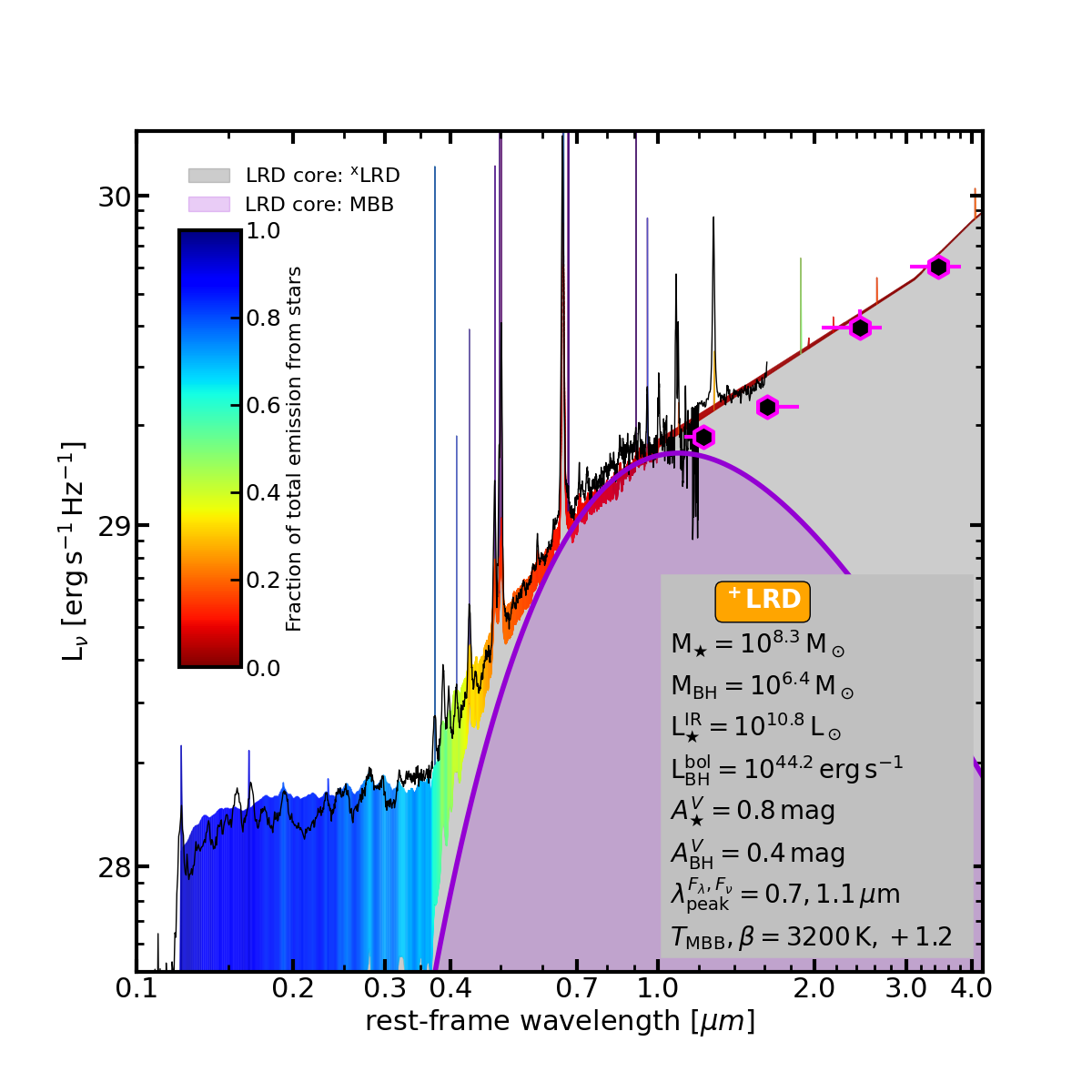}
\includegraphics[clip, trim=1.5cm 1.5cm 2.0cm 0.0cm,width=9cm,angle=0]{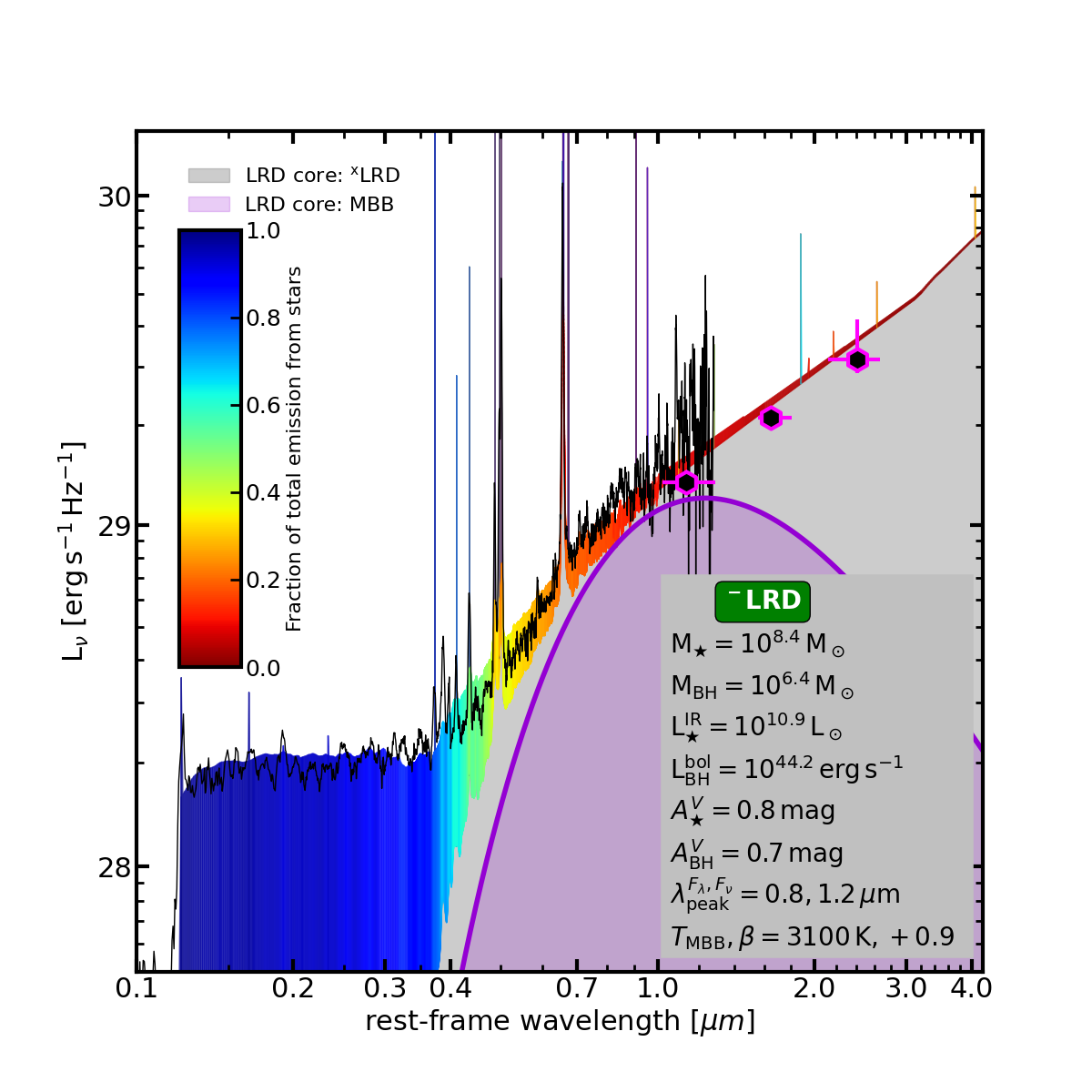}
\includegraphics[clip, trim=1.5cm 1.5cm 2.0cm 0.0cm,width=9cm,angle=0]{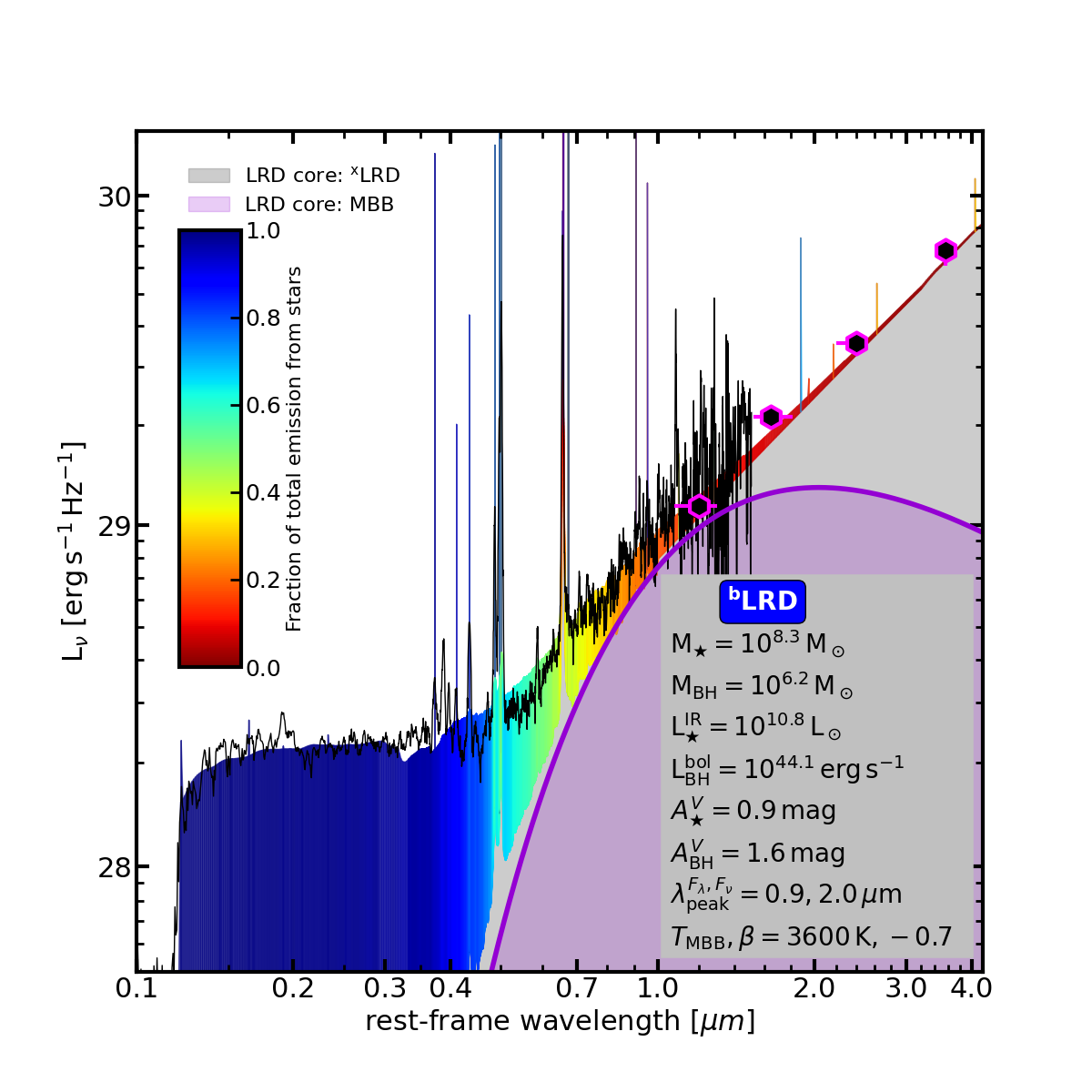}
\caption{\label{fig:sedfit}SED fitting results for the LRD subtypes. In all panels, we show in black the stacked spectrum and the MIRI stacks with black data points and errors in magenta. The shaded areas depict the LRD core component modelled with the two SED-fitting methods outlined in the text, i.e., using the $\mathrm{^xLRD}$ stack as the LRD core (gray), or using an MBB (magenta). The rainbow-palette colored regions represent the contribution from stellar emission as a function of wavelength (as given in the scale on the left of each panel). We provide the main physical properties obtained with this SED fitting exercise, namely the stellar mass M$_\bigstar$, the BH mass M$_\mathrm{BH}$, the power emitted by the stars and absorbed by dust L$^\mathrm{IR}_\bigstar$, the bolometric luminosity of the AGN obtained by integrating the LRD core component L$^\mathrm{bol}_\mathrm{BH}$, and the $A^V$ attenuation applied to the BH and stellar components, $A^V_\mathrm{BH}$ and $A^V_\bigstar$, respectively, the peak of the MBB emission, and its temperature and emissivity ($\beta$). }
\end{figure*}

Additional insight is provided by the comparison with the high-SNR NIRSpec analysis of \citet{2025arXiv250316595R}, who derived black hole masses for a sample of 12 LRDs using $R=1000$–2700 spectroscopy and a detailed H$\alpha$ profile decomposition that includes an intrinsic narrow component and line broadening due to electron scattering. In the right panel of Figure~\ref{fig:xLRDs}, the ten sources in common between their sample and ours are shown in green, normalized by their black hole mass estimates. The SMBH-mass normalized LRD-core empirical template lies in between the spectra in \citet{2025arXiv250316595R} and those for LRDs normalized by the single-epoch SMBH masses obtained by \citet{2025arXiv251007376J}. This might reflect uncertainties in Equation~\ref{eq:bhmass}, for example, linked to some non-negligible contribution to the LRD-core emission from processes not directly related to AGN accretion (e.g., a recent intense starburst). The latter would imply that our SMBH masses would be overestimated by a factor of 0.5~dex. As caveats to this statement, we can mention that additional uncertainties arise from our assumptions, such as the uniform $\lambda_\mathrm{Edd}\sim0.6$, the radiative efficiency and/or SED assumption \citep[following ][]{2026ApJ...996..129G}. Nevertheless, also the alternative premise about virial equilibrium and size-luminosity scalings underlying single-epoch calibrations \citep[e.g.,][]{2005ApJ...630..122G,2015ApJ...813...82R} may not be warranted in LRDs (or even for high-redshift AGN in general, see \citealt{2015ApJS..216....4S,2016ApJ...825..126D,2020ApJ...899...73F}).

\subsection{Best-fit BH and stellar properties of the LRD subtypes}
\label{sec:sedfit_results}

Figure~\ref{fig:sedfit} shows the results of the SED fitting for the stacked spectra of the individual subtypes. The full sample stack analysis provides similar results to those obtained for the $\mathrm{^+LRD}$ and $\mathrm{^-LRD}$ subtypes. We note that the $\mathrm{^xLRD}$ subtype is adopted as our template “naked” LRD core and is therefore not fitted with an explicit stellar component. Integrating its SED yields a characteristic black hole mass of $\mathrm{M}_\mathrm{BH}\sim10^{6.6\pm0.3}$~M$_\odot$.

We examine trends in the best-fit properties from the reddest to the bluest LRD subtypes (top-left to bottom-right panels), spanning $\mathrm{L_{5100}/L_{2500}}=10.4$, 3.8, 2.5, and 1.4, corresponding to optical slopes (in f$_{\lambda}$) $\beta_{\rm OPT}=1.52$, 0.53, 0.0, and $-0.6$, respectively.

%The bluest LRDs formally fall outside the \citet{2025ApJ...986..126K} selection criterion ($\beta_{\rm OPT}>0$), but are robustly identified using the more complete photometric selection of \citet{2024arXiv241201887B}, as discussed in Section~\ref{sec:selection}.

The color scale, which traces the fractional stellar contribution as a function of wavelength, shows that the primary driver of the declining $\mathrm{L_{5100}/L_{2500}}$ ratios and optical slopes is the increasing contribution of the host galaxy, strongest in the UV but extending into the optical. The host-to-core luminosity ratio at 5100~\AA\ increases sharply from $\mathrm{L}_{\bigstar}/\mathrm{L}_{\rm core}=0.3$ to 0.6 and 2.0 from the $\mathrm{^+LRD}$ to the $\mathrm{^bLRD}$ subtypes. For the $\mathrm{^xLRD}$, adopted as our ``naked" core template, the ratio is set to zero by construction, as its UV emission is assumed to arise entirely from the BH or its immediate surroundings.

The overall SED shape of the LRD subtypes also evolves systematically with decreasing $\mathrm{L_{5100}/L_{2500}}$. The UV continua become progressively bluer (flatter), with slopes $\beta_{\rm UV}=-1.6$, $-1.8$, and $-2.0$, while the optical continua become more power law like compared to the intrinsically curved LRD-core template (black line in Figure~\ref{fig:xLRDs}). As noted in \citet{2025arXiv251215853B}, this behavior is driven primarily by the growing contribution of the host-galaxy continuum (see also \citealt{2026arXiv260120929S}), but also by increasing attenuation of the LRD core, with $A^V_{\rm BH}$ rising from 0.4 to 1.6~mag toward the bluest subtypes.

In terms of intrinsic properties, the SED-fit black hole masses exhibit a mild decrease toward bluer subtypes, with $\mathrm{M}_\mathrm{BH}\sim10^{6.6}$–$10^{6.2}$~M$\odot$ and $L^{\rm bol}_{\rm BH}=10^{44.4}$–$10^{44.1}$~erg~s$^{-1}$, a change consistent with the smaller optical luminosities for bluer LRDs.

The stellar masses are quite constant across LRD subtypes, $\mathrm{M}_\bigstar\sim10^{8.3}$~M$\odot$. The stellar attenuation is also very similar for all LRD subtypes, $A^V_\bigstar\sim0.8$~mag.

Additional support for enhanced dust attenuation comes from the steep MIRI fluxes extending the NIRSpec SEDs, which follow a similar power-law trend and suggest the presence of dust re-emission (Section~\ref{sec:nir}). Comparing the level of reddening of the core and stellar emission, the   absorbed luminosities for the LRD core correspond to infrared luminosities in the range $10^{9.7}$ to $10^{10}$~L$_\odot$. These values are a factor 6--13 smaller than the total infrared luminosities linked to dust attenuation of the stellar emission, typically slightly below the luminous infrared galaxy (LIRG) limit ($\mathrm{L}_\mathrm{IR}^\bigstar\sim10^{11}$~L$_\odot$). The LRD core absorbed luminosities we calculate with the SED-fitting exercise are consistent with torus emission models dominated by hot dust. For example, only the models in the right panel of Figure~\ref{fig:stacked_all_miri} with peak dust emission around 20--30~$\mu$m provide low-enough luminosities. Those models are characterized by very compact tori and the lowest optical depths in \citet{2015A&A...583A.120S} models.

% consistent with the observed flattening of the UV slopes

%As a complement to our default host+LRD-core fits, Figure~\ref{fig:xLRDs} also shows two-component fits combining a stellar host with an MBB, enabling direct comparison with the modeling approaches of \citet{2025arXiv251121820D} and \citet{2025arXiv251204208U}. As expected, the MBB component closely resembles the LRD core and exhibits a similar decline in $\mathrm{L}_{\star}/\mathrm{L}_{\rm MBB}$ at 5100~\AA, reinforcing the conclusion that the increasing contribution of the host galaxy is the primary driver of the observed color diversity among LRD subtypes.

As a complement to our default host+LRD-core fits, Figure~\ref{fig:sedfit} presents two-component models combining a stellar host with an MBB core, allowing direct comparison with the approaches of \citet{2025arXiv251121820D} and \citet{2025arXiv251204208U}. The MBB component closely resembles the $\mathrm{^xLRD}$ template and shows a similar decline in $\mathrm{L}_{\rm MBB}/\mathrm{L}_{\bigstar}$ at 5100~\AA\ toward bluer subtypes, consistent with host-galaxy growth driving the observed color diversity. However, as discussed in Section~\ref{sec:nir}, the MBB models fail to reproduce the steepening of the rest-frame near-IR SED traced by the MIRI photometry, which suggests the presence of an additional dust re-emission component in all LRD subtypes, or a mix of emitting components with a variety of temperatures.

From the MBB fits we also estimate the wavelength of the SED peak in $f_\lambda$, which varies across the subtypes from $\lambda_{\rm peak}=0.7$ to 0.9~$\mu$m (corresponding to 1.1 to 2.0~$\mu$m in $F_\nu$). This range is consistent with \citet{2025arXiv251121820D}, who find $\lambda_{\rm peak}\sim0.60$–0.80~$\mu$m and report a mild anti-correlation between $\lambda_{\rm peak}$ and Balmer-break strength. Agreeing with this trend, the LRD core, which has the highest $\mathrm{L_{5100}/L_{2500}}$ and strongest Balmer break among the subtypes, peaks just bluewards of $\lambda_{\rm peak}\simeq0.7$~$\mu$m, while bluer subtypes peak at progressively longer wavelengths. 

%The $\mathrm{^bLRDs}$ show a modest deviation from this trend. In these systems, the host-galaxy contribution becomes dominant, reaching $\gtrsim50$\% in the 0.6–0.8~$\mu$m range, making the weaker LRD-core peak poorly constrained.

\subsection{The \texorpdfstring{$\mathrm{M}_\mathrm{BH}/\mathrm{M}_\bigstar$}~ relation for LRD subtypes}

\begin{figure*}[htp!]%[htp!]%[ht!]
\begin{center}
\includegraphics[clip, trim=1.0cm 0.5cm 1.5cm 0.0cm,width=16cm,angle=0]{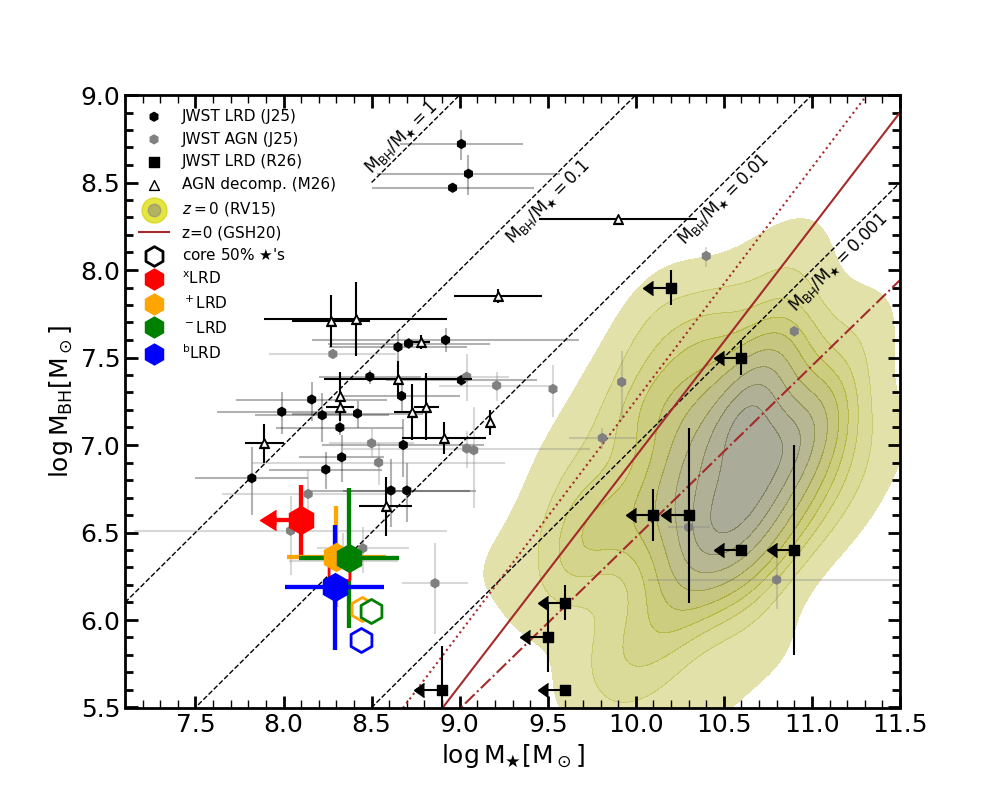}
\caption{\label{fig:bh_vs_stars}Comparison of the BH and stellar masses for our stacks  are shown with filled hexagons in colors, each one representing a different LRD subtype. Open hexagons show results if we consider that 50\% of the bolometric luminosity of the LRD core is linked to stellar emission (see text for details). Open hexagons show results for our stacks using a calibration based on BH masses from single-epoch Balmer line data. We also show with a shaded area the $z=0$ AGN in \citet{2015ApJ...813...82R}, and with brown lines the relationships found by \citet{2020ARA&A..58..257G} for all (continuous line), early-type (dotted), and late-type (dash-dot) $z=0$ galaxies. The dots in black and gray depict the AGN in \citet{2025arXiv251007376J}, LRDs and non-LRDs, respectively. The squares are the AGN in \citet{2025arXiv250316595R}. The open triangles depict the AGN in \citet{2026arXiv260115962M}. Dashed oblique lines mark specific BH-to-stellar mass ratios.}
\end{center}
\end{figure*}

% with a black line and gray shaded area the correlation found by \citet{2013ARA&A..51..511K}

The stellar and BH masses are compared in Figure~\ref{fig:bh_vs_stars}. As also obtained by \citet{2025arXiv251204208U}, the BH-to-stellar mass ratio for LRDs are above the local relationship but only moderately. Typically BH masses are 0.8--3\% of the stellar values. This ratio is more similar to the values obtained for AGN which do not qualify as LRDs \citep[e.g.,][]{2024A&A...691A.145M,2025arXiv251007376J,2025arXiv251119609B}, and significantly smaller than the ratios obtained by combining the observed width of the broad lines with single-epoch BH mass estimations. Figure~\ref{fig:bh_vs_stars} also shows a comparison of masses if 50\% of the bolometric emission of the $\mathrm{^xLRD}$ spectrum, assumed to be the LRD core and identified with a BH surrounded by dense gas, had instead a stellar origin, linked to a massive young burst, consistent with the WR features tentatively identified in all LRD subtypes, including the most extreme ones. The open hexagons shows the results after applying this correction. BH-to-stellar mass ratios would decrease by a factor of 2--3 compared to our fiducial estimations (filled hexagons).

We compare our masses to the aforementioned \citet{2025arXiv250316595R}, who use single-epoch estimations based on the \textit{intrinsic} line widths. The BH masses of our stacks are slightly larger than the values obtained for their 12 individual LRDs (most of their sources would be classified as $\mathrm{^xLRD}$s and $\mathrm{^+LRD}$s). They report relatively large upper limits ($10^{9-11}$~M$_\odot$) for the stellar masses for these sources, given that they fit the whole UV-to-near-infrared SED with stellar population, but they acknowledge the optical range might be dominated by an AGN, so most of their host masses can be regarded as upper limits.

%so the combination of lower BH masses compared to single-epoch estimations and larger stellar masses than what we obtain in this paper ($\sim10^{8.5}$~M$_\odot$),  result in their  BH-to-stellar mass ratio being consistent with the classical AGN in, e.g., \citet{2013ARA&A..51..511K} and \citet{2015ApJ...813...82R}

\section{Discussion}
\label{sec:discussion}

We organize the discussion around eight different open questions, presenting our results in the context of our current knowledge about LRDs.

\subsection{Are LRDs AGN?}
\label{sec:diss1}

Evidence for AGN activity in LRDs is primarily based on the detection of broad Balmer emission lines at high incidence across the population, most notably H$\alpha$ \citep[e.g.,][]{2023ApJ...954L...4K,2024ApJ...964...39G}, with reported FWHM values of 2000$-$4000~km~s$^{-1}$ \citep{2025ApJ...986..126K,2025arXiv251215853B}. The stacked spectra presented here similarly reveal broad H$\beta$, H$\alpha$, and He\,II~$\lambda4687$ emission, with H$\beta$ FWHM values of 3000$-$5000~km~s$^{-1}$ that systematically decrease toward bluer LRD subtypes. Interestingly, for H$\alpha$ we find the opposite trend. i.e., bluer LRDs present larger values for the H$\alpha$ FWHM, although this could arise in part from [NII] contamination, which we cannot measure reliably with the spectral resolution of the prism. %The H$\beta$ FWHM is  20\% larger than that for H$\alpha$ for the reddest LRDs, but $\sim60\%$ smaller for the two median and the bluest LRD subtypes. 

In this paper, we explicitly test whether the observed diversity of LRD spectra (bright and faint emission lines and continuum) and optical-to-UV luminosity ratios can be reproduced by combining the core LRD template and varying host-galaxy contributions. We base our analysis on SED fitting using the {\sc synthesizer-AGN} code, and on the assumption that the LRD core can be modeled with the $\mathrm{^xLRD}$ stack, which is weak in the UV. We find that stellar emission contributes non-negligibly to the continuum, and to some emission lines, including those of hydrogen. The stellar contribution is strongest in the UV continuum, exceeding 50\% bluewards of 0.4~$\mu$m and 80\% at  $<0.25$~$\mu$m for all but the extremely red LRDs, which comprise $\sim$15\% of the population. This is consistent with the findings in \citet{2026arXiv260115962M}, based on image decomposition (see also \citealt{2026arXiv260202702B}). At optical continuum wavelengths, stars contribute $\sim$40\% of the flux at 0.5~$\mu$m for the median LRD, reaching up to 60\% for the bluest half of the population (Figure~\ref{fig:sedfit}). Our analysis do not consider that different types of LRDs could still present some UV-bright AGN component, especially in the near-UV, where the emission from the accretion disk and the BLR peak (in $F_\nu$), and in contrast with the far-UV.
%, where C\,III] and He\,II emission point to a dominant role of star formation. 

Additional evidence for the presence of a luminous AGN is provided by Fe\,II emission, which is ubiquitous in nearby AGN and QSOs (see Section~\ref{subsec:ironlines} for references). Our stacked spectra show that optical Fe\,II emission, in the 4100$-$5500~\AA\ region as well as around 6500~\AA, is common in nearly all LRDs, consistent with previous detections in both the UV \citep{2024arXiv241204557L} and the optical \citep{2025arXiv251000103T,2025arXiv251000101D, 2025arXiv250909607L,2025ApJ...994L...6T}. In the UV, Fe\,II emission is also detected in all but the bluest LRDs.

%e\,II features are not clearly detected in the bluest $\sim$15\% of the population, where young stellar populations dominate these spectral regions, whereas

The location of LRDs in diagnostic diagrams based on emission-line ratios further supports the presence of AGN activity in most systems. The bulk of the population exhibits high [NeIII]$_\mathrm{3870}$/[OII]$_\mathrm{3728}$ ratios in the OHNO diagram \citep[][see measurements in our stacks in Table~\ref{tab:stats_specobs}]{2011ApJ...742...46T,2015ApJ...798...29Z,2022ApJ...926..161B}, consistent with ionization by an intrinsically hard source such as an AGN. The reddest LRDs, however, occupy a different region of this diagram, showing low [OIII]/H$\beta$ ratios (Figure~\ref{fig:ratios}), which may reflect lower metallicities (see discussion in \citealt{2025arXiv250522567M} about near pristine gas in an AGN) or additional competing effects (e.g., linked to dense gas clouds). Further evidence for a powerful ionizing source is provided by the widespread detection of He\,II and HeI lines discussed in this work. However, we find that UV lines such as CIII] and He\,II$\lambda$1640 point to a significant contribution, if not a dominant one, of stars in this spectral regime (consistent with measurements for sources identified as super star clusters or globular clusters in lensed galaxies, \citealt{2014MNRAS.445.3200S,2017ApJ...842...47V}). In contrast, Fe\,II and Mg\,II emission (the latter more easily produced in star-forming galaxies) in the near-UV might indicate that this range, where the ($F_\nu$) SED of QSOs typically peak (see Figure~\ref{fig:stacked_all}), could be more dominated by an AGN (in particular, the accretion disk and BLR).

\subsection{Do LRDs host overmassive SMBHs?}
\label{sec:diss2}

SMBH masses in LRDs have been estimated using single-epoch spectroscopy based on relations calibrated on local AGN and QSOs, linking broad emission-line widths and optical luminosities \citep{2013ARA&A..51..511K,2015ApJ...813...82R}. Such estimates have been reported for both individual sources \citep[e.g.,][]{2024A&A...691A..52K,2025arXiv250316596N,2025A&A...701A.168D,2025NatCo..16.9830T} and statistical samples \citep{2024ApJ...963..129M,2024ApJ...964...39G,2025arXiv251007376J,2025arXiv251119609B}. Claims of overmassive black holes at high redshift, with SMBH-to-stellar mass ratios of order 0.1 or higher, are common in the literature \citep{2023ApJ...959...39H,2023ApJ...957L...3P,2024A&A...691A.145M}, with some LRDs inferred to reach even larger ratios \citep{2025ApJ...983...60C,2025arXiv251007376J,2025arXiv250821748J}.

Using the bolometric luminosities derived from the SED modeling of the
stacked NIRSpec spectra from UV to near-IR wavelengths
(Section~\ref{sec:sedfit}), and the Eddington ratios inferred from
Fe\,II emission as discussed in Section~\ref{subsec:ironlines}, we
estimate black hole masses following \citet{2025arXiv251204208U}, as well as with a mass-normalized empirical template. The
resulting bolometric luminosities,
$\mathrm{L}_\mathrm{BH}^\mathrm{bol}=10^{44.1-44.4}$~erg~s$^{-1}$, combined
with typical Eddington ratios of $\lambda_\mathrm{Edd}\sim0.6$, imply
black hole masses of $\mathrm{M}_\mathrm{BH}\sim10^{6.0-6.5}$~M$_\odot$. When
compared to typical stellar masses of $\mathrm{M}_\bigstar\sim10^{8.3}$~M$_\odot$,
these values correspond to SMBH-to-stellar mass ratios of $\sim$0.07–3\%.
This is at least an order of magnitude higher than observed in nearby
AGN at similar stellar masses, and a factor of $\sim$3$-$4 above
the highest ratios seen in massive local galaxies
\citep{2020ARA&A..58..257G}.

We remark several points about this comparison. First, we are comparing total stellar masses for galaxies with BH masses. If LRDs probe the early formation of bulges, we should compare with results in, e.g., \citet{2013ARA&A..51..511K}, which obtain larger BH-to-stellar mass ratios (as high as 0.7\%). These ratios would be more consistent with our measurements for LRDs, especially if the LRD core is not completely dominated by an AGN but has some stellar contribution (cf. open hexagons in Figure~\ref{fig:bh_vs_stars}). 
%\citep{2013ARA&A..51..511K}.

Indeed, our calculations assume that the LRD core emission is powered exclusively by a SMBH. However, we find spectral features consistent with massive WR stars even in the most luminous and reddest LRDs, which otherwise appear AGN-dominated in the UV. This would therefore introduce a non-negligible correction to the inferred bolometric luminosities and SMBH masses, which would be smaller. We would move, consequently, into the intermediate-mass regime ($\mathrm{M}_\mathrm{BH}<10^{6}$~M$_\odot$; see \citealt{2020ARA&A..58..257G}) . On the opposite direction, an underestimation of the bolometric luminosities (i.e., if the intrinsic SED has considerable missing flux), that might depend on LRD subtype, would make BH masses larger than our estimations. The effect of Eddington ratios (assumed to be constant for all LRDs) is also a caveat to consider.

%Interpreting the emission around He\,II$_\mathrm{4687}$ as a WR blue bump implies a star-forming component with a mass of order $10^8$~M$_\odot$. The bolometric luminosity of such a young burst could exceed $10^{44.5}$~erg~s$^{-1}$, depending primarily on its age, and especially if the SFH is extended, and would therefore introduce a non-negligible correction to the inferred bolometric luminosities and SMBH masses. 

Considering all the evidence -- and under the assumptions outlined above -- we conclude that the BHs in LRDs could be mildly overmassive. But within the uncertainties the BH-to-stellar mass ratios appear to be similar to those of other AGN at similar redshifts. In addition, LRDs, as AGN in general, that are luminous enough for detection are likely to sample the high-scatter side of the BH-to-stellar-mass relation \citep{2007ApJ...670..249L}. 

\subsection{How can we reconcile single-epoch BH mass estimates with measurements based on bolometric luminosities?}
\label{sec:diss3}

Several explanations may account for the discrepancy between BH mass estimates obtained with measurements based on bolometric luminosities compared to single-epoch spectroscopy. The latter are based on empirical relations between optical (5100~\AA) continuum luminosity and H$\alpha$ or H$\beta$ line emission calibrated for local AGN (and, in particular, mostly based on a limited, biased sample of Sy~1 galaxies) and QSOs \citep{2005ApJ...630..122G,2013ApJ...775..116R}. However, those relationships provide predicted line luminosities which are 0.2–0.6~dex lower than those observed in LRDs, even when the full line profiles are considered (i.e., not only the broad components) and considering all the emission in the optical is associated with the AGN (i.e, there is a negligible contribution from other emitting sources such as a nuclear starburst). If the bolometric-luminosity–based masses are closer to the true values, this would imply that black holes in LRDs are unusually efficient at producing Balmer-line emission, or that a significant fraction of the observed SED is not directly associated with AGN accretion \citep{2026arXiv260206954C}. Notably, this tension largely disappears for bluest LRD subtypes, for which black hole masses derived from emission-line widths and from bolometric luminosities agree to within $\sim$0.2~dex. Jointly with the higher efficiency in producing line photons, part of the widening of emission lines might be associated with radiative transfer effects, rather than with virial motions \citep{2025arXiv250316595R,2026MNRAS.545f2131C}. 

%The discrepancy in scaling relations found for QSOs also applies to Fe\,II emission, as we mentioned in Section~\ref{sec:eddington}.

\subsection{Physical size scales of the central engine in LRDs}

Within the BH$^{\star}$ or BHE framework, the radius of the emitting photosphere can be estimated from the Stefan–Boltzmann law \citep{2025arXiv251204208U}. Using the measurements for the different LRD subtypes presented here, together with the temperature–luminosity relation from \citet{2025arXiv251121820D}, we infer photospheric radii of $\sim$7$-$13 light-days (1200$-$2300 AU) from the reddest to the bluest LRDs, with a typical value of $\sim$9 light days, consistent within a factor of $\sim$2 with independent estimates \citep{2025arXiv251121820D,2026arXiv260212548P} and theoretical work \citep{2025MNRAS.544.3407K,2026arXiv260302317L}.

A size estimate for the BLR can be obtained by applying the reverberation-mapping relation between optical luminosity and BLR size found in \citet{2000ApJ...533..631K}, after correcting the optical luminosities using our SED fitting results. This yields BLR sizes of $\sim$37–7 light days (6500--1200
AU) from the reddest to the bluest LRDs, with a typical value
of $\sim$16 light days.

%If the asymmetry observed in the H$\alpha$ line profile traces inflowing gas in the inner BLR \citep{2023Symm...15.1859M}, then the inflow velocity must remain below the free-fall velocity. For a black hole mass of $\mathrm{M}_\mathrm{BH}\sim10^{6.5}$~M$_\odot$, appropriate for $\mathrm{^xLRD}$s, this implies that the inflowing material originates at radii smaller than $\sim$3–4 light days from the BH.

%1191.86 to ld
%0.006 to 0.011~pc for LRD subtypes from the reddest to the bluest, the average being 0.008~pc

\subsection{Are LRDs quasi-stars (aka black hole stars)?}

The red optical slopes of LRDs and the strong Balmer breaks observed in the $\sim$16\% of the LRD population with the highest $\mathrm{L_{5100}/L_{2500}}$ values ($\mathrm{^xLRD}$s) can be reproduced by models invoking a very luminous source embedded in dense gas, as proposed in several studies
\citep{2025arXiv250709085B,2025arXiv250316596N,2025A&A...701A.168D,2025arXiv251121820D,2025arXiv251204208U,2025arXiv251203130I}.
However, a number of observational features are difficult to reconcile with a stable black-hole star scenario.

To start, the high-EW emission lines such as H$\alpha$ cannot arise from within the blackbody, which by definition needs to be optically thick at all wavelengths. To power these lines, some models invoke intense star formation outside the BH$^\star$ envelope, but this scenario makes it hard to reconcile the correlation between the continuum and line luminosity \citep{2025arXiv251121820D}. Alternatively, a clumpy medium (see, e.g., \citealt{2026arXiv260403563T}) could allow leakage of the H$\alpha$  photons from the inner regions through some sightlines (and explain self-absorptions).

In addition, several LRDs, all belonging to the $\mathrm{^xLRD}$ subtype, show clear evidence for self-absorption in the brightest emission lines \citep[e.g.,][see also \citealt{2025arXiv251215853B}]{2024ApJ...963..129M,2025arXiv251000103T,2025MNRAS.tmp.1995D}. While not in direct
contradiction with BH$^\star$ models, these features point to the presence of warm, dense gas with complex kinematics, including signatures of outflowing \citep{2025arXiv251000103T} or possibly inflowing material -- sometimes both blueshifted and redshifted absorption are seen simultaneously in the same object \citep[e.g.,][]{2025arXiv251000101D}. This complexity is indicative of a dynamically evolving rather than stable configuration (see section~\ref{sec:timescales} for a discussion about timescales; see also \citealt{2023ApJ...948....5F} for a comparison with QSOs), clumpy gas, and/or a non-isotropic configuration.

%The observed self-absorption may also be related to the asymmetric H$\alpha$ profiles reported here for $\mathrm{^xLRD}$s (Figure~\ref{fig:halpha}), an effect that is much weaker or absent in other LRD subtypes, and which is difficult to reconcile with a scenario lacking an accretion disk component.

The near-UV spectral region (i.e., bluewards of the Balmer break) shows emission lines that are characteristic  of typical AGN, such as strong Mg\,II and Fe\,II emission, generally associated with the BLR (especially with regions near the accretion disk). This would suggest that there is not an isotropic envelope around the BH, but rather the covering factor is lower than one, due to holes/channels in the dense envelope or a more classical accretion disk (maybe with a thick inner region, see \citealt{2018FrASS...5....6M}) and torus structure. Emission in the far-UV is also present in the most extreme, reddest LRDs, including Ly$\alpha$ emission and flux bluewards of it, which is difficult to reconcile with a pure BH$^{\star}$/BHE scenario, and is typically identified with star formation (again, even in the most extreme LRDs). This would mean that pure naked SMBHs do not exist or are very rare in LRDs. 

Apart from the fact that our stacked spectra is consistent with a young stellar population dominating the UV emission in the majority of LRDs, we have found indications that massive stars may also contribute at optical wavelengths (see also \citealt{2026arXiv260120929S}), where black-hole star models predict the embedded source should outshine stellar emission. We remark that this is also true for our fiducial LRD core model, i.e., the $\mathrm{^xLRD}$ stack also shows features possibly linked to WR stars. These stellar contributions further challenge a purely black-hole star or black hole envelope interpretation, or rather demonstrate that more exhaustive modeling is needed to account for the non-negligible contributions of several emitting components at all wavelengths. We further discuss this topic in the following subsection.

Other weaknesses of the BH$^{\star}$/BHE hypothesis have been presented in this paper. 
%First, the larger H$\alpha$ width compared to H$\beta$ for virtually all LRDs except for $\mathrm{^xLRD}$s, except for the reddest, points to Raman scattering playing a role, something that would need some strong UV source \citep{2026MNRAS.545f2131C} near the dense gas emitting cloud \citep{2026arXiv260206954C}. 
The most important of these is that a single BH$^{\star}$ characterized by a MBB is not able to reproduce the emission in the near-IR, as probed by our MIRI stacks.

%%the infall hypothesis should be discussed in terms of a net infall radial velocity component lower than the free-fall velocity and associated with the innermost BLR (the shifts amplitudes are larger toward the line base), co-planar or nearly so with the accretion disk.

%6600 AA wing implies 0.01c velocity. 

\subsection{Do LRDs have dust?}

Our SED modeling indicates an increasing contribution from young stellar populations to the UV emission of LRDs toward bluer subtypes. The fits favor very young ages ($<10$ Myr) to reproduce part of the observed emission-line fluxes, consistent with the possible presence of Wolf$-$Rayet stars discussed in this work. To match the observed flat UV slopes, these models typically require moderate dust reddening ($A^{\rm V}_{\bigstar}\sim1$~mag). Alternatively, some studies fit the UV continuum using constant star formation models with ages of $\sim$10$-$20 Myr, which intrinsically produce flatter spectra and therefore require less dust ($A^V \sim 0.2$ mag; e.g., \citealt{2025arXiv251215853B}).

In either case, the stellar populations dominating the UV are expected to contribute non-negligibly to the optical continuum, particularly around H$\beta$ and even H$\alpha$. Thus, given that some fraction of the emission at 0.4--0.5~$\mu$m would come from the stars dominating the UV flux, and their contribution to the redder (0.5--1.0~$\mu$m) range would be much smaller, the spectral shape of the component dominating the optical (what we call the LRD core) should be different for the distinct types of LRDs. The LRD core is modeled in this work with the $\mathrm{^xLRD}$ stacked spectrum plus some amount of reddening, which increases as we move to bluer LRDs. In our SED-fitting strategy, this effect is parameterized through dust attenuation, although it could also reflect increased opacity or a lower effective temperature of the source responsible for the red optical SED in progressively bluer subtypes.

As shown in Figure~\ref{fig:sedfit}, the inferred reddening of the core component increases systematically across the LRD subtypes, from $A^{\rm V}_{\rm BH} = 0$~mag in $\mathrm{^xLRD}$s, the subtype used as a template (however, the $\mathrm{^xLRD}$s might have some dust themselves) to $A^{\rm V}_{\rm BH} \sim 1.5$~mag in the bluest LRDs, in agreement with similar BH$^{\star}$ modeling presented by \citet{2025arXiv251215853B}. If the core is instead modeled as a modified blackbody (MBB), the best-fit temperatures are around $T \sim 3000-4000$ K, but with varying $\beta$ values to reproduce the SED behavior up to $\sim1$~$\mu$m. Single-component MBB models are not able to reproduce the emission redwards of that value in any case. From SED modeling alone, it is difficult to determine whether attenuation or temperature primarily drives this evolution. However, there is a clear systematic change in SED shape, from strong curvature for the reddest LRDs to more power-law–like behavior in $F_\nu$  for blue ones \citep{2025arXiv251215853B}, or equivalently a shift in the peak of the SED in $F_\lambda$, as reported by \citet{2025arXiv251204208U} and \citet{2025arXiv251121820D}.

%Assuming the reddening effect is due to dust, the typical attenuations for the LRD core are moderate, $A^V\sim1$~mag, very similar to the values obtained for the stars. This might mean that both components are surrounded by a dust screen, which is, in fact, inherent to the \citet{2000ApJ...533..682C} law, so there is a circular argument there. 

While SED fitting provides indirect constraints that depend on modeling assumptions, MIRI photometry offers a direct probe of dust re-emission, as discussed in Section~\ref{sec:nir}. The stacked LRD SED shows a relatively flat rest-frame near-infrared continuum with a mild upturn beyond $\sim2$~$\mu$m, inconsistent with pure stellar populations or $T \sim 4000$~K MBBs, both of which predict declining spectra and negative $[1-3\,\mu\mathrm{m}]$ colors. Figures~\ref{fig:xLRDs} and \ref{fig:sedfit} show that this behavior persists across all LRD subtypes. Notably, the rest-frame near-infrared slope becomes progressively redder from $\mathrm{^xLRD}$s to $\mathrm{^bLRD}$s, with $[1-3\,\mu\mathrm{m}]$ color increasing from $\sim$0.6 to $\sim$1.6~mag. This trend supports increasing attenuation in the core component and suggests that changes in SED shape are more naturally driven by reddening rather than by decreasing temperature. It is also consistent with the findings of \citet{2024arXiv241201887B}, who reported redder $[1-3\,\mu\mathrm{m}]$ colors toward bluer UV-to-optical LRDs, and with those of \citet{2025A&A...704A.313D}, who demonstrate the need for hot ($\sim800$~K) dust to explain stacks of photometry for LRDs up to 3~$\mu$m, with colder dust dominating at longer wavelengths (see also \citealt{2024MNRAS.535..853J} and \citealt{2026arXiv260122214B}). We note that, under the hypothesis of increasing host contribution from $\mathrm{^xLRD}$ to $\mathrm{^bLRD}$, a natural prediction is the increased presence of dust as we move to bluer systems.

Independent evidence for dust in LRDs comes from their Balmer decrements, which typically imply $A^{\rm V} \gtrsim 0.5$~mag across all subtypes. For the reddest LRDs, however, interpreting these decrements purely in terms of dust is not straightforward, as their line emission may deviate from Case B recombination \citep{2025arXiv251006362N,2025arXiv251000101D}, for example, due to collisional excitation or resonant scattering in dense gas environments \citep{2026MNRAS.545f2131C, 2025arXiv251000103T}.

Finally, the presence of dust is consistent with the clear non-negligible metal enrichment observed in LRDs. All subtypes show strong oxygen, neon, and carbon emission lines, and many also exhibit features from iron, nitrogen, and sulfur, indicating chemically evolved gas even in these compact, high-redshift systems. However, the very low [OIII]/H$\beta$ ratios for the reddest LRDs might imply low metallicities, which would conflict with the presence of large amounts of dust.

\subsection{What timescales govern the evolution of the LRD population?}
\label{sec:timescales}

There are five types of inferred physical properties for LRDs that can be used as a clock to estimate the timescales of the LRD phenomenon and even establish an evolutionary sequence within the different LRD subtypes defined in this paper. These are:

\begin{itemize}
\item The stellar mass, which, leaving aside yields and mass loss by winds, and disregarding the effect of interactions (not likely to affect the compact LRDs), should  increase with time.
\item The supermassive black hole mass, which should also increase with time.
\item The presence of WR features, which point to a specific range of ages. And within those features, the evolution from nitrogen-rich WR to carbon-rich WR, which are typically claimed to follow a time sequence \citep{2007ARA&A..45..177C}.
\item Number density considerations, which should be linked to duty-cycle timescales.
\item The redshift distribution of the different subtypes of LRDs.
\end{itemize}

Out of the five clocks above, the mass of the supermassive black hole is the most uncertain, since it depends on the accretion mode as discussed in Sections~\ref{sec:diss2}~and~\ref{sec:diss3}. With this in mind, Figure~\ref{fig:bh_vs_stars} shows no clear evolutionary sequence among the LRD sub-populations; all of them present very similar BH masses. 

At least for the bulk of the LRD population, the stellar mass may be a more reliable measure of evolution. The stellar masses for all types except for the $\mathrm{^xLRD}$s are directly linked to the rest-frame UV emission. The relative contribution of stars and AGN to the continua in this spectral region is uncertain, but the line ratios (C\,III] over He\,II) point to a non-negligible contribution from stars. In addition, the SEDs are always flat but present emission lines, so the stars should be young (if some flux in the emission lines is linked to stars). The only way to make young stars (ages younger than $\sim10$~Myr) have a flat SED is dust. Older stars (ages around 50~Myr; see, for example, Figure~9 in \citealt{2025A&A...697A..90B}) can also explain flat UV slopes. Our SED fitting calculations point to modest amounts of dust, with $A^V\sim1$~mag. This dust attenuation would translate to mid-to-far infrared dust thermal emission typical of a LIRG ($\mathrm{L}_\mathrm{IR}^\bigstar\sim10^{11}$~L$_\odot$), which is consistent with upper limits imposed by ALMA and Herschel data \citep[see,e.g.,][]{2020A&A...643A...8G,2024ApJ...968...34W,2024ApJ...975L...4C,2025ApJ...991...37A,2025ApJ...991L..10S}.

Our stellar mass estimation for $\mathrm{^xLRD}$ is an upper limit, calculated by imposing that only the UV might have some contribution from stellar emission. In such case, $\mathrm{^xLRD}$ would correspond to an early evolutionary phase, and then the evolution would then progress to bluer LRDs. We note that there are not large statistically relevant differences in the stellar mass of the four different LRD subtypes. The most worrisome caveat for this discussion about evolution based on stellar masses is the unknown contribution of stars to the rest-frame optical and near-IR, and to a lesser extent to the near-UV (with respect to the far-UV). If the optical/near-IR spectral regions are dominated by or present a significant contribution from highly embedded stars or supermassive stars (see Section~\ref{sec:sms}), instead or jointly with an accreting SMBH, the stellar masses presented in Figure~\ref{fig:bh_vs_stars} should be considered as lower limits. However, the amount of mass hidden there would not be large compared to the mass calculated from the UV emission, as we will discuss next, and consistent with clustering arguments, which give stellar masses around $10^{8}$~M$_\odot$, \citep{2025ApJ...988..246M,2025MNRAS.539.2910P}.

Figures~\ref{fig:wr} and \ref{fig:wrall} show the spectral region around He\,II$\lambda4687$. Our fits of this region in terms of the WR blue bump has interesting implications. All subtypes of LRDs present a broad He\,II component. That would mean that LRDs should have ages around 3-6~Myr. In general, all LRD subtypes present brighter nitrogen compared to carbon lines, which would imply a spectrum dominated by WN stars, with ages around 3-5~Myr \citep{2007ARA&A..45..177C}. The WR signatures are weakest in $\mathrm{^bLRD}$s, indicating a more evolved state, dominated by more evolved stars. The link between WR and LRD subtype would imply a strong synchronization between the SFH and the black-hole accretion activity and/or black-hole evolutionary stage.

%The most different subtype in this sequence is $\mathrm{^bLRD}$s. Their properties are quite different from those of the other subtypes in several features. They present NV, [NIV], and [OIV] emission in the UV. Those observational properties, jointly with the higher stellar and SMBH masses, seem to indicate that they are a very evolved state of the LRD phase more similar to regular blue AGN, thus corresponding to a transition stage to classical QSOs.

Not much can be inferred from the redshift distribution of the different types of LRDs except for the bluest, whose redshift distribution is much broader and peaks at lower values, $z\sim5$ compared to $z\sim7$ for the rest of subtypes. The timescale for their appearance is then in the hundred Myr scale, rather than a few Myr.

A further piece of evidence about the timescales governing LRDs (and their subtypes) comes from their number densities. Leaving aside the $\mathrm{^bLRD}$, which can be identified with a more evolved stage given their redshift distribution, LRDs present a relatively flat redshift distribution covering the $4<z<9$ range, and a number density around $5-8\times10^{-5}$~Mpc$^{-3}$ \citep[e.g.,][]{2024ApJ...968...38K,2025ApJ...986..126K}. Considering a stellar mass of around $10^{8.3}$~M$_\odot$,  and comparing with the stellar mass function at the same redshifts \citep[e.g.,][]{2024MNRAS.533.1808W}, which provides number densities around $3-6\times10^{-3}$~Mpc$^{-3}$, LRDs account for 1--3\% of the general galaxy population. If all galaxies experience an LRD phase, this would mean that the timescale is around 20~Myr. For the  $\mathrm{^xLRD}$, which are $\sim16$\% of the global population, the timescale would be $\sim3$Myr, consistent with the WR discussion.

There are two more important points. After the WR phase, one could expect a high abundance of red supergiants. These stars could contribute significantly to the optical and near-infrared spectrum of LRDs. Indeed, the typical temperature of red supergiants is $\sim4000$~K, which is similar to the temperatures found with fits to modified black bodies by \citet{2025arXiv251121820D}. The second point is that all subtypes of LRDs present significant iron emission. This would be consistent with an enrichment due to the most massive stars. 

Drawing on all the information, all LRD subtypes except the bluest would be coeval, with mild indications of a possible evolutionary sequence from $\mathrm{^xLRD}$ to $\mathrm{^+LRD}$s and $\mathrm{^-LRD}$s, and finally $\mathrm{^bLRD}$s. The separation between $\mathrm{^xLRD}$ and $\mathrm{^+LRD}$s and $\mathrm{^-LRD}$s, however, seems to be dominated by BH mass (Figure~\ref{fig:bh_vs_stars}) and, most probably, by gas density differences (Figure~\ref{fig:ratios}). If all galaxies experience an LRD phase, the typical timescale is 20~Myr for the full LRD population, 3~Myr for the most extreme cases.

\subsection{Is there a credible alternative scenario to LRDs ``being" AGN?}
\label{sec:sms}

As we mentioned in the Introduction, several alternative scenarios have been proposed to explain the observational characteristics of LRDs. The solution that dominates the literature is the identification of LRDs with a very unique phase of SMBH evolution with distinct AGN signatures. The most direct evidence for the LRD core being dominated by an AGN is the broad hydrogen lines which are identified with a BLR and also a narrow component identified with a NLR or, most probably, surrounding star formation. The other distinctive spectral feature of LRDs, the V-shape SED, has been explained with the presence of a dense gas envelope, possibly in the form of a black hole star, or quasi-star.

However, the two main observational characteristics (i.e., V-shape and broad emission lines) of LRDs can still have a stellar origin. In this paper, we present some evidence of the presence of massive hot stars in virtually all LRDs, including the most extreme, ($\mathrm{^xLRD}$s), whose SED seems to be dominated by the LRD core at all wavelengths. This contrasts with bluer LRDs, which present significant emission in the UV with less prominent spectral features, more similar to star formation with relatively low dust content. If the blue bump in the $\mathrm{^xLRD}$ stacked spectrum is identified with WR stars, if their luminosity is similar to those observed in nearby galaxies, and if the IMF is universal and follows a \citet{2003PASP..115..763C} law, a nearly $10^8$~M$_\odot$ burst would be embedded in $\mathrm{^xLRD}$s. 

The problem is then how to explain the red spectral slope in the optical. Dust has been invoked in a number of papers \citep{2024ApJ...968....4P,2024ApJ...968...34W}, i.e., part of the recent starburst could be enshrouded by dust. Typical attenuations should exceed $A^V\sim2$~mag and present a gray attenuation law or differential extinction \citep{2024arXiv241201887B}. The presence of dust also implies that stellar masses should reach values around $10^{9.5}$~M$_\odot$ or even higher which, jointly with the small sizes of LRDs, would imply very high densities. Although these densities could explain the broadening of the emission lines (see \citealt{2024ApJ...977L..13B}), their values are similar to those observed for the regions with the highest mass concentrations in nearby galaxies (e.g., globular clusters). Those mass densities could be even exceeded if the stellar masses of LRDs are significantly higher than $10^{9.5}$~M$_\odot$, something that was claimed in the first papers about LRDs (e.g., \citealt{2023Natur.616..266L}), although now the mass estimates are 1.0-1.5~dex lower, more consistent with dynamical constraints \citep[e.g.,][]{2024MNRAS.535..853J,2025arXiv250821748J,2026MNRAS.545f2117D}. Another problem is the thermal emission of the dust associated with that intense star formation, which should be radiated in the mid-IR rather than the far-IR, i.e., the dust should be warmer than typical (U)LIRGs.

Although the presence of warm/hot dust is supported by the SEDs at wavelengths longer than $\sim0.7$~$\mu$m for some LRDs (in fact, the bluest, see next subsection), the $\mathrm{^xLRD}$ stack is  bluer than the other stacks, which would point to a lower dust content. However, the Balmer decrement is quite high for $\mathrm{^xLRD}$s, implying attenuations of $A^V\gtrsim3$~mag, which is not consistent with the $A^V$ values from SED fitting and is best reproduced with radiative transfer effects in dense gas clouds \citep{2026MNRAS.545f2131C}. 

If dust is not the main culprit for the red optical spectrum of LRDs, the alternative is some type of red star. Supermassive stars (SMS) have been claimed to produce the spectral signatures of LRDs \citep{2025arXiv250712618N,2025arXiv250722014Z,2026arXiv260215935C}, and in fact were claimed to be detectable by JWST prior to launch \citep{2018ApJ...869L..39S,2020A&A...633A...9M}. The presence of supermassive stars in LRDs does not rule out their identification with SMBHs, since the seeds for the latter have been claimed to come from the remnants of these red supermassive stars \citep[e.g.][]{2003ApJ...596...34B,2009MNRAS.396..343R,2019PASA...36...27W}

Taking the median LRDs ($\mathrm{^+LRD}$ and $\mathrm{^-LRD}$ subtypes) as example, our calculations point to the presence of a $\sim10^{7.7}$~M$_\odot$
 starburst in the core of LRDs, which, being identified with the $\mathrm{^xLRD}$ subtype, also presents features that can be linked to WR stars in a young starburst. A 4~Myr old burst with that mass would emit $L^\mathrm{bol}=10^{44.1}$~erg~s$^{-1}$, which is similar to the median bolometric luminosity $L^\mathrm{bol}=10^{44.2}$~erg~s$^{-1}$. The supermassive star should then be less luminous than $10^{10}$~L$_\odot$. Using the evolution in the HR diagram from \citet{2025arXiv250712618N}, the LRD emission would then need a moderate mass SMS, around $10^{5}$~M$_\odot$. For $10^{5}$~M$_\odot$, the temperature would be around 6000-7000~K, which is higher than the temperatures obtained from directly fitting the LRD spectra to modified black bodies (e.g., \citealt{2025arXiv251121820D,2025arXiv251204208U}), ranging from 3000 to 6500~K (typically, $\sim5000$~K). That range of temperatures would instead favor a number of $10^{3-4}$~M$_\odot$ stars rather than a more massive single object. However, SMSs can be as cool as 4000--5000~K (similar to supergiants), depending on their formation process \citep{2018MNRAS.478.2461G}, so $10^{3-5}$~M$_\odot$ could be consistent with an SED similar to a $\sim4000$~K blackbody, more massive (and hot) in more extreme, reddest LRDs.

Several pieces of evidence could suggest the presence of a SMS. The $\mathrm{^xLRD}$s present very low [OIII]/H$\beta$ ratios, which could indicate the infall (consistent with the H$\alpha$ profile) of low metallicity gas, favoring the formation of very massive stars (also direct collapse BHs, see \citealt{2026arXiv260114368P}). The presence of prominent nitrogen lines and large N/O ratios have also been interpreted to favor the presence of massive stars while forming proto globular clusters in the early Universe \citep{2024A&A...681A..30M}.

\begin{figure*}%[htp!]%[ht!]
\centering
%\fbox{
%\includegraphics[clip, trim=0.cm 0.cm 0.0cm 0.0cm,width=18.0cm,angle=0]{Figures/lrd_cartoon_v1.2.pdf}
\includegraphics[clip, trim=0.cm 0.cm 0.0cm 0.0cm,width=18.0cm,angle=0]{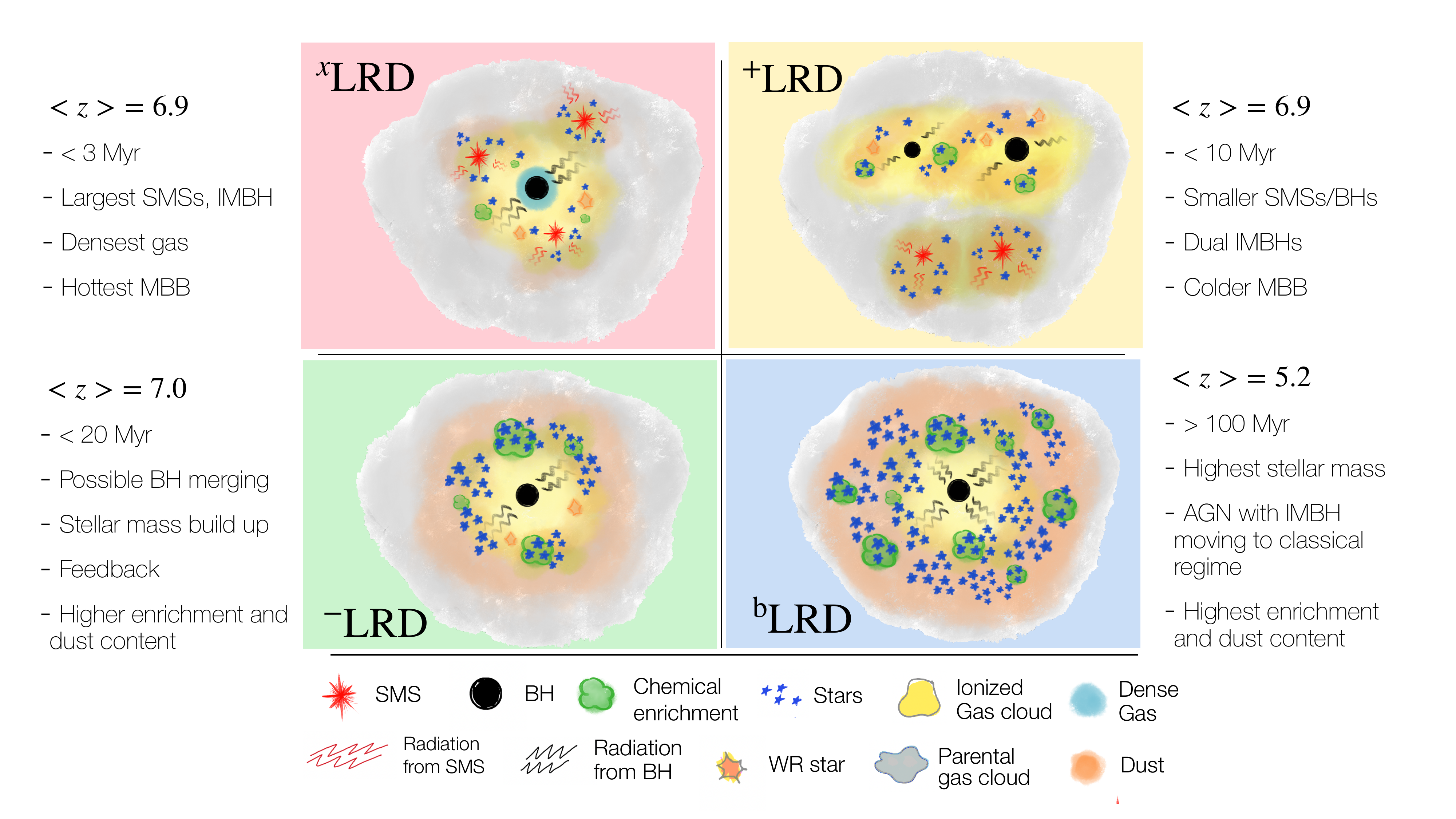}
%}
\caption{\label{fig:cartoon}Cartoon of the different subtypes of LRDs and the emitting components discussed in the paper. Each panel represent a different subtype of LRD: $\mathrm{^xLRD}$ with the background in red, $\mathrm{^+LRD}$ in orange, $\mathrm{^-LRD}$ in green, and $\mathrm{^bLRD}$ in blue. Star formation complexes are shown with yellow circles, with regular stars in blue, massive stars (rapidly evolving into red supergiants) in red, WR stars in cyan, black holes in black. The whole compact structure is embedded in a dense gas environment (gray cloud), with some dust content which gets higher as we move to bluer LRDs. Some relevant points introduced in the Discussion section are given in text boxes by each cartoon.}
\end{figure*}

As the [OIII]/H$\beta$ ratio increases toward bluer LRDs, likely reflecting higher metallicity and chemical enrichment, the total bolometric luminosity decreases, even for similar LRD core masses. If the LRD core is a SMBH, its mass increases with time, naturally implying lower Eddington ratios. At the same time, enriched star formation becomes progressively more important, building stellar mass and dominating the
UV emission, while conditions become less favorable for the formation of SMSs, leading to a decrease in the optical-to-UV luminosity ratio. Feedback from massive stars (including WR stars and possible SMSs), the SMBH, or both, together with gas exhaustion, would then quench further star formation. In parallel, dust production increases and the global attenuation rises. 

We summarize this evolutionary picture and its main physical ingredients in the schematic cartoon in Figure~\ref{fig:cartoon}. We note that the cartoon includes some relevant information about each subtype of LRDs, resulting from the analysis in this paper. In particular, concerning the median redshifts, they come from photometric measurements, given that we showed in Section~\ref{sec:classes} that the spectroscopic sample is biased. We remark that, given that uncertainties in photometric redshifts, the differences in median redshifts between the three reddest subtypes (values between $<z>=6.9$ and $<z>=7.0$) are not statistically relevant, but the difference with respect to the bluest LRDs is large enough to consider an evolutionary difference (as discussed in Section~\ref{sec:timescales}). We also note that the cartoon should not be interpreted as a detailed description of the internal structure of LRDs, but about the relevant ingredients, which might be organized in a complex spatial configuration.

The evolutionary timescales also constrain the nature of any putative SMS component. Supermassive stars evolve rapidly. According to \citet{2025arXiv250712618N}, the ages for which SMS would reach optical luminosity densities similar to the most extreme LRDs would be $\sim1$~Myr for $10^{4}$~M$_\odot$ objects and even shorter for higher masses. For stars more massive than 300--500~M$_\odot$, the lifetime of stars becomes independent of mass. The typical main sequence timescale for stars up to 150~kM$_\odot$ is 1.5~Myr \citep{2020MNRAS.494.2236W}. If SMSs accrete, their lifetime can be significantly longer \citep{2018MNRAS.478.2461G}.
 
According to our analysis, the presence of WR features implies ages of $\lesssim5$–6~Myr, and number-density arguments suggest LRD phases lasting several Myr. This indicates that the LRD phenomenon cannot be instantaneous and favors either moderately massive ($\lesssim10^{4}$~M$_\odot$) SMSs or clusters hosting multiple such objects, consistent with the tendency of massive stars to form in binaries or multiple systems \citep[e.g.,][]{2012Sci...337..444S}. We note that an intense and very compact star formation event extending along several Myr is consistent with the formation of globular clusters (see \citealt{2018ARA&A..56...83B} and references therein; see also the connection between LRDs and globular clusters in \citealt{2026arXiv260215935C}).

Recently, \citet{2026arXiv260106015Y} reported the discovery of two LRDs, separated by $\approx 70$ pc and embedded in a star-forming galaxy with $M_\mathrm{UV}\sim -19$~mag at $z\sim 7$, dubbed ``The Red Eyes". The galaxy is highly magnified ($\mu \sim 20$) by the gravitational lensing field of cluster PLCKG004.5-10.5, observed as part of the Cycle 4 JWST program VENUS (GO-6882; PIs: S. Fujimoto \& D. Coe), and thanks to the increased resolution, they could establish that the LRD pair is not located in the center of the hosting galaxy but offset by roughly $r_e$, the effective radius. In other words, the LRDs are not "nuclear". This might point to merging as playing a significant role in the early assembly of AGN. But it might also provide evidence that LRD are not necessarily always "galactic nuclei", and that they might form around black holes of suitable mass.  This favors a scenario where LRDs are young galaxies observed early in their evolution with massive star formation involving WR phenomena, possibly SMSs, and rapid evolution to red supergiants (with cold atmospheres), quickly leading to the appearance of IMBHs which might then merge to form SMBHs. This scenario would be consistent with the formation of the most massive, earliest globular clusters, also taking into account the nitrogen enrichment mentioned above (see \citealt{2026arXiv260215935C}).

\section{Summary and conclusions}
\label{sec:conclusions}

We have compiled all publicly available NIRSpec/prism spectroscopy ($R\sim100$) and MIRI imaging for a sample of 249 LRDs distributed across six independent JWST fields, spanning redshifts $2.3<z<9.3$ (median and quartiles $z=5.7^{7.4}_{4.0}$). Using high SNR stacked spectra\footnote{Downloadable from \href{https://doi.org/10.5281/zenodo.18679061}{this link}.} covering the rest-frame wavelength range 0.1–4~$\mu$m, we investigate the physical properties and nature of LRDs across their full observed diversity.

The sample encompasses the entire heterogeneity of LRD properties, from the reddest and most extreme systems to the bluest LRDs, whose characteristics approach those of classical AGN and QSOs while still retaining the defining observational features of the LRD class, namely, their compact morphology and distinctive V-shaped UV–optical SED. To capture this diversity quantitatively, we construct stacked spectra for four subtypes defined by their optical-to-UV luminosity density ratios, $\mathrm{L_{5100}/L_{2500}}$: the most extreme $\mathrm{^xLRD}$s, the above-median $\mathrm{^+LRD}$s, the below-median $\mathrm{^-LRD}$s, and the bluest $\mathrm{^bLRD}$s.

The spectroscopic sample is broadly representative of the photometrically selected LRD population identified with NIRCam. Nevertheless, current follow-up programs show a modest bias toward lower redshifts, particularly for the reddest subtypes (median $z\sim5.2$ in the NIRSpec sample compared to $z\sim6.9$ in the corresponding NIRCam-selected sample).

The stacked spectra allow us to identify a set of robust observational characteristics of LRDs, summarized below.

\begin{itemize}

\item All stacked spectra of LRDs, independent of $\mathrm{L_{5100}/L_{2500}}$, show spectral features consistent with a stellar origin. 

\item Stellar emission has a non-negligible contribution to the far-UV spectral region for the $\mathrm{^xLRD}$s (including extreme examples such as The Cliff and MoM-BH*-1) and rest of LRDs, as inferred from C\,III] equivalent widths and C\,III]/He\,II$\lambda1640$ ratios.

\item Optical spectral signatures associated with young massive stars in a Wolf–Rayet phase (e.g., He\,II$\lambda4687$, carbon and especially nitrogen features) are detected in all LRD subtypes.

\item An absorption feature near 4550~\AA\ of uncertain origin (stellar atmospheres of cold stars -e.g., supergiants- or interstellar medium lines, most plausibly, Fe\,II absorption) is present in all LRD stacks.

\item Clear AGN signatures are found in all LRDs. In addition to broad H$\alpha$ and H$\beta$ components (the broad He\,II$\lambda$4687 could also be linked to an AGN), we detect ubiquitous Fe\,II emission in both the optical (notably around $\sim5300$~\AA) and the UV (2200–-3000~\AA).

\item Using Fe\,II emission jointly with Mg\,II and Balmer-line measurements, and assuming an AGN origin, we infer high accretion rates with $\lambda_\mathrm{Edd}\sim0.6\pm0.2$ for all LRD subtypes, except for the bluest systems, where reliable constraints are not possible.

\item Compared to classical AGN, LRDs exhibit enhanced H$\alpha$/optical and Fe\,II/optical luminosity ratios, indicating a higher efficiency in producing line emission relative to the optical continuum and suggesting a distinct structure and/or accretion mode relative to local AGN and QSOs.

\item Emission-line diagnostics such as [O\,III]/H$\beta$, Balmer decrements (H$\alpha$/H$\beta$, H$\beta$/H$\gamma$, H$\beta$/H$\delta$), Paschen-to-Balmer ratios (H$\beta$/Pa$\gamma$), and Balmer break strength correlate systematically with $\mathrm{L_{5100}/L_{2500}}$. Redder LRDs show lower [O\,III]/H$\beta$, stronger Balmer decrements, weaker Paschen-to-Balmer ratios, and larger Balmer breaks.

\item The observed line and flux ratios likely arise from multiple physical effects. The Balmer and Paschen decrements are broadly consistent with radiative-transfer processes (e.g., collisional excitation) in high-density gas enshrouding a SMBH or intense nuclear burst. In contrast, the systematic variation in [O\,III]/H$\beta$ with $\mathrm{L_{5100}/L_{2500}}$ is also naturally explained by changes in the relative contribution of the host galaxy and the LRD core, with stronger narrow-line emission emerging toward bluer, more host-dominated systems.

%\item We detect differences in the width of H$\alpha$ and H$\beta$ which depend on the LRD subtype. $\mathrm{FWHM(H\beta)/FWHM(H\alpha)}=1.2\pm0.1$ for $\mathrm{^xLRD}$ (consistent with the average for local AGN and QSO), and $\mathrm{FWHM(H\beta)/FWHM(H\alpha)}=0.6\pm0.2$ for the rest of subtypes. This can be interpreted as Raman scattering, which would point to an intense radiation source present in all LRDs except the reddest.

\item He\,I line ratios also point to significant effects linked to dense gas, with densities $n_e>10^5$~cm$^{-3}$, optically thick material, and temperatures around 5000-7000~K being necessary to explain the large measured values for the reddest LRDs.

\end{itemize}

We also perform SED fitting to the NIRSpec+MIRI stacks using a two-component model consisting of an LRD core and a stellar host. The LRD core is modeled empirically after the $\mathrm{^xLRD}$ stacked spectrum, which includes intrinsic (non-stellar) UV emission, or a MBB without it. In both approaches, the shape of the LRD core is allowed to vary through a free attenuation parameter to account for dust and/or gas obscuration. The core emission is interpreted as arising from an accreting SMBH and/or a compact nuclear starburst embedded in dense gas. The stellar host includes a very young ($<5$~Myr) population to reproduce the emission-line features and is modeled with its own independent attenuation. This analysis provides the following main results:

\begin{itemize}

\item The observed diversity in continuum slopes is largely explained by variations in the host-to-core luminosity ratio. At $\lambda=5100$~\AA, $\mathrm{L}_{\bigstar}/\mathrm{L}_{\rm core}$ increases from 0 in the ``naked'' $\mathrm{^xLRD}$ template to 0.3, 0.6, and 2.0 toward progressively bluer subtypes, indicating a transition from BH-dominated to host-dominated optical emission. This trend is accompanied by a modest increase in attenuation of the BH core, with $A^V_{\rm BH}$ rising from 0.4 to 1.6~mag toward the bluest LRDs. 
\item The stellar masses are remarkably uniform across LRD subtypes, with $\mathrm{M}_\bigstar\sim10^{8.3}$~M$_\odot$ and a scatter of $\sim0.3$~dex. 
%The bluest LRDs show slightly higher stellar masses (by $\sim0.2$~dex).

\item Stellar emission contributes more than 50\% of the flux at $\lambda<0.4~\mu$m for most subtypes (and more than 80\% for the bluest LRDs), while remaining non-negligible ($\sim20$\%) at optical and near-infrared wavelengths.

\item The BH masses of all subtypes are $\mathrm{M}_\mathrm{BH}\sim10^{6.0-6.5}$~M$_\odot$. This means that the BH-to-stellar mass ratio is around 1\%, with up to 2\% for the reddest LRDs, and 0.7\% for the bluest. These ratios are similar to the ones found for other AGN at high redshift. They are more than an order of magnitude larger than what is observed for galaxies of the same stellar mass at $z=0$, and a factor of 3$-$4 larger than the ratios observed for the most massive galaxies in the local Universe. 

\item Smaller BH masses, within the intermediate-mass range ($\mathrm{M}_\mathrm{BH}<10^{6}$~M$_\odot$), would be obtained if the LRD core emission is not completely dominated by nuclear activity.

\item A modified blackbody with $T\sim4000$~K reproduces the rest-frame optical SED reasonably well, but fails to account for the excess emission in the rest-frame 1--4~$\mu$m range probed by MIRI. This near-infrared excess becomes progressively stronger toward bluer LRDs, suggesting the presence of an additional hot (1000–1500~K) dust component and/or emission from cooler dense gas clouds.

\item The near- to mid-infrared SED of LRDs is very similar to that observed for nearby Sy~1 galaxies in terms of colors and peaking wavelength, but it is very different to the SED of obscured AGN (e.g., Sy~2). A comparison with torus models point to large viewing angles (or high covering factors, but not necessarily unit), small torus radii, and relatively low optical depths of the dust clouds. 

\item The presence of dust increasing towards bluer LRDs would be consistent with a decreasing BH-to-stellar mass ratio, as the galaxy's ability to create more dust grows. The dust would then need to be channeled to the LRD core, with mechanisms that depend how far away the stars are from the nucleus.

\item Based on the SED-fitting analysis, as well as number-density arguments and the relative abundance of LRD subtypes, we estimate that the LRD phase lasts $\lesssim20$~Myr, with the most extreme (reddest, $\mathrm{^xLRD}$) phase lasting only $\sim3$~Myr, consistent with age constraints from Wolf–Rayet features.

\end{itemize}

We discuss all spectro-photometric properties together, jointly with results in the literature, in terms of nuclear activity linked to super-massive (or even intermediate-mass) BHs, intense and compact starbursts with the possible presence of super-massive stars (with a mass around $10^{5}$~M$_\odot$), and a link to globular clusters (e.g., based on strong nitrogen emission). Further spectro-photometric analysis of large samples of LRDs will be needed to constrain their nature more securely, taking into account their heterogeneity.

\begin{acknowledgments}
We thank the referee for their constructive comments to our original manuscript.
We thank Roberto Maiolino, Kimihiko Nakajima, Ismael Garc\'{\i}a-Bernete, and Cristina Ramos-Almeida for useful comments on the manuscript. We want to thank the PIs and co-Is of all the programs listed in Table~\ref{tab:obs}, and acknowledge funding linked to some of those programs for many of the co-authors of the paper. PGP-G acknowledges support from grant PID2022-139567NB-I00 funded by Spanish Ministerio de Ciencia, Innovaci\'on y Universidades MCIU/AEI/10.13039/501100011033,
FEDER {\it Una manera de hacer
Europa}. S.C.\ and G.V.\ acknowledge support by European Union's HE ERC Starting Grant No.\ 101040227 - WINGS.
F.D.\ and X.J.\ acknowledge support by the Science and Technology Facilities Council (STFC), by the ERC through Advanced Grant 695671 ``QUENCH'', and by the UKRI Frontier Research grant RISEandFALL. S.A.\ and B.R.P.\ acknowledge support from the research project PID2021-127718NB-I00 funded by the Spanish Ministry of Science and Innovation/State Agency of Research (MICIN/AEI/10.13039/501100011033). A.J.B.\ and J.C.\ acknowledge funding from the ``FirstGalaxies'' Advanced Grant from the European Research Council (ERC) under the European Union's Horizon 2020 research and innovation programme (Grant agreement No.\ 789056). E.C.L.\ acknowledges support of an STFC Webb Fellowship (ST/W001438/1). D.J.E.\ is supported as a Simons Investigator and by JWST/NIRCam contract to the University of Arizona, NAS5-02105. Support for program \#3215 was provided by NASA through a grant from the Space Telescope Science Institute, which is operated by the Association of Universities for Research in Astronomy, Inc., under NASA contract NAS 5-03127. J.S.\ acknowledges support by the Science and Technology Facilities Council (STFC) and ERC Advanced Grant 695671 ``QUENCH''. The research of C.C.W.\ is supported by NOIRLab, which is managed by the Association of Universities for Research in Astronomy (AURA) under a cooperative agreement with the National Science Foundation. C.T.D.\ acknowledges support from JWST program number GO-6368, provided through a grant from the STScI under NASA contract NAS5-03127. The work of GHR  was supported in part by grant 80NSSC18K0555, from NASA Goddard Space Flight Center to the University of Arizona. H\"U acknowledges funding by the European Union (ERC APEX, 101164796). Views and opinions expressed are however those of the authors only and do not necessarily reflect those of the European Union or the European Research Council Executive Agency. Neither the European Union nor the granting authority can be held responsible for them. JW gratefully acknowledges support from the Cosmic Dawn Center through the DAWN Fellowship. The Cosmic Dawn Center (DAWN) is funded by the Danish National Research Foundation under grant No. 140. C.N.A.W. and Y.Z. acknowledge support from the JWST/NIRCam contract to the University of Arizona NAS5-02105. ST acknowledges support by the Royal Society Research Grant G125142. JSD and DJM acknowledge the support of the Royal Society through the award of a Royal Society University Research Professorship to JSD. B.R.P acknowledges support from grant PID2024-158856NA-I00 funded by Spanish Ministerio de Ciencia e Innovación MCIN/AEI/10.13039/501100011033 and by ''ERDF A way of making Europ''. This work is based on
observations made with the NASA/ESA/CSA James Webb
Space Telescope. The data were obtained from the Mikulski
Archive for Space Telescopes at the Space Telescope Science Institute, which is operated by the Association of Universities for Research in Astronomy, Inc., under NASA contract NAS5-03127 for JWST.
\end{acknowledgments}

\vspace{5mm}
\facilities{JWST (NIRCam), JWST (MIRI), HST (ACS), HST (WFC3). All the JWST data used in this paper can be found in MAST: \dataset[10.17909/jmxm-1695]{http://dx.doi.org/10.17909/jmxm-1695} and \dataset[ 10.17909/8tdj-8n28]{http://dx.doi.org/10.17909/8tdj-8n28}. Spectro-photometric stacks can be downloaded from \href{https://doi.org/10.5281/zenodo.18679061}{this link}.}

\software{astropy \citep{2013A&A...558A..33A,2018AJ....156..123A},  
          Cloudy \citep{2013RMxAA..49..137F}, {\sc prospector} \citep{2017ApJ...837..170L, 2019ApJ...876....3L, 2021ApJS..254...22J}, {\sc synthesizer} \citep{2003MNRAS.338..508P,2008ApJ...675..234P}.
          }

\appendix
\section{Scatter of SED shapes for the LRD subtypes}
\label{app:sedexamples}

In this Appendix, we further discuss the properties of the different LRD subtypes defined in the main text (see Sections~\ref{sec:classes} and \ref{sec:subtypes}), especially focusing on the scatter of SEDs for each subtype.

\begin{figure*}[h!]%[htp!]%[ht!]
\centering
\includegraphics[clip, trim=0.2cm 0.2cm 1.5cm 2.0cm,width=8.cm,angle=0]{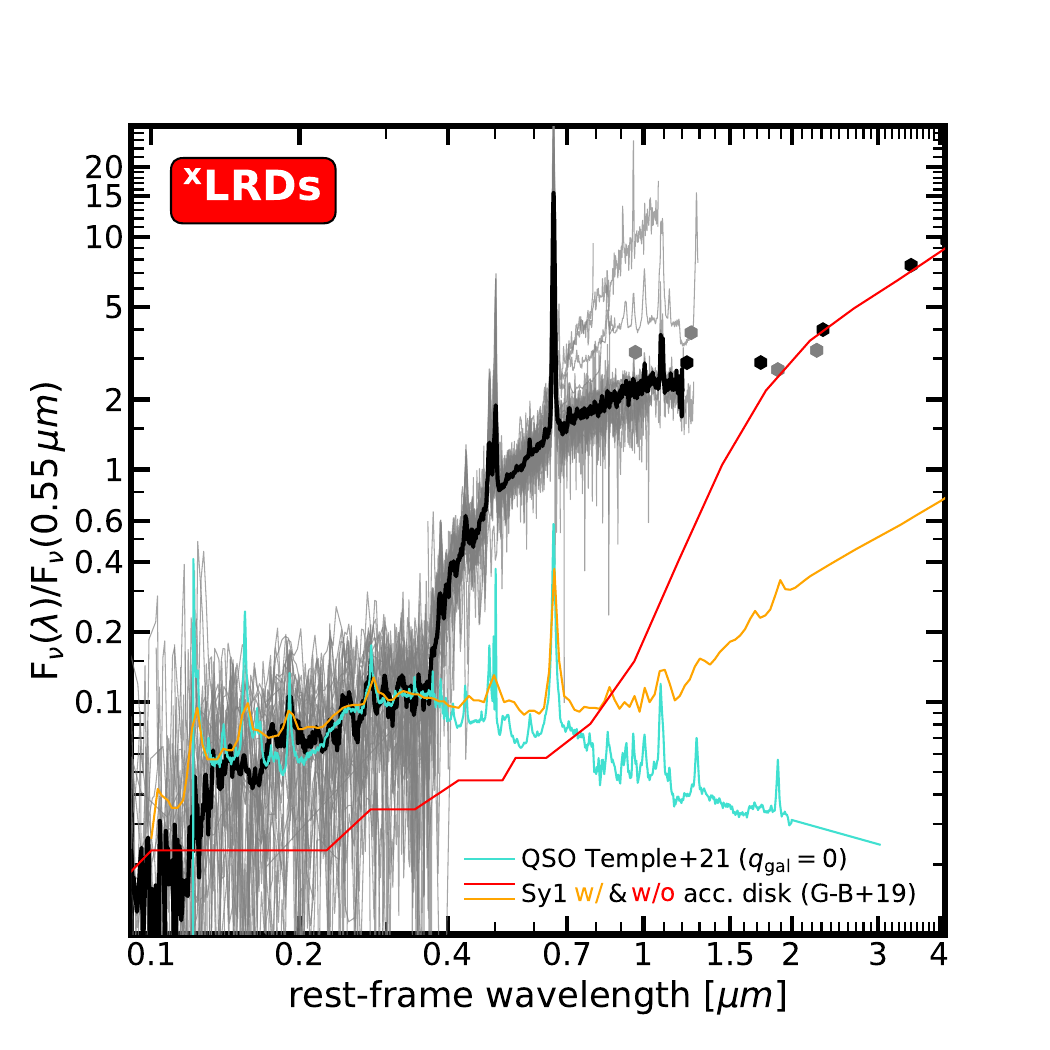}
\includegraphics[clip, trim=0.2cm 0.2cm 1.5cm 2.0cm,width=8.cm,angle=0]{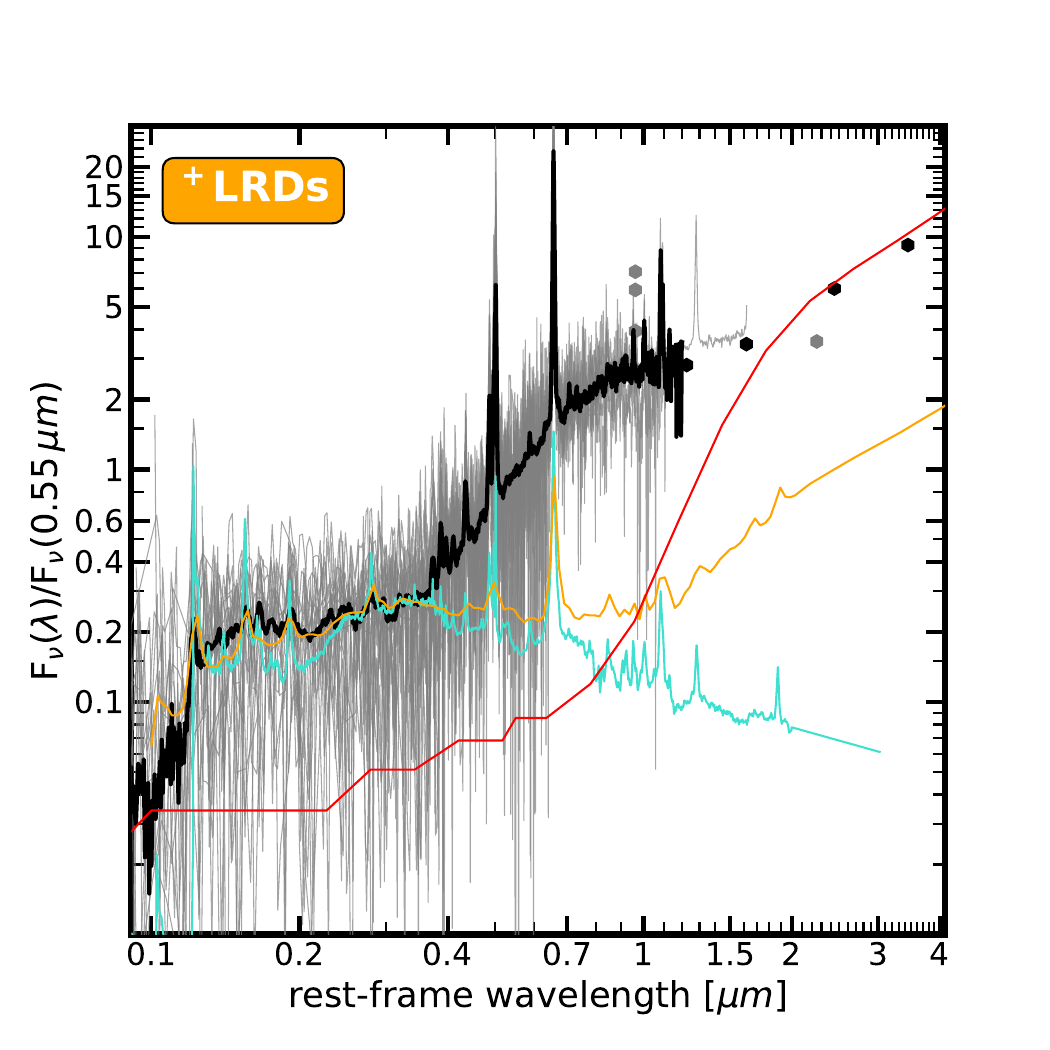}
\includegraphics[clip, trim=0.2cm 0.2cm 1.5cm 2.0cm,width=8.cm,angle=0]{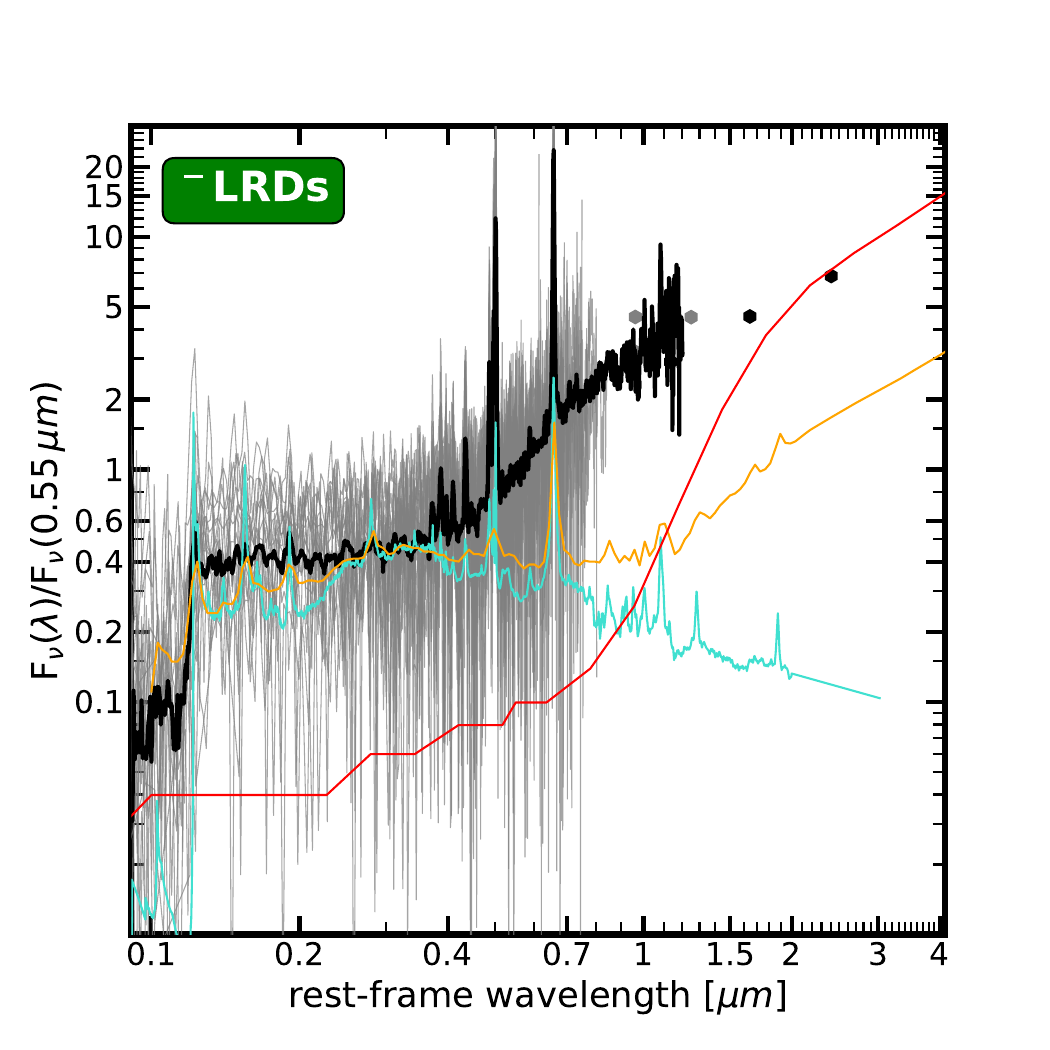}
\includegraphics[clip, trim=0.2cm 0.2cm 1.5cm 2.0cm,width=8.cm,angle=0]{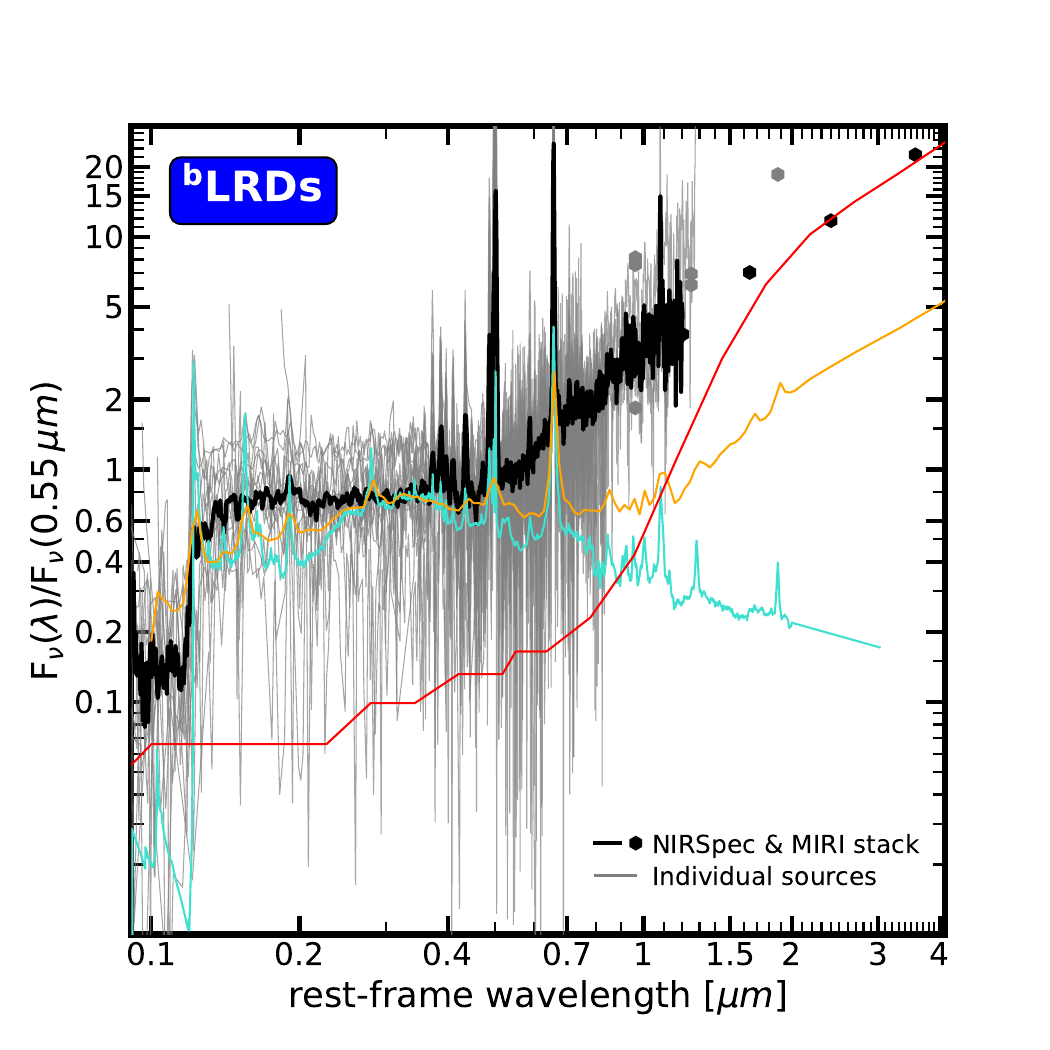}
\caption{\label{fig:scatter}Spectroscopic stacks (black lines), MIRI imaging stacks (black hexagons) and spectra for ten individual LRDs (gray lines) randomly selected from the parent subsamples for the four LRD subtypes defined in Section~\ref{sec:classes}. The plots compare these LRD spectra to the QSO SED from \citet[][normalized to the Mg~II region; turquoise line]{2021MNRAS.508..737T} and the Sy~1 stacked spectra and models from \citet{2019MNRAS.486.4917G}, including (orange; normalized to the Mg~II region) and excluding (red; normalized to the 3~$\mu$m spectral region, as in Figure~\ref{fig:stacked_all_miri}) the accretion disk emission.}
\end{figure*}

Figure~\ref{fig:scatter} shows the stacks for different LRD subtypes presented in Section~\ref{sec:subtypes} compared to  ten individual LRDs randomly selected from the parent spectroscopic sample. We also compare the LRD spectra with some AGN models and stacks.

This figure shows that the UV spectral region for the $\mathrm{^xLRD}$s (upper left panel), which we take in the paper as the LRD core and use to calculate BH masses based on bolometric luminosities, resembles the emission of an accretion disk and BLR, as represented by the QSO template from \citet{2021MNRAS.508..737T} or the Sy~1 stack from \citet{2019MNRAS.486.4917G}, normalized in this plot to the Mg~II region, where the AGN templates peak. The LRD core spectrum only deviates significantly from those templates in the far-UV, at wavelengths bluer than 0.2~$\mu$m, where the $\mathrm{^xLRD}$ stack is redder than the AGN templates. The fluxes in the far-UV for the LRD core stack are a factor 1.5--2.0 fainter than those for the AGN templates. These AGN templates, however, fail to reproduce the optical emission of LRDs by up to a factor of 20.

Concerning the dust torus template (for a Sy~1), if it is set to reproduce the mid-IR emission of LRDs (as shown in Figure~\ref{fig:scatter}) by normalizing at 2--4~$\mu$m, it would not reproduce the LRD core spectrum bluewards of 1.5~$\mu$m. We notice, however, that some contribution to the UV can be expected from light scattered by the torus, as also shown by the \citet{2015A&A...583A.120S} models plotted in Figure~\ref{fig:stacked_all_miri}.

A similar comparison between LRD stacks and AGN templates is provided in the rest of panels in Figure~\ref{fig:scatter} for the other subtypes. The main result from this comparison is that the UV emission from all LRD subtypes is bluer than the AGN templates (dominated at these wavelengths by the accretion disk and BLR). If the near-UV were dominated by the AGN emission, some other component (e.g., star formation, as assumed in our SED fitting exercise) should contribute non-negligibly (at least 30-50\% of the total emission). Related to this point, our comparison in Figure~\ref{fig:scatter} is carried out by normalizing the AGN templates to the emission of LRDs in the Mg~II region. At these wavelengths, the LRD spectra and AGN templates are very similar for the reddest LRDs, presenting both very resemblant continuum shapes and line equivalent widths. However, the comparison worsens as we move to bluer LRDs, which could be explained by a dilution of the contribution of the AGN to the total emission. This would translate to higher relative importance of other components (such as star formation, and maybe also some scattered light from the AGN central engine) in both the near- and specially the far-UV.

%\startlongtable
\begin{deluxetable*}{lcccc}
%\centerwidetable 
%\movetableright=-1in
\tabletypesize{\scriptsize}
\caption{Summary of spectroscopic and photometric observations used in this paper.}
\tablehead{\colhead{Field} &  \colhead{Number of spectra} &\colhead{Number of photometric LRDs} & \colhead{NIRSpec PIDs} & \colhead{MIRI PIDs} %& \colhead{References}
}
\startdata
GDS    & 29 &  63 & 1180, 1210, 1212, 1286, 1287, 2198, 5997, 6541, 8060 & 1180, 1207, 1283, 2516, 3954, 4498, 5279, 5407,
6511 \\
GDN    & 31 &  109 & 1181,  8018 & 1181,
1264, 2926, 4530, 4586, 4762, 5407 \\
UDS    & 84 &  294 & 1215, 2565, 4233, 5224, 6368 & 1657, 1837,  7814 \\
COSMOS & 36 & 159 & 2565, 5224, 6368, 6585 & 1284, 1727, 1762, 1837, 2417, 2775, 3954, 5893, 6595\\
EGS    & 52 &  141 & 1345, 2750, 4106, 4233, 6368 & 1345, 1448, 3794, 4586 \\
A2744  & 17 &   21 & 2561, 3073 & 4530, 6743 \\
\enddata
\tablecomments{\label{tab:obs}Table with information about the data used in the paper, including program IDs (PIDs) for NIRSpec and MIRI observations.}
\end{deluxetable*}
%\startlongtable
\begin{deluxetable*}{lcccccc}
%\centerwidetable 
%\movetableright=-1in
\tabletypesize{\scriptsize}
\caption{Statistics of observational imaging properties of  photometric and spectroscopic samples of LRDs.}
\tablehead{\colhead{Sample (N$_\mathrm{gal}$)} &  \colhead{F444W} &\colhead{F277W-F444W} & \colhead{F115W-F200W} & \colhead{F200W-F444W} &  \colhead{F444W-F770W} & \colhead{F770W-F1800W} 
%\\
%\colhead{-}  &\colhead{-} & \colhead{-} & \colhead{-} & \colhead{-} & \colhead{-} & \colhead{-} & \colhead{-}
}
\startdata
Photometric NIRCam (NC) sample (787) & $25.9_{25.0}^{26.6}$ & $1.2_{0.9}^{1.6}$ & $0.3_{0.0}^{0.6}$ & $1.7_{1.3}^{2.2}$  & $0.3_{-0.1}^{0.7}$ (450) & $1.5_{0.6}^{2.2}$ (275)\\
Spectroscopic NIRSpec (NS) sample: NS (249) & $25.9_{25.0}^{26.5}$ & $1.2_{0.8}^{1.6}$ & $0.3_{0.0}^{0.7}$ & $1.7_{1.4}^{2.3}$  & $0.3_{0.0}^{0.7}$ (143) & $1.2_{0.5}^{2.0}$ (84)\\
\hline
NC $\mathrm{^xLRD}$, $\mathrm{L_{opt}/L_{UV}}>6.3$ (127)         & $25.3_{24.4}^{25.9}$ & $1.6_{1.2}^{2.1}$ & $0.4_{0.0}^{0.8}$ & $2.6_{ 2.5}^{3.0}$ & $0.3^{-0.1}_{0.7}$ (72) & $1.0^{0.2}_{1.7}$ (51)\\
NC $\mathrm{^+LRD}$, $3.1<\mathrm{L_{opt}/L_{UV}}<6.3$ (254) & $25.7_{24.9}^{26.4}$ & $1.4_{1.0}^{1.7}$ & $0.3_{0.0}^{0.8}$ & $2.0_{ 1.8}^{2.2}$ & $0.3^{0.0}_{0.6}$ (145) & $1.1^{0.5}_{1.7}$ (76)\\
NC $\mathrm{^-LRD}$, $1.8<\mathrm{L_{opt}/L_{UV}}<3.1$ (281) & $26.1_{25.3}^{26.7}$ & $1.1_{0.9}^{1.3}$ & $0.2_{0.0}^{0.5}$ & $1.4_{ 1.2}^{1.6}$ & $0.3^{-0.2}_{0.7}$ (152) & $1.7^{1.0}_{2.4}$ (89)\\
NC $\mathrm{^bLRD}$, $\mathrm{L_{opt}/L_{UV}}<1.8$ (130)         & $26.4_{25.2}^{27.0}$ & $1.1_{0.7}^{1.2}$ & $0.3_{-0.1}^{0.5}$ & $1.0_{0.9}^{1.5}$ & $0.5_{-0.1}^{1.2}$ (81) & $2.1_{1.6}^{2.5}$ (59)\\
NS $\mathrm{^xLRD}$, $\mathrm{L_{opt}/L_{UV}}>6.3$ (29)         & $24.5_{23.6}^{25.2}$ & $1.6_{0.7}^{2.0}$ & $0.7_{0.4}^{1.3}$ & $2.5_{2.3}^{3.1}$ & $0.1_{0.0}^{0.6}$ (21) & $0.6_{0.0}^{1.1}$ (11)\\
NS $\mathrm{^+LRD}$, $3.1<\mathrm{L_{opt}/L_{UV}}<6.3$ (74) & $25.6_{24.9}^{26.4}$ & $1.3_{1.0}^{1.7}$ & $0.3_{0.1}^{0.7}$ & $2.1_{1.6}^{2.6}$ & $0.4_{0.1}^{0.8}$ (46) & $0.9_{0.4}^{1.6}$ (26)\\
NS $\mathrm{^-LRD}$, $1.8<\mathrm{L_{opt}/L_{UV}}<3.1$ (77) & $26.3_{25.4}^{26.6}$ & $1.2_{0.9}^{1.4}$ & $0.2_{-0.1}^{0.4}$ & $1.6_{1.3}^{2.0}$ & $0.1_{-0.2}^{0.6}$ (37) & $0.6_{0.3}^{2.4}$ (23)\\
NS $\mathrm{^bLRD}$, $\mathrm{L_{opt}/L_{UV}}<1.8$ (69)         & $26.1_{25.4}^{26.6}$ & $1.1_{0.7}^{1.3}$ & $0.2_{-0.1}^{0.7}$ & $1.4_{ 1.2}^{1.7}$ & $0.4_{-0.1}^{0.8}$ (39) & $2.0_{1.5}^{2.3}$ (24)\\
\enddata
\tablecomments{\label{tab:stats_photobs}Table with statistical observational properties of the sample of LRDs in this paper. We provide medians and quartiles for the F444W magnitude and several colors involving NIRCam and MIRI bands, for the whole sample of LRDs selected in the six fields considered in this paper, the subsample observed by NIRSpec, as well as for several subtypes of LRDs (see main text for definitions) constructed with cuts in optical-to-UV luminosity ratio.}
\end{deluxetable*}

%\startlongtable
\begin{deluxetable*}{lcccc}
%\tabletypesize{\scriptsize}
\caption{Spectroscopic properties of LRDs}
\tablehead{\colhead{Property} & \colhead{$\mathrm{^xLRD}$} & 
\colhead{$\mathrm{^+LRD}$} &
\colhead{$\mathrm{^-LRD}$} & 
\colhead{$\mathrm{^bLRD}$} }
\startdata
Spec-z NS & $5.2_{3.8}^{7.1}$ &  $5.7_{4.0}^{7.4}$ & $6.2_{4.2}^{7.3}$ & $5.9_{4.0}^{7.9}$ \\
Photo-z NC & $6.9_{5.3}^{7.6}$ &  $6.9_{5.3}^{7.8}$ & $7.0_{5.3}^{7.5}$ & $5.2_{3.0}^{7.3}$ \\
$\log\mathrm{L}_\mathrm{5100}$ NS [erg\,s$^{-1}$\,Hz$^{-1}$] & $29.2_{28.9}^{29.5}$ & $28.8_{28.5}^{29.2}$ & $28.7_{28.3}^{29.1}$ & $28.5_{28.1}^{28.9}$\\
$\log\mathrm{L}_\mathrm{5100}$ NC [erg\,s$^{-1}$\,Hz$^{-1}$] & $29.2_{28.9}^{29.5}$ & $29.0_{28.8}^{29.3}$ & $28.9_{28.6}^{29.1}$ & $28.5_{28.2}^{28.9}$\\
%$\nu\mathrm{L}_\mathrm{5100}$ & & & &\\
$\mathrm{L}_\mathrm{5100}/\mathrm{L}_\mathrm{2500}$ NS & $10.4_{7.8}^{18.6}$ & $3.8_{3.3}^{5.1}$ & $2.5_{2.0}^{3.0}$ & $1.4_{0.7}^{1.6}$ \\
$\mathrm{L}_\mathrm{5100}/\mathrm{L}_\mathrm{2500}$ NC & $9.4_{7.3}^{11.8}$ & $4.2_{3.6}^{5.1}$ & $2.4_{2.1}^{2.9}$ & $1.3_{0.9}^{1.6}$ \\
\hline
EW$_\mathrm{0}$(C\,III]$\lambda$1908) [$\AA$] & $18.9\pm5.2$ & $13.3\pm4.6$ & $20.3\pm3.9$ & $11.6\pm4.1$ \\
C\,III]$\lambda$1908/He\,II$\lambda$1640 & $>5.1$ & $1.5\pm0.6$ & $3.2\pm1.7$ & $2.9\pm0.9$\\
%EW (4550\AA) [\AA] & & & &\\
$\log\mathrm{M_T^{WR}}$ [M$_\odot$] & $8.0\pm0.3$ & $7.7\pm0.2$ & $7.8\pm0.2$ & $7.4\pm0.2$\\
He\,I$\lambda$7067/He\,I$\lambda$5877 & $1.7\pm0.3$ & $1.4\pm0.2$ & $1.2\pm0.2$ & $1.0\pm0.1$\\
He\,I$\lambda$10833/He\,I$\lambda$5877 & $10.0\pm1.5$ & $12.0\pm1.1$ & $7.1\pm1.0$ & $7.0\pm0.7$ \\
$[$O\,III]$\lambda$5008/H$\beta$ & $1.1\pm0.1$ & $3.2\pm0.1$ & $4.4\pm0.3$ & $4.9\pm0.5$ \\
H$\alpha$/H$\beta$ & $10.1\pm0.8$ & $8.0\pm0.3$ & $6.4\pm0.3$ & $5.5\pm0.5$ \\
H$\beta$/H$\gamma$ & $8.5\pm2.8$ & $4.4\pm1.0$ & $3.4\pm0.9$ & $3.0\pm1.0$ \\
H$\beta$/H$\delta$ & $13.4\pm3.7$ & $8.3\pm1.3$ & $7.1\pm1.3$ & $5.2\pm1.3$ \\
H$\beta$/Pa$\gamma$ & $2.3\pm0.9$ & $2.6\pm0.5$ & $3.6\pm2.3$ & $3.0\pm1.0$ \\
Balmer break & $3.3\pm0.7$ & $1.4\pm0.3$ & $1.2\pm0.2$ & $1.1\pm0.2$ \\
$[$O\,III]$\lambda$4363/H$\gamma$ & $1.1\pm0.0$ & $1.0\pm0.5$ & $0.6\pm0.0$ & $1.0\pm0.5$ \\
$[$O\,III]$\lambda$5008/$[$O\,II]$\lambda$3728 & $14.8\pm4.6$ & $21.8\pm6.6$ & $23.6\pm4.7$ & $17.5\pm5.4$ \\
$[$Ne\,III]$\lambda$3870/H$\gamma$ & $1.9\pm0.6$ & $1.1\pm0.3$ & $2.3\pm0.5$ & $1.5\pm0.5$ \\
%He\,II$\lambda$4687/H$\beta$ & & & &\\
%L(H$\alpha$) [L$_\odot$] & & & &\\
%L(H$\beta$) [L$_\odot$] & & & &\\
\hline
M$_\bigstar$ [M$_\odot$] & $<8.1$ & $8.3\pm0.3$ & $8.4\pm0.3$ & $8.3\pm0.3$ \\
M$_\mathrm{BH}$ [M$_\odot$] & $6.6\pm0.3$ & $6.4\pm0.3$ & $6.4\pm0.4$ & $6.2\pm0.4$ \\
$\log\mathrm{L}\mathrm{^{IR}_{\bigstar}}$ [L$_\odot$] & -- & $10.8\pm0.2$ & $10.9\pm0.2$ & $10.8\pm0.2$ \\
$\log\mathrm{L}\mathrm{^{bol}_{BH}}$ [erg\,s$^{-1}$] & $44.4\pm0.2$ & $44.2\pm0.3$ & $44.2\pm0.3$ & $44.1\pm0.3$ \\
$A^V_\bigstar$ [mag] & -- & $0.8\pm0.2$ & $0.8\pm0.2$ & $0.9\pm0.2$ \\
$A^V_\mathrm{BH}$ [mag] & -- & $0.4\pm0.1$ & $0.7\pm0.2$ & $1.6\pm0.6$ \\
$\lambda_\mathrm{peak}^{F_\nu}$ [$\mu$m] & $1.1\pm0.1$ & $1.1\pm0.1$ & $0.8\pm0.1$ & $0.9\pm0.1$ \\
$\lambda_\mathrm{peak}^{F_\lambda}$ [$\mu$m] & $0.7\pm0.1$ & $0.7\pm0.1$ & $1.2\pm0.1$ & $2.0\pm0.1$ \\
T$_\mathrm{MBB}$ [K] & $4100\pm100$ & $3200\pm100$ & $3100\pm100$ & $3600\pm200$ \\
$\beta_\mathrm{MBB}$ & $+0.4\pm0.1$ & $+1.2\pm0.1$ & $+0.9\pm0.2$ & $-0.7\pm0.3$ \\
\enddata
\tablecomments{\label{tab:stats_specobs}Table with  spectroscopic information and physical properties of the sample of LRDs in this paper. The first set of properties provide statistical information (median and scatter given as 16\%--84\% quantiles) about the spectroscopic (NS) and photometric (NC) samples. The second set gives line ratios and equivalent widths measured (with their uncertainties) in the stacks (only for spectroscopic samples). The third set provides physical properties inferred from analysis of the stacks (only for spectroscopic samples).}
\end{deluxetable*}

\bibliography{lrd_spectroscopy}{}
\bibliographystyle{aasjournalv7}

\end{document}